\documentclass[titlepage,a4paper,11pt]{book}

\usepackage[lofdepth,lotdepth]{subfig}
\usepackage{graphicx}

\usepackage{algorithm}
\usepackage{algpseudocode}

\usepackage{cite}
\usepackage[cmex10]{amsmath}
\usepackage{amssymb}
\usepackage{gensymb}

\usepackage[T1]{fontenc}

\newcommand{\dif}{{\rm d}}

\newcommand{\dsur}{{\rm d}{\bf s}}

\newcommand{\vJ}{{\bf J}}
\newcommand{\vE}{{\bf E}}
\newcommand{\ve}{{\bf e}}
\newcommand{\vu}{{\bf u}}
\newcommand{\vB}{{\bf B}}
\newcommand{\vD}{{\bf D}}
\newcommand{\vX}{{\bf X}}
\newcommand{\vA}{{\bf A}}
\newcommand{\va}{{\bf a}}
\newcommand{\vH}{{\bf H}}
\newcommand{\vF}{{\bf F}}
\newcommand{\vT}{{\bf T}}
\newcommand{\vS}{{\bf S}}

\newcommand{\vl}{{\bf l}}
\newcommand{\rotT}{\nabla\times{\bf T}}
\newcommand{\rotDT}{\nabla\times\Delta{\bf T}}
\newcommand{\rotDTp}{\nabla' \times\Delta{\bf T}'}
\newcommand{\vr}{{\bf r}}
\newcommand{\vg}{{\bf g}}
\newcommand{\half}{\frac{1}{2}}

\newcommand{\dvoln}{{\rm d}^nr}
\newcommand{\fui}{f^{(u_i)}}
\newcommand{\fuia}{f^{ ( u_i^{(\alpha)} ) }}
\newcommand{\fuij}{f^{(u_iu_j)}}
\newcommand{\fuiajb}{f^{( u_i^{(\alpha)}u_j^{(\beta)} )}}
\newcommand{\fuijb}{f^{( u_iu_j^{(\beta)} )}}
\newcommand{\fuip}{f^{(u_i')}}
\newcommand{\fuiap}{f^{ ( {u_i'}^{(\alpha)} ) }}
\newcommand{\hui}{h^{(u_i)}}
\newcommand{\huia}{h^{ ( u_i^{(\alpha)} ) }}

\begin{document}
\title{Three-dimensional electromagnetic modelling of practical superconductors for power applications}

\author{M. Kapolka% <-this % stops a space
\thanks{}%
}

\markboth{}%
% \markboth{Extended Abstract XX-XX-XX distributed to participants of EUCAS 2015, Lyon, France, September, 2015.}%
{Shell \MakeLowercase{\textit{et al.}}: Bare Demo of IEEEtran.cls for Journals}

%%%%%%%%%%%%%%%%%%%%%%%%%%%%%%%%%%%%%%%%%%%%%%%%%%%%%%%%%%%%%%%%%%%%%%%%%%%%%%%%%%%%%%%%%%%%%%%
%%%%%%%%%%%%%%%%%%%%%%%%%%%%%%%%%%%%%%%%%%%%%%%%%%%%%%%%%%%%%%%%%%%%%%%%%%%%%%%%%%%%%%%%%%%%%%%

\begin{titlepage}
   \begin{center}
       \vspace*{-3cm}
			 {\Large \textbf{SLOVAK UNIVERSITY OF TECHNOLOGY IN BRATISLAVA}}

       \vspace*{0.25cm}
        {\Large \textbf{FACULTY OF ELECTRICAL ENGINEERING AND INFORMATION TECHNOLOGY}}

       \vspace*{1.0cm}
        {\raggedright \textbf{Evidence number: FEI-104400-76956}}

\begin{figure}[hbt!]
\centering
{\includegraphics[trim=0 0 0 0,clip,width=2.5 cm]{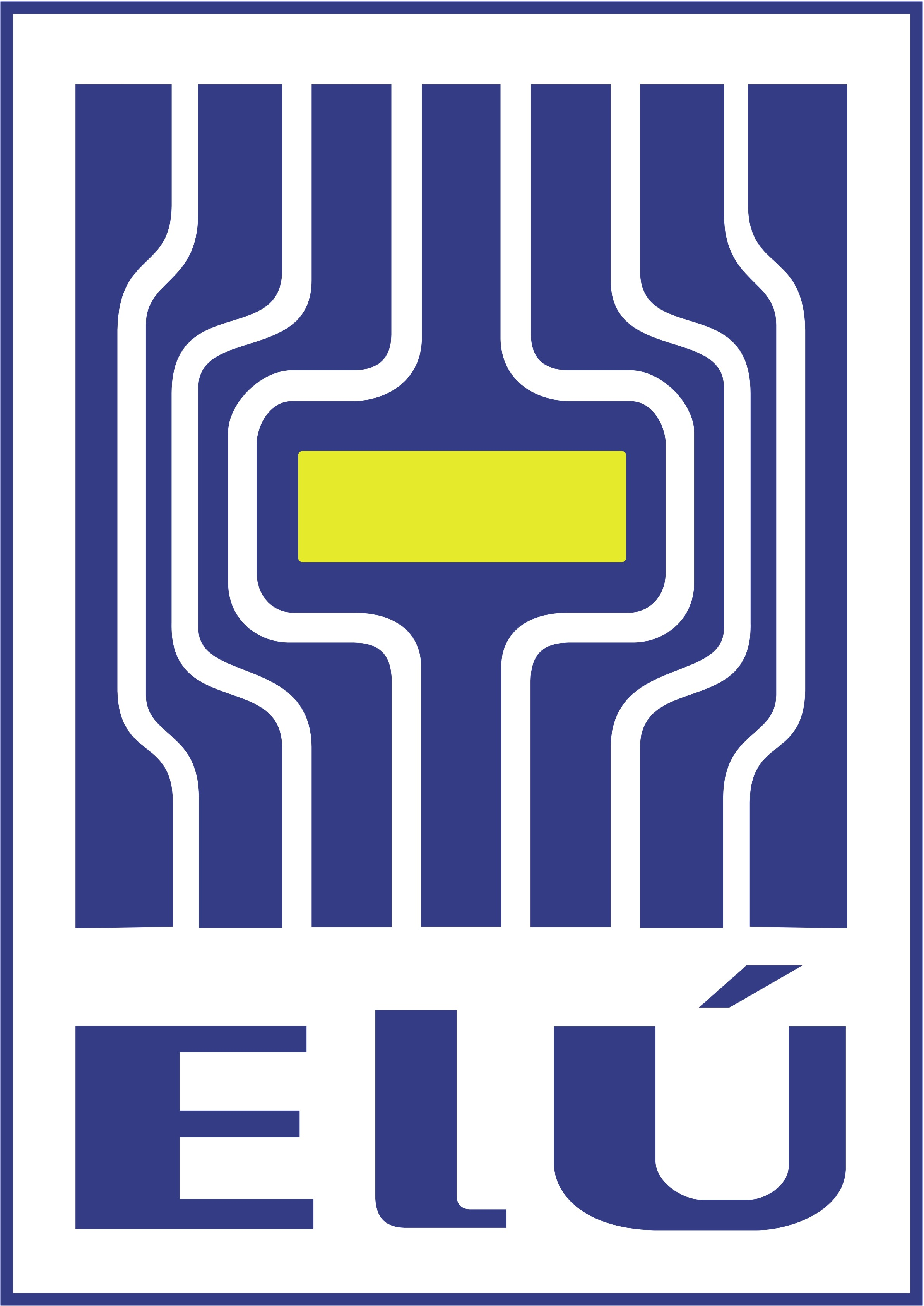}}
\end{figure}

       \vspace*{3.0cm}
 				{\Large \textbf{Three-dimensional electromagnetic modelling of practical superconductors for power applications}}

       \vspace{1.0cm}
       \textbf{Dissertation thesis}

       \vspace{4.0cm}
        \textbf{Study Programme: Physical Engineering}

       \vspace{0.1cm}
        \textbf{Field of Study: 5.2.48 Physical Engineering}

       \vspace{0.1cm}
        \textbf{External Supervising Institution: Institute of Electrical Engineering,}

        \vspace{0.1cm}
        \textbf{Slovak Academy of Sciences}

        \vspace{0.1cm}
        \textbf{Thesis Supervisor: Mgr. Enric Pardo, PhD}

        \vspace{2cm}
				{\Large \textbf{BRATISLAVA 2018 \hspace{20 mm} Ing. Milan Kapolka}}
       \vfill

       \vspace{0.8cm}

   \end{center}
\end{titlepage}

%%%%%%%%%%%%%%%%%%%%%%%%%%%%%%%%%%%%%%%%%%%%%%%%%%%%%%%%%%%%%%%%%%%%%%%%%%%%%%%%%%%%%%%%%%%%%%%
%%%%%%%%%%%%%%%%%%%%%%%%%%%%%%%%%%%%%%%%%%%%%%%%%%%%%%%%%%%%%%%%%%%%%%%%%%%%%%%%%%%%%%%%%%%%%%%

\tableofcontents
\newpage

%%%%%%%%%%%%%%%%%%%%%%%%%%%%%%%%%%%%%%%%%%%%%%%%%%%%%%%%%%%%%%%%%%%%%%%%%%%%%%%%%%%%%%%%%%%%%%%

\chapter*{Acknowledgments}

I would like to express my gratitude to my supervisor Mgr. Enric Pardo, PhD. For leading, help, comments and remarks for this Thesis as well as during the whole PhD study. I also thank Jan Srpcic, Difan Zhou, Mark Ainslie, Anthony Dennis, Francesco Grilli, Victor Zermeno, Shengnan Zou, Antonio Morandi and Leonid Prigozhin for very important and constructive comments and helpful cooperation. 

The Dissertation Thesis was financially supported by the use of computing resources provided by the project SIVVP, ITMS 26230120002 supported by the Research \& Development Operational Programme funded by the ERDF, the financial support of the Grant Agency of the Ministry of Education of the Slovak Republic and the Slovak Academy of Sciences (VEGA) under contract No. 2/0126/15. and No.2/0097/18, and the support by the Slovak Research and Development Agency under the contract No. APVV-14-0438.

%%%%%%%%%%%%%%%%%%%%%%%%%%%%%%%%%%%%%%%%%%%%%%%%%%%%%%%%%%%%%%%%%%%%%%%%%%%%%%%%%%%%%%%%%%%%%%%

\chapter*{Abstrakt}

V\'yvoj vysokoteplotn\'ych supravodi\v cov umo\v znil n\'astup supravodiv\'ych v\'ykonov\'ych aplik\'acii ako s\'u gener\'atory, motory, transform\'atory a v\'ykonov\'e prenosov\'e vedenia. Supravodi\v ce umo\v z\v nuj\'u mnohon\'asobn\'e zv\'y\v senie pr\'udovej hustoty a generovanie magnetick\'eho po\v la v porovnan\'i s norm\'alnymi vodi\v cmi ako je me\v d a hlin\'ik a preto s\'u supravodi\v ce ich jedin\'a alternat\'iva. Pre mnoho v\'ykonov\'ych aplik\'acii ako s\'u ``large-bore" magnety s\'u po\v zadovan\'e vysok\'e v\'ykony, viac ne\v z 10 MW pre vetern\'e turb\'iny a viac ne\v z 1MW pre leteck\'e pohonn\'e motory. Ke\v d\v ze chladiaci syst\'em mus\'i chladi\v t supravodi\v ce medzi teplotami kvapaln\'eho h\'elia a dus\'ika je jeho \'u\v cinnos\v t n\'izka. Kryog\'enny syst\'em spotreb\'uva 10 a\v z 100 kr\'at viac elektrickej energie ako je odvod tepla zo supravodiv\'eho mater\'alu. Supravodi\v ce v striedavom (AC) re\v zime generuj\'u tepeln\'e straty, preto \'urove\v n AC str\'at je d\^{o}le\v zit\'y parameter vo v\'ykonnov\'ych zariadeniach. AC straty z\'avisia na mnoh\'ych faktoroch ako s\'u pr\'udov\'e profily, magnetick\'e polia, geometria supravodiv\'eho vynutia a magnetick\'y material.

Po\v c\'ita\v cov\'e modelovanie je d\^{o}le\v zit\'e, aby odhalilo v\v setky vlastnosti supravodiv\'ych zariadeni, materi\'alove vlastnosti a ich optimaliz\'aciu. Supravodi\v ce s\'u vysoko neline\'arne materi\'aly komplikuj\'uce modelovanie. Modelovac\'i n\'astroj mus\'i by\v t r\'ychly a presn\'y, aby predv\'idal v\v setky efekty. 2D modely vyu\v z\'ivaj\'u symetriu na zni\v zenie stup\v nov vo\v lnosti v prierezovej rovine. Av\v sak pr\'ave preto nem\^{o}\v zu zahrn\'u\v t efekty kone\v cn\'ych rozmerov. Z tohto d\^{o}vodu s\'u potrebn\'e plne 3-rozmern\'e (3D) modelovacie met\'ody, ktor\'e zvl\'adnu obrovsk\'y po\v cet elementov v 3D mrie\v zke. Na\v sa 3D modelovacia met\'oda m\^{o}\v ze modelova\v t \v lubovo\v ln\'u z\'avislos\v t elektrick\'eho po\v la $\vE$ na pr\'udovej hustote $\vJ$, $\vE(\vJ)$, \v co je obrovsk\'a v\'yhoda.

T\'ato pr\'aca je zameran\'e na v\'yvoj plne 3D modelovacieho n\'astroja. Modelovacia met\'oda je zalo\v zen\'a na novej varia\v cnej met\'ode nazvanej ``Minimum Electro Magnetic Entropy Production in 3D", MEMEP 3D, ktor\'a pou\v z\'iva efekt\'ivnu magnetiz\'aciu $\vT$ ako nezn\'amu veli\v cinu. Modelovac\'i n\'astroj je naprogramovany v jazyku C++ so \v strukt\'urou paraleln\'eho programovania. V\'ypo\v ctov\'a met\'oda je overen\'a 2D analytick\'ymi predpovedami a \v dal\v s\'imi 3D modelovac\'imi met\'odami. Modelovacie pr\'ipady s\'u zameran\'e na tenk\'y film a objemov\'u vzorku.   

Efektivita paraleln\'eho programovania je 80\% po otestovan\'i na po\v c\'ita\v covom klastri. V\'ysledky modelovanie na jednoduch\'ych tvaroch potvrdili 2D analytick\'e predpovede pr\'udov\'ych profilov a hyster\'eznych slu\v ciek. MEMEP 3D dosiahol len 3\% odch\'ylku predpoved\'i AC str\'at na dvoch sp\'ajkovan\'ych supravodiv\'ych p\'askach. Pomocou met\'ody sme na\v sli 3D pr\'udov\'e cesty v objemovej vzorke a objavili nenulov\'u $z$ zlo\v zku pr\'udovej hustoty pri nesaturovanom stave. 3D kr\'i\v zov\'y demagnetiza\v cn\'y proces predpovedal asymetriu zachyten\'eho po\v la, ktor\'u potvrdilo meranie. Mo\v znos\v t \v lubovo\v lnej $\vE(\vJ)$ z\'avislosti dovolila \v studova\v t ``force-free" efekty v tenk\'ych filmoch a hranoloch pri nato\v cenom externom magnetickom poli. Modelovanie tak odhalilo v\v setky efekty kone\v cn\'ych rozmerov supravodi\v ca a ``force-free" efekty. 

Celkovo, MEMEP 3D modelovanie potvrdilo u\v zito\v cnos\v t celej met\'ody s vysokou \'u\v cinnos\v tou a kr\'atkym v\'ypo\v ctov\'ym \v casom. Varia\v cn\'a met\'oda zvl\'adla ve\v lmi vysok\'y po\v cet stup\v nov volnosti s overen\'im a\v z do 1 milliona a odhalila nov\'e nezn\'ame supravodiv\'e efekty.

%%%%%%%%%%%%%%%%%%%%%%%%%%%%%%%%%%%%%%%%%%%%%%%%%%%%%%%%%%%%%%%%%%%%%%%%%%%%%%%%%%%%%%%%%%%%%%%

\chapter*{Abstract}

The development of high temperature superconductors opens the road for superconducting power applications such as generators, motors, transformers and power transmission lines. Superconductors allow to drastically increase the current density and generated magnetic field compared to normal conductors like coper or aluminium, and hence they became the only alternative to them for many applications like large-bore magnets, $>$10 MW wind generators and $>$1MW airplane propulsion motors. Since the cryogenic system needs to cool down the superconductor between liquified helium and nitrogen temperature their efficiency is low. The cryogenic system consumes 10-100 times more energy as it removes heat from the superconducting material. Superconductors dissipate energy in AC regime, and hence dissipation of the entire power device is an important feature. The dissipation depends on many factors, such as current distribution, magnetic fields, geometry of the superconducting winding, magnetic materials and others.    

Computer modelling predictions are necessary, in order to reveal all features of superconducting devices or the material properties and optimize them. Superconductors are highly non-linear materials, which complicates modelling. The modelling tool needs to be fast and accurate, in order to predict all effects. 2D models use symmetry, and thence they reduce the degrees of freedom to the cross-sectional planes. However, they cannot include finite size effects. Therefore, full 3D models are needed, which can handle a huge number of elements in the complete 3D mesh. Our 3D modelling method can model any relation between the electric field $\vE$ and current density $\vJ$, $\vE(\vJ)$, which is a big advantage.   

This thesis aims to develop a full 3D modelling tool. The modelling method is based on a novel variational method called Minimum Electro Magnetic Entropy Production MEMEP 3D that uses the effective magnetization $\vT$ as state variable. The modelling tool is written in C++ programming language with parallel computing structure. The modelling method is verified by 2D analytical predictions and other 3D modelling methods. Modelling cases are focused on thin film and bulk samples. 

The parallel computing efficiency is checked by a computer cluster and reached 80\% efficiency. The modelling results on simple samples confirmed 2D analytical predictions of current profiles and hysteresis loops. The MEMEP 3D method presents only 3\% error of AC loss prediction of two soldered superconducting tapes. The method finds 3D current paths of a cubic sample and reveals non-zero $J_z$ component at non-saturated state. The full 3D cross-field demagnetization process predicts asymmetry of the trapped field, which measurements confirmed. The possibility of including any $\vE(\vJ)$ relation allows to study force-free effects in thin films and prisms with tilted applied field angles, which reveals all finite size and force-free effects.       

In conclusion, MEMEP 3D proved usefulness of the entire method with high accuracy and low calculation time of the results. The variational method can handle huge number of degrees of freedom, checked up to 1 million, and reveals new unknown superconducting effects.

%%%%%%%%%%%%%%%%%%%%%%%%%%%%%%%%%%%%%%%%%%%%%%%%%%%%%%%%%%%%%%%%%%%%%%%%%%%%%%%%%%%%%%%%%%%%%%%
%%%%%%%%%%%%%%%%%%%%%%%%%%%%%%%%%%%%%%%%%%%%%%%%%%%%%%%%%%%%%%%%%%%%%%%%%%%%%%%%%%%%%%%%%%%%%%%
%%%%%%%%%%%%%%%%%%%%%%%%%%%%%%%%%%%%%%%%%%%%%%%%%%%%%%%%%%%%%%%%%%%%%%%%%%%%%%%%%%%%%%%%%%%%%%%
%%%%%%%%%%%%%%%%%%%%%%%%%%%%%%%%%%%%%%%%%%%%%%%%%%%%%%%%%%%%%%%%%%%%%%%%%%%%%%%%%%%%%%%%%%%%%%%

\chapter{Introduction}

The continuous improvement of Low and High Temperature Superconductors (LTS and HTS) by increasing critical current density $J_c$ for lower costs opens the room for superconducting power applications such as generators, motors, transformers, wind-turbines or power transmission lines. The superconductor in power applications needs to be cooled down below a certain critical temperature $T_c$, and hence the cryogenic system has to be well optimized. Superconductors present AC loss in AC regime, and thence they dissipate energy. Even small dissipation of energy at helium or nitrogen temperature is problematic, since cryocooling units are inefficient at such temperatures and can cause malfunction of the entire power device. In addition, the magnetization response is useful for material characterization.

Computer modelling can predict the current distribution in superconductors and AC loss. 2D cross-sectional models based on analytical solutions of the Critical State Model CSM or numerical calculations reached maturity. There are many models based on integral methods \cite{brandt95PRL,brandt95PRB,brandt96PRB,rhyner98PhC,costa04SST,morandi15SST,vestgarden08PRB}, Finite Element Method \cite{amemiya16SST,russenschuck99rep,kurz02IES}, variational method \cite{prigozhin96JCP,prigozhin97IES,prigozhin98JCP,prigozhin11SST,HacIacinphase,pancaketheo,pardo15SST,sanchez06JAP,via15SST,ruuskanen14IES,zhangY15SST} or circuit method \cite{vannugteren16IES}. However, analytical solutions can predict current profiles only in simple samples without combinations of transport current and applied magnetic fields. Cross-sectional models cannot include finite size effects, which are important in power devices of finite size. Therefore, full 3D numerical models are required.

3D models include all finite size effects but 3D mesh requires a huge number of degrees of freedom. Therefore, the computation time needs to be low with accurate results. There are many 3D numerical methods like variational methods in $H$ formulation \cite{bossavit94IEM,elliott06JNA,kashima08MNA} or finite element method by $H$ formulation \cite{grilli13Cry,zermeno14SSTa,escamez16IES,stenvall14SST}, $\vA$-$\phi$ vector and scalar potential \cite{lousberg09SST,fagnard16SST,campbell09SST} and $\vT$-$\ohm$ vector and scalar current potential \cite{grilli05IES} or cohomology \cite{stenvall14SST}.

This thesis is focused on the development of a novel 3D modelling tool based on an original variational method. The method is called the Minimum Electro Magnetic Entropy Production MEMEP 3D, which is suitable for tasks of huge number of elements. The MEMEP 3D model is verified by analytical predictions of 2D cross-sectional models and comparison to measurements. The modelling results are focused on the study of thin films and bulk samples, regarding both fundamental and application research. The fundamental study is about the ``force-free" effects \cite{Vlasko15FL,mishev15SST}, since MEMEP 3D can include anisotropic power law for the case of parallel current density and local magnetic field. We also model isotropic rectangular prisms in applied magnetic fields, obtaining an unexpected behavior of the 3D current lines. The application study is of cross-field demagnetization for bulk superconductor magnets \cite{Durrell14SST}. The estimation of AC loss in thin films is important for characterization of tapes.  

%%%%%%%%%%%%%%%%%%%%%%%%%%%%%%%%%%%%%%%%%%%%%%%%%%%%%%%%%%%%%%%%%%%%%%%%%%%%%%%%%%%%%%%%%%%%%%%
%%%%%%%%%%%%%%%%%%%%%%%%%%%%%%%%%%%%%%%%%%%%%%%%%%%%%%%%%%%%%%%%%%%%%%%%%%%%%%%%%%%%%%%%%%%%%%%
%%%%%%%%%%%%%%%%%%%%%%%%%%%%%%%%%%%%%%%%%%%%%%%%%%%%%%%%%%%%%%%%%%%%%%%%%%%%%%%%%%%%%%%%%%%%%%%
%%%%%%%%%%%%%%%%%%%%%%%%%%%%%%%%%%%%%%%%%%%%%%%%%%%%%%%%%%%%%%%%%%%%%%%%%%%%%%%%%%%%%%%%%%%%%%%

\chapter{Background}

The discovery of superconductivity by Kamerlingh Onnes, who liquified helium for the first time and reached temperatures of around 3K, leads to the beginning of a new research field. Later in decades, there have been investigated many superconducting materials, being the most interesting for applications NbTi, NbSn, MgB$_2$, YBCO, Bi2232 and Bi2212. Superconductors have many advantages compared to usual conductors, and hence superconductors are essential for magnets and superconducting power applications started to be promising for commercial use. 

The large-bore magnets can only be made by superconductors. Most magnets are made of Low-Temperature superconductors like NbTi and Nb$_3$Sn, which need liquid helium temperatures to operate (4.2 K) but high-field magnets, generating more than 20 T, require High-Temperature superconductors like REBCO. The critical temperature of several type I and type II superconductors are in table \ref{t.temp}.  Nowadays, commercial tapes of type II and High-Temperature superconductors (with critical temperature above 77 K) are more promising for power applications like generators (such as wind turbines), motors, transformers, power transmission lines and fault-current limiters. However, type II superconductors dissipate energy in AC regime. The cryogenic system is less complex with better efficiency at nitrogen temperature (77 K) than liquid helium temperature (4.2 K). 

In order to improve and explain superconductivity, many theories appeared. Microscopic theories explained the origin of superconducting phenomena. The BCS theory postulated that the electrons form Cooper pairs that condensate into the superconducting state, while the Ginzburg-Landau theory predicted type II superconductors and vortices \cite{tinkham}. The microscopic theories cannot explain phenomenological effects in real size superconductors, since the calculation time of such scale is not feasible. Therefore, macroscopic theories with many assumptions to simplify the calculation are required like the Critical State Model. Analytical predictions are not always possible to find, and hence numerical methods play a significant role to model superconducting power applications. 

There are many numerical formulations, which have been proposed like Finite Element Methods. The variational method is another approach of numerical method. The 2D variational method showed accuracy and short calculation time. However, it needs improvement to the 3D case.   

\begin{table}[tpb]
\begin{center}
\begin{tabular}{llllllll}
\hline
\hline
{Type I} &{Hg} &{Pb} &{In} &{Mo} &{Al} &{Ga} &{Ta} \\
\hline
T$_c$ [K] & 4.2 & 7.2 & 3.4 & 0.92 & 1.2 & 1.1 & 4.48 \\
\hline
\hline
{Type II} &{NbTi} &{Nb$_3$Sn} &{MgB$_2$} &{YBCO} &{Bi2232} &{Bi2212} &{HBCCO}\\
\hline
T$_c$ [K] & 9 & 18 & 39 & 90 & 110 & 85 & 134\\
\hline
\hline
\end{tabular}
\caption{Critical temperatures of several type I and type II superconducting materials.}
\label{t.temp}
\end{center}
\end{table}

%%%%%%%%%%%%%%%%%%%%%%%%%%%%%%%%%%%%%%%%%%%%%%%%%%%%%%%%%%%%%%%%%%%%%%%%%%%%%%%%%%%%%%%%%%%%%%%
%%%%%%%%%%%%%%%%%%%%%%%%%%%%%%%%%%%%%%%%%%%%%%%%%%%%%%%%%%%%%%%%%%%%%%%%%%%%%%%%%%%%%%%%%%%%%%%

\section{Superconducting power applications}
\label{s.power_ap}

Superconductors are an alternative to normal conductors with great potential for power applications. The big advantage of superconductors compared to coper is that they enable high current density. Nowadays, tapes can carry currents up to around $550$ A for a tape width of 10 mm \cite{fujikura}. The width is in the range of 4-12 mm and thickness from 1 to 1.6 $\mu$m. High current density in windings can induce magnetic fields of several tesla \cite{larbalestier14NaM,Kim17SST,Park18IES,Liu16IES}, which is difficult to reach by copper conductor. However, the world record of the highest magnetic field of 45 T is by hybrid magnets (superconducting and resistive) with the coper winding cooled by pressurized water and power consumption around 30 MW\cite{magnet45T}. 

Another aspect of the high current density is the possibility of reducing magnetic iron parts in power devices like motors and generators, and hence reducing the weight of the power device. The weight of the power device is crucial in mobile applications like ship \cite{Yanamoto17IES,Gamble11IES} or aircraft propulsion \cite{Masson07IES,masson13IES}, as well as for superconducting wind turbines for offshore applications \cite{Jeong17IES,abrahamsen10SST,suprapower}, since they have to be as light as possible in order to reduce cost of the tower in the sea. Another area where weight is important are space applications like passive shielding or propulsion, where any additional weight highly increases the cost to bring any spacecraft to the Earth's orbit.

Superconducting power applications reduce the dimensions of the devices. In order to increase the power in the power network in big metropolises, smaller power devices are necessary. Superconducting transformers \cite{Hellmann17IES,Glasson17IES,Schwenterly99IES,Mehta11PhC} can increase the reliability and safety of the electric network and include fault current limiters \cite{Pascal17IES,Xin13IES,Morandi13PhC,souc12SST,Kozak16SST}, either as part of the transformers or as an independent device. Superconducting power lines can reduce the voltage by increasing the transport current. This can simplifying the power network by reducing the number of the step-down transformers. Low-voltage cables can be placed with low distance between them in corridors, and hence increase power-line density even higher. Superconducting cables are also promising for High Voltage DC lines, enabling lower dissipation and higher power. There are many studies of power line cables, such as those in \cite{Volkov16PCS,Yagi15IES}. 

Magnet applications like commercial Magnetic Resonance Imaging MRI magnets \cite{Minervini18SST}, particle accelerators like the LHC in CERN \cite{CERN}, or tokomaks for plasma and fusion research like ITER \cite{iter} put many requirements on superconducting tapes producers. The tapes have to have long length of several 100 m up to km with homogeneous critical current density along the tape and low cost, which for 2G HTS tapes it is in present around 200 EUR per kAm at self-field and 77 K. At present, there are many producers of 2G HTS tape \cite{Chepikov17SST,Lijima15IES,Lee14IES,Rosii16SST,Xu17IES,Lin17AIM,Miura17SST}. 

The disadvantage of superconductors is that they need to be cooled below certain critical temperature $T_c$ (section \ref{s.LHTS}) to become superconducting, and even lower temperature to have useful properties. Low Temperature Superconductors (LTS) usually operate at liquefied helium temperature 4.2 K, while High Temperature Superconductors (HTS) can operate up to 77 K by liquid nitrogen. Middle Temperature Superconductors like MgB$_2$ operate up to 20 K. The cost of liquefied nitrogen is 10 times lower than liquefied helium, and hence HTS superconductors open the door for commercial use of superconducting power applications. Cryocoolers can cool down superconductors to any temperature down to 4 K, but the efficiency at such temperature is very low. In order to remove 1 W of heat at 70 K, cryocoolers need around 20 W of electricity. The efficiency at 4 K is even lower, for 1 W of heat cryocoolers have 100 W of electricity consumption. Therefore, the AC loss (section \ref{s.film_2filament}) have to be as low as possible. Superconducting power applications are promising for the near future but they still are complex systems. 

%%%%%%%%%%%%%%%%%%%%%%%%%%%%%%%%%%%%%%%%%%%%%%%%%%%%%%%%%%%%%%%%%%%%%%%%%%%%%%%%%%%%%%%%%%%%%%%

\section{Type I and type II superconductors}
\label{s.typeI_II}

\begin{figure}[tbp]
\centering
 \subfloat[][]
{\includegraphics[trim=0 0 -20 0,clip,width=5.5 cm]{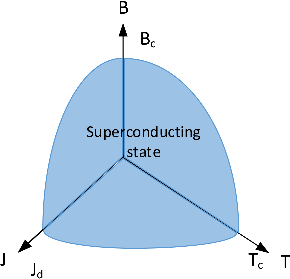}} 
 \subfloat[][]
{\includegraphics[trim=0 0 0 0,clip,width=5.0 cm]{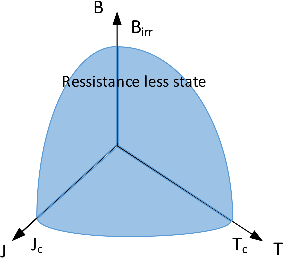}} 
\caption{(a) Type I superconducting state conditions of depairing current density $J_d$, critical temperature $T_c$ and critical magnetic field $B_c$. (b) Type II resistance-less conditions, where $J_c$ is the critical-current density and $B_{\rm{irr}}$ is the irreversibility magnetic field.}
\label{f.typeI}
\end{figure}

The first discovered type of superconductors (type I) has to accomplish three conditions, in order to reach the superconducting state. The superconductor needs to carry transport current or shielding current below a depairing current density $J_d$, the material has to be cooled down below certain critical temperature $T_c$, and the applied magnetic field needs to be below the critical current density $H_c$. Only under the previous three conditions, type I superconducting materials present superconductivity, as it is shown on figure \ref{f.typeI}(a). There are two signs of perfect superconductivity. The superconductor is a perfect diamagnetic with $\mu_r\approx-1$, and hence it shields completely any applied magnetic field below $H_c$. The Meissner shielding current penetrates only a few $\mu$m into the sample with London penetration depth $\lambda$. Measurements show that these Meissner currents are permanent. A superconducting ring with induced shielding current shows no decrease of current density at least for $10^5$ years \cite{tinkham}.

Superconductors are divided among type I and type II superconductors. Another type of superconductors, type II superconductors, have two limits of magnetic field $H_{c1}$ and $H_{c2}$. The applied field below $H_{c1}$ induces Meissner screening current in $\lambda$ penetration depth and the superconductors behave like type I. Applied fields $H_a$ of value $H_{c1}<H_a<H_{c2}$ cause penetration of vortices into the superconductor. Vortices are predicted by Ginzburg-Landau theory inspired by quantum physics. Vortices enable to flow magnetic flux along the normal vortex zone, and hence type II superconductors are still superconducting even under high magnetic fields. The theoretical value of the upper limit of $H_{c2}$ is around 120 T (for REBCO), which is far beyond $H_c$ of typical the type I superconductors, being around 80 mT for Pb. An applied field higher than $H_{c2}$ kills the Cooper pairs and the material looses superconductivity. 

%%%%%%%%%%%%%%%%%%%%%%%%%%%%%%%%%%%%%%%%%%%%%%%%%%%%%%%%%%%%%%%%%%%%%%%%%%%%%%%%%%%%%%%%%%%%%%%

\section{Vortices and vortex pinning}
\label{s.vortices}

\begin{figure}[tbp]
\centering
{\includegraphics[trim=0 0 0 0,clip,width=7.0 cm]{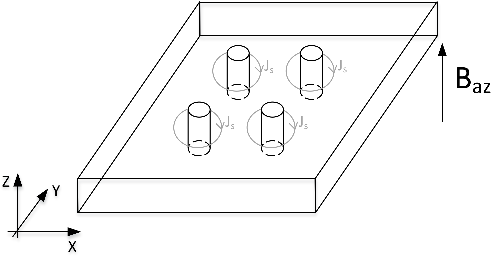}} 
\caption{Votices with normal zone of a tube structure with radius $r\approx\xi$, where $\xi$ is the coherence length, and with superconducting shielding current around it.}
\label{f.vortex}
\end{figure} 

Superconducting vortices in superconductors are explained by the microscopic Ginzburg-Landau theory. The vortex structure consists mainly on a tube with central non-superconducting zone and superconducting current flowing around it. The applied or self magnetic field penetrates through the normal zone in the vortex with value of one fluxon $\theta_0=hc/2e$, where $h$ is Planck's constant, $c$ is the speed of the light and $e$ is the charge of the electron. The vortex line is parallel with the induced magnetic flux. 

In the case when $H_a>H_{c1}$, where $H_a$ is the applied field and $H_{c1}$ is the first critical magnetic field, the vortices start to enter into the sample from the sides oriented parallel to the flux lines. The screening current density $|\vJ|$ creates a driving force on the vortices $\vF_d=\vJ\times\vB$, and hence the vortices start to move. The movement of the vortices induces electric field, which creates dissipation $P=\vJ\cdot\vE$ \cite{Grilli14IES}. 

Pinning centers with non-superconducting zone anchor or pin the vortices to a certain position, and hence pinning centers cancel dissipation. The pinning force $F_p$ is against the driving force $F_d$. If the pinning force is $F_p>F_d$ the vortices do not move, and hence there is no dissipation. When the driving force is $F_d>F_p$ the vortices move and dissipate. The pinning centers improve the superconducting properties. Then, there is a big effort to increase the pinning force in tapes by various additives. The pining centers come from voids, nano-particles, in-homogenities, and defects in crystal lattice like dislocations and twin planes.   

The type II materials are still superconducting for $|\vJ|>J_c$, being $J_c$ the critical current density. For $|\vJ|>J_c$, there is vortex flux flow, which causes AC loss and an effective resistance. The resistance-less conditions are on figure \ref{f.typeI}(b). However, the material is still internally superconducting. The irreversibility field, $B_{\rm{irr}}$, is the maximum magnetic field where vortices are pinned. In LTS, and Mg$_2$B $B_{\rm{irr}}=B_{c2}$. In type II superconductors, there is still superconductivity for $J_d>J>J_c$ and $B_{c2}>B>B_{irr}$ but vortices move for under transport current, creating an effective resistance.

%%%%%%%%%%%%%%%%%%%%%%%%%%%%%%%%%%%%%%%%%%%%%%%%%%%%%%%%%%%%%%%%%%%%%%%%%%%%%%%%%%%%%%%%%%%%%%%

\section{Low- and high- temperature superconductors}
\label{s.LHTS}

Superconductors are classified as well according the critical temperature $T_c$. Low Temperature Superconductors (LTS) are metals and alloys with $T_c$ below liquid hydrogen (20 K). Middle Temperature Superconductors (MTS) are Mg$_2$ or iron-based superconductors with $T_c$ between 20 K and liquid nitrogen temperature (77 K). The last type is High Temperature Superconductors (HTS) with $T_c$ above liquid nitrogen temperature (77 K), which are perovskites containing CuO planes.

After studying several LTS compounds, such as NbCr, NbMo, NbW, NbTa, VNb \cite{rogalla}, in 1961 it was shown that the best two candidates were of NbTi and Nb$_3$Sn. Their critical temperatures are 9 K and 18 K. These LTS are the main workhorses of superconducting magnets up to date. The NbTi superconducting wires were embedded in Cu or Ni stabilization matrix. Further research on NbTi resulted in fine wires produced by Powder-in-Tube (PIT) method. The wires showed performance of $J_c=5\cdot 10^9$ A/m$^2$ at 5 T and 4.2 K, which were used for superconducting magnets up to 12 T. The second important LTS material, Nb$_3$Sn, reached high magnetic fields of 23 T at 4.2 K. The wires are prepared by the Internal Thin (IT) process and results in $J_c=3\cdot 10^9$ A/m$^2$ at 12 T and 4.2 K. The critical magnetic field increased to 26 T by addition of Ti. Although there have been discovered other LTS with $T_c$ up to 25 K, NbTi and Nb$_3$Sn are still dominant due to their high performance at 4.2 K and ease of production.

In 1987, another important discovery was the superconducting material YBa$_2$Cu$_3$O$_7$ with $T_c$=90 K \cite{Jin87APL}, which set the new group of High Temperature Superconductors (HTS). YBCO was prepared by the Oxide Powder in Tube (OPIT) method in bulk samples. Bednorz and Muller already found the first superconductor of the same family in 1986 \cite{Bednorz86PhysB} but it had a critical temperature of only around 35 K. Another HTS material is Bi$_2$Sr$_2$CaCu$_2$O$_8$ (Bi2212) found in 1989 with $T_c=85$ K and up to 45 T \cite{Trociewitz05FL}. Even higher $T_c=110$ K reached (Bi,Pb)$_2$Sr$_2$Ca$_2$Cu$_3$O$_{10}$ (Bi2223), a material first presented in \cite{Michel87PB}. Bi(2212) round wires reached high performance at 4.2 K with $J_c=1.5\cdot 10^8$ A/m$^2$ at 26 T \cite{Heine89APL}. It is a promising material for high field magnets in the range of 30-50 T. Bi2212 wires and Bi2223 tapes are the first generation HTS conductors. 

The second generation 2G tapes are coated conductors of REBCO, where RE is a rare-earth such as Y,Gd or Sm with the following structure: metallic substrate, multifunctional oxide barier, buffer layer, superconducting layer and silver or copper stabilization. The production rate of growing layers is slow with a several meters per hour. Later improvemnt of REBCO structure increased the productiuon rate to 816,4 m of 1 cm width tape with $I_c=572$ A \cite{rogalla}. The production of superconducting tape is done by various growing methods, such as Pulse Laser Deposition (PLD), Atomic Laser Deposition (ALD), Metal Organic Chemical Vapour Deposition (MOCVD), Electron Bean Depositon (EBD) and others. 

A more recent discovery was the MgB$_2$ superconductor in 2001 \cite{Nagamatsu01NTR} with $T_c=39$ K, which sets a new group of middle temperature superconductors. MgB$_2$ wires are prepared by ex-situ, in-situ and Internal Magnesium Diffusion (IMD) process. The ex-situ method mixes the powders of Mg and B and puts them into a metal tube. However, the in-situ method with not reacted powders has shown higher performance by $J_c=10^8$ A/m$^2$ at 13,2 T and 4.2 K \cite{Braccini07SST}. The last production method is IMD. IMD puts boron powder around a magnesium rod and embeds it in an outer metal tube. This method shows the highest performance by $J_c=1.3\cdot 10^7$ A/m$^2$ at 3 T/10 K and $J_c=5\cdot 10^9$ A/m$^2$ at 10 T/4.2 K \cite{Brunner14PCS,kovavc15SST}. The results led the MgB$_2$ as a potential material for power applications of low applied fields, since MgB$_2$ is easy to produce with fine wires and low cost.      

At the end, there are many families of iron-based superconductors such as LaFeAsOF, SmFeAsOF, SrKFeAs and many more, which are promising. However, they did not reach the advantages of REBCO superconductors regarding high $T_c$ and $H_{c2}$.

%%%%%%%%%%%%%%%%%%%%%%%%%%%%%%%%%%%%%%%%%%%%%%%%%%%%%%%%%%%%%%%%%%%%%%%%%%%%%%%%%%%%%%%%%%%%%%%

\section{${\vE(\vJ)}$ relations}
\label{s.EJ}

The most typical $\vE(\vJ)$ relation of practical superconductors can be explained by the collective thermal flux creep \cite{brandt97PRB,blatter94rmp}, which is a microscopic theory. This $\vE(\vJ)$ relation is 
\begin{equation}
{\vE(\vJ)=E_ce^{\frac{-u(\vJ)}{kT}}}
\label{}
\end{equation}  
and
\begin{equation}
{u(\vJ)=u_0\left[\left(\frac{J_c}{|\vJ|}\right)^\alpha-1\right]^{\frac{1}{\alpha}}}, 
\label{}
\end{equation}  
where $E_c$ is the critical electric field with usual value $1\cdot 10^{-4}$ V/cm, $k$ is the Boltzmann constant, $T$ is the temperature, $u$ is the activation energy, $J_c$ is the critical current density, and $u_0$ and $\alpha$ are a constants. Experiments show that $\alpha\ll 1$ for many experimental situations, as is the case of technical superconductors for power and magnet applications \cite{Dekker92PRL}, and hence $u(\vJ)$ and $\vE(\vJ)$ become 
\begin{equation}
{u(\vJ)\approx u_0\ln\left(\frac{J_c}{|\vJ|}\right)} 
\label{}
\end{equation}  
and
\begin{equation}
{\vE(\vJ)=E_c\left(\frac{|\vJ|}{J_c}\right)^n\frac{\vJ}{|\vJ|}},  
\label{e.EJ}
\end{equation}
where $J_c$ is the critical current density and ${n=u_0/kT}$, which depends on the superconducting material. The power law exponent $n$ smoothly bends the $\vE(\vJ)$ curve from $E\approx 0$ at $|\vJ|<J_c$ to $E>>E_c$ at $|\vJ|>J_c$. The $\vE(\vJ)$ relation above is an isotropic power law. 

The smooth power law can include several dependences like $n=n(\vB)$, $J_c(\vr)$ and $J_c(\vB)$, where $\vr$ is a position within the superconductor. These dependencies are necessary in order to make accurate predictions compared to experiments. In the sample, $J_c$ depends on the local magnetic field (see section \ref{s.CSM}), which is not constant along the sample. The superconducting material is not homogeneous, containing defects and cracks, and hence $J_c$ changes with the position. 

\begin{figure}[tbp]
\centering
{\includegraphics[trim=0 0 0 0,clip,width=5.5 cm]{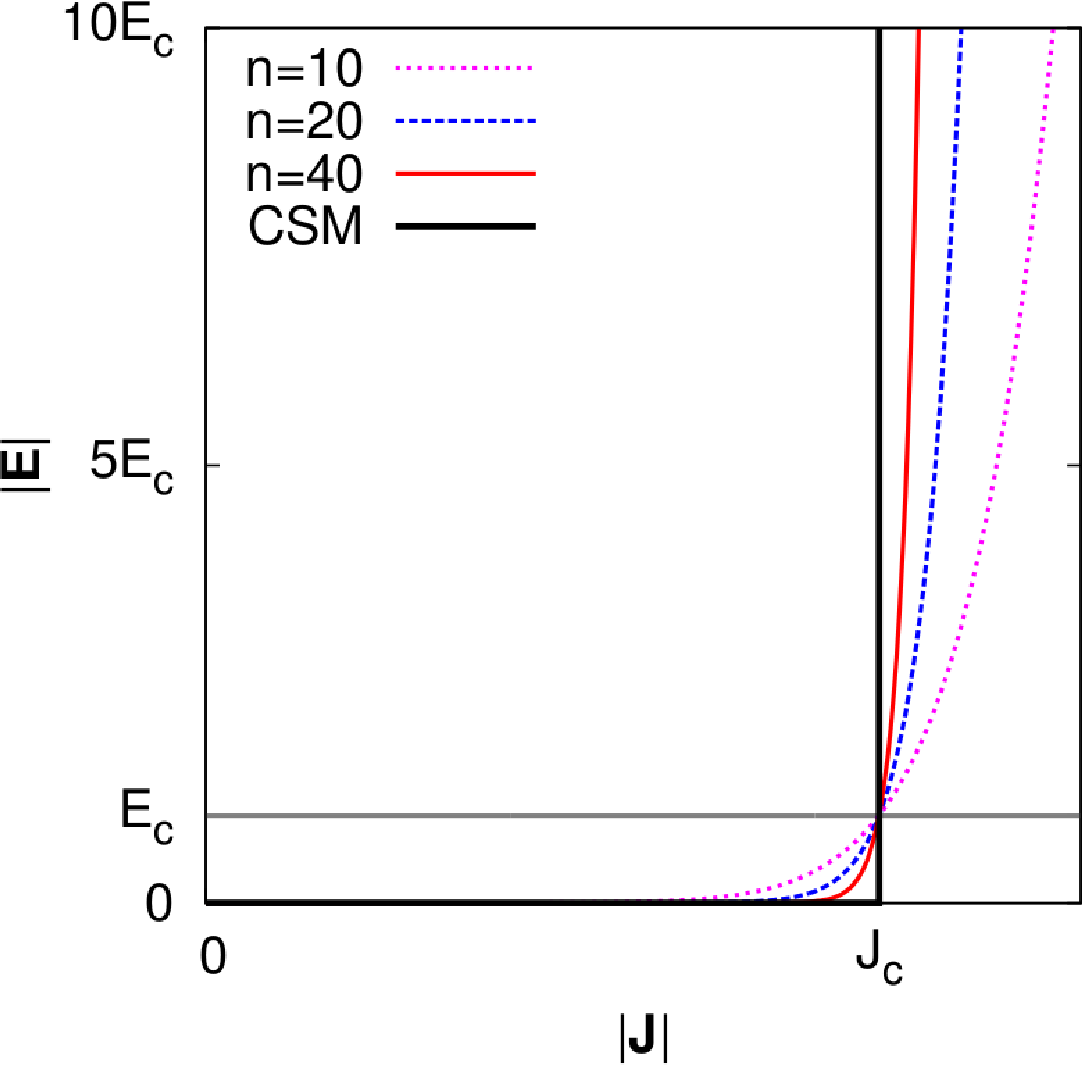}}
\caption{The isotropic power law with various $n$ values and with limit $n\to\infty$, which is like Critical State model approximation.} 
\label{f.EJ_curve}
\end{figure}

The $\vE(\vJ)$ relation is on figure \ref{f.EJ_curve}. The sharpness of the curve depends on the $n$-value, being between $n\to\infty$ and $n=1$. The $\vE(\vJ)$ relation with $n\to\infty$ corresponds to the Critical-State Model CSM approximation (see section \ref{s.CSM} below). The power law with $n=1$ becomes Ohm's law $\vE(\vJ)=(E_c/J_c)\vJ$, where $E_c/J_c=\rho$, which is the resistivity of the linear material. The superconductor models based on the power law are more realistic compared to the CSM model. 

%%%%%%%%%%%%%%%%%%%%%%%%%%%%%%%%%%%%%%%%%%%%%%%%%%%%%%%%%%%%%%%%%%%%%%%%%%%%%%%%%%%%%%%%%%%%%%%

\section{Critical State Model (CSM)}
\label{s.CSM}

The Critical State Model CSM proposed by Bean \cite{bean64RMP} predicts the electromagnetic response of superconductors under uniform applied magnetic fields. This thesis is focused on a thin film and a bulk modelling sample, and therefore we outline the CSM on similar geometries such as infinite thin film and slab. The model proposes analytical solutions for current density and magnetic field profiles inside the sample. The modelling examples shown below are without transport current. However, there exist analytical formulas including transport current and more general geometries of the sample \cite{zeldov94PRB}. The case of finite size samples or the combination of the applied magnetic field and transport current requires numerical calculation. 

\begin{figure}[tbp]
\centering
{\includegraphics[trim=0 0 0 0,clip,width=5.5 cm]{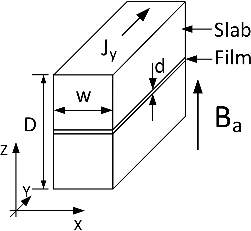}}
\caption{The geometry of the infinitely long thin film and slab with thickness $d$ for thin film, thickness $D$ for slab and width $w$. The applied field is along the $z$ axes.} 
\label{f.CSM_geometry}
\end{figure}

The original CSM is a macroscopic theory with the statement that ``Any electromagnetic force induces a current with constant critical current density $J_c$". The $\vE(\vJ)$ relation (section \ref{s.EJ}) with $n\to\infty$ approaches to the CSM, being $n=100$ often sufficient. The general CSM relation is 
\begin{eqnarray}
 |\vJ| & = & J_c,\qquad {\rm if} |\vE|>0 \nonumber \\ 
       & \le & J_c,\qquad {\rm if} |\vE|=0.
\label{}
\end{eqnarray} 
As stated by Bossavit and Prigozhin \cite{Bossavit94IEG,prigozhin96JCP} the CSM enables $|\vJ|\le J_c$ at $|\vE|=0$, although in bulk samples of translation- or cylindrical-symmetry $|\vJ|=0$ or $J_c$ only. The cause is that long bulks, or cylinders can shield completely the applied magnetic field by currents with $|\vJ|=J_c$. The geometry of the slab and thin film is on figure \ref{f.CSM_geometry}, where $D$ is the slab thickness, $d$ is the film thickness and $w$ is the width of both samples. The applied magnetic field is parallel to the $z$ axis and samples are infinitely long along the $y$ axes. The current density for a slab at the initial magnetization curve \cite{bean64RMP,zeldov94PRB} is           
\begin{eqnarray}
 J_{y}(x) & = & J_{c},\qquad -w/2<x<-a  \nonumber \\ 
          & = & 0,\qquad |x|<a  \nonumber \\
          & = & -J_{c},\qquad a<x<w/2,   
\label{}
\end{eqnarray} 
where ${a=w\left(1-\frac{H_a}{H_p}\right)}$, $H_a$ is the applied magnetic field and $H_p$ is the penetration field $H_p=J_cw/2$.
We use in this thesis terms as ``magnetic flux density" and ``magnetic field" indistinctively, since we do not take magnetic materials into account. We assume that the magnetic field $\vB$ created by superconductors is always only due to superconducting current, and hence $\vB=\mu_0\vH$.   

\begin{figure}[tbp]
\centering
\centering
% \subfloat[][]
{\includegraphics[trim=40 0 40 0,clip,width=5.0 cm]{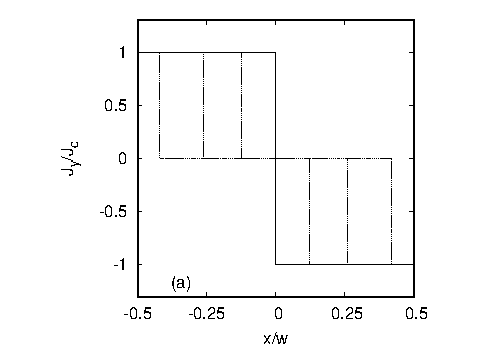}}
% \subfloat[][]
{\includegraphics[trim=40 0 40 0,clip,width=5.0 cm]{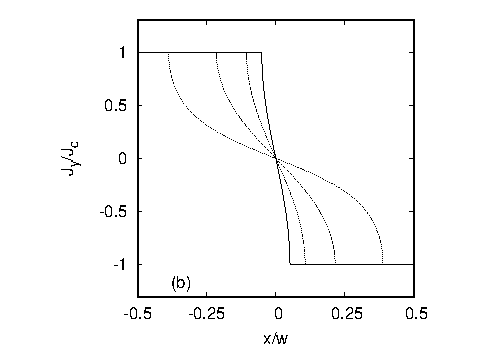}}\\ 
% \subfloat[][]
{\includegraphics[trim=40 0 40 0,clip,width=5.0 cm]{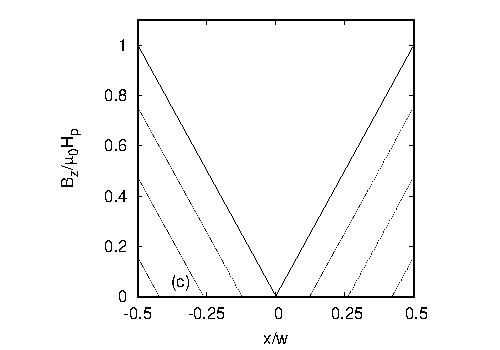}} 
% \subfloat[][]
{\includegraphics[trim=40 0 40 0,clip,width=5.0 cm]{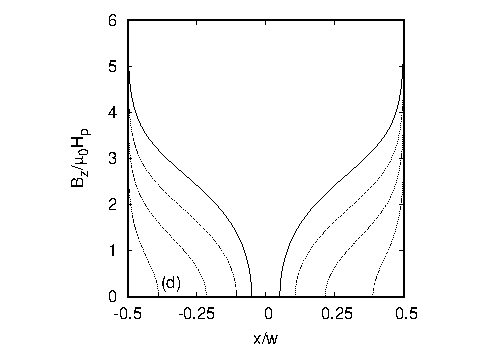}}\\ 
% \subfloat[][]
{\includegraphics[trim=40 0 40 0,clip,width=5.0 cm]{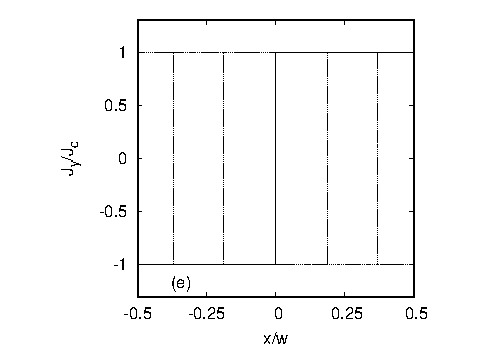}}
% \subfloat[][]
{\includegraphics[trim=40 0 40 0,clip,width=5.0 cm]{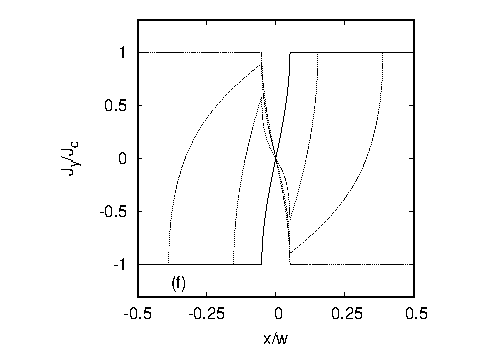}}\\
% \subfloat[][]
{\includegraphics[trim=40 0 40 0,clip,width=5.0 cm]{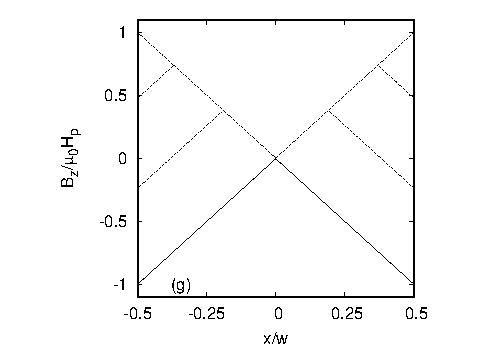}}
% \subfloat[][]
{\includegraphics[trim=40 0 40 0,clip,width=5.0 cm]{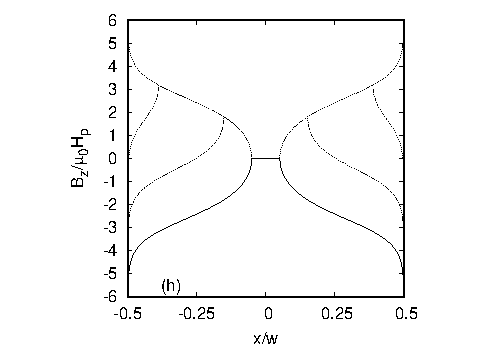}} 
\caption{Critical State Model for a slab (a,c,e,g) and thin film (b,d,f,h). The penetration front of $J_y$ current density (a) in the slab with applied field $B_z/\mu_0H_p=0.25,0.5,75,1$ and (b) in the thin film with $B_z/\mu_0H_p=0.75,1.5,2.25,3$. The penetration of the magnetic flux is with the same applied fields as for the current density for (c) slab and (d) film. The current (e,f) and magnetic field (g,h) penetration at the decreasing curve of the applied fields erase the previous penetrated front with the new front of opposite sign.} 
\label{f.CSM_profile}
\end{figure}

The current penetration to the slab under the applied magnetic field is on figure \ref{f.CSM_profile}(a). The current density penetrates into the sample from the edges with positive sign on the left and negative on the right side. The reason is that the screening current shields the applied field following the ``right hand rule" and opposing to the applied field. The screening current with $|\vJ|=J_c$ in the slab completely shields the applied field in the region with no current density, since the slab thickness allows to induce the necessary $|\vJ|$. Therefore, the penetration of the flux density [figure \ref{f.CSM_profile}(c)] is in the same penetration depth as the screening current density [figure \ref{f.CSM_profile}(a)]. The saturation field fully saturates the slab with the screening current and flux density. Ramping down the applied field induces screening current density with opposite sign, which penetrates into the slab again from the edges. The penetration front rewrites the previous value of the screening current with current of the opposite sign, and hence at remanent state the penetration front reaches only half of the penetration depth. The cause is that the change in current density of the newly induced current is twice $J_c$ instead of only $J_c$, as in the initial curve. The applied field has to reach minus saturation field, in order to fully saturate the slab with screening current and erase the previous current [figure \ref{f.CSM_profile}(e)]. The same behaviour shows the penetration of flux density on figure \ref{f.CSM_profile}(g).        

The AC loss per cycle and sample length \cite{Bean62PRL,Grilli14IESa} is 
\begin{eqnarray}
 Q & = & Dw\frac{2\mu_0H^3_a}{3H_p},\qquad H_a<H_p \nonumber \\ 
   & = & Dw\frac{\mu_0H_p}{3}(6H_a-4H_p),\qquad H_a<H_p  \nonumber \\
   & = & Dw2\mu_0H_pH_a,\qquad H_a>>H_p. 
\label{} 
\end{eqnarray}

The current density of thin film \cite{zeldov94PRB,halse70JPD,brandt93PRBa} is 
\begin{eqnarray}
 J_y(x) & = & \frac{2J_c}{\pi}\arctan{ \frac{cx}{\sqrt{(b^2-x^2)}}},  |x|<b,  \nonumber \\ 
          & = & J_{c}\frac{x}{|x|},   b<|x|<w/2,   
\label{}
\end{eqnarray}
where 
\begin{equation}
{b=\frac{w/2}{\cosh{\frac{H_a}{H_c}}}},   
\label{} \\
\end{equation}  
\begin{equation}
{c=\tanh{\frac{H_a}{H_c}}},   
\label{}
\end{equation}  
and
\begin{equation}
{H_c=\frac{J_cd}{\pi}}.  
\label{}
\end{equation}

The thin film sample shows similar response to the applied magnetic field as the slab. The screening current penetrates into the sample under the applied magnetic field [figure \ref{f.CSM_profile}(b)]. Since the film is very thin, inducing screening current of only value $J_c$ does not generate a uniform magnetic field on the zone with $J=0$. Therefore, there exists screening current density below $J_c$, which penetrates further into the center of the sample. In that zone, named sub-critical zone, the magnetic field is zero. Further increase of the applied field moves the penetration front with $|\vJ|$ around $J_c$ deeper into the sample center. The thin film completely saturates with $|\vJ|=J_c$ with infinite applied field. The thin film penetrates with $|\vJ|=J_c$ in 90\% of the width already with applied field $B_a=\mu_0H_c\cdot\cosh^{-1}(10)=\mu_0 H_c\cdot 2.99$. The flux density penetrates into the sample with the same penetration depth as the screening current density with magnitude $J_c$ [figure \ref{f.CSM_profile}(d)]. The decrease of applied field creates penetration of new screening current with the opposite sign. The new penetration front reaches only half of the penetration depth at the remanent state and saturates the sample under minus saturation field [figure \ref{f.CSM_profile}(f)]. The flux density presents the same penetration behaviour, such as erasing and rewriting the previous flux density, with the flux density of opposite sign [figure \ref{f.CSM_profile}(h)].

The AC loss formula for the thin film per cycle is \cite{halse70JPD}
\begin{equation}
{Q =\frac{8\mu_{0} J_{c}^{2}w^{2}}{\pi} \left[\ln \cosh\left(\frac{\pi H_{a,m}}{J_{c}}\right)-\frac{\pi H_{a,m}}{2J_{c}}\tanh\left(\frac{\pi H_{a,m}}{J_{c}}\right) \right]}, 
\label{} \\
\end{equation}  
where $H_{a,m}$ is the maximum applied magnetic field.

\begin{figure}[tbp]
\centering
{\includegraphics[trim=40 0 50 0,clip,width=6 cm]{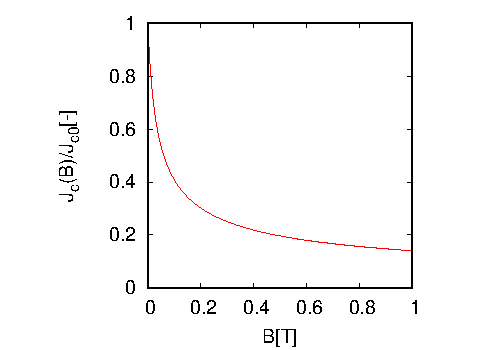}} 
\caption{$J_c(B)$ Kim-like dependence on the local magnetic field with parameters $B_0=$20 mT, $m$=0.5 and $J_{c0}=3.615\cdot 10^{10}$ A/m$^2$.} 
\label{f.CSM_Kim}
\end{figure}

More realistic than the CSM is the Kim model, which introduces a $J_c(B)$ dependence. The critical current density $J_c$ is not constant, but it is reduced by the local magnetic field $\vB$, as it is in real superconductors. The Kim dependence\cite{Kim62PRL,Kim63PR} is 
\begin{equation}
{J_{c}(B)} = \frac{J_{co}}{\left(1+\frac{|\vB|}{B_{0}}\right)^{m}}, 
\label{e.Kim}
\end{equation}  
where $J_{c0}$ is the critical current density at zero local magnetic field, $m$ is a parameter and $B_0$ is a characteristic magnetic field. In this thesis, we choose the following arbitrary parameters as an example of dependence: $B_0=$20 mT, $m$=0.5 and $J_{c0}=3.615\cdot 10^{10}$ A/m$^2$. The reduction of $J_c$ is more than 80\% with the local magnetic field at 0.5 T [figure \ref{f.CSM_Kim}]. 

%%%%%%%%%%%%%%%%%%%%%%%%%%%%%%%%%%%%%%%%%%%%%%%%%%%%%%%%%%%%%%%%%%%%%%%%%%%%%%%%%%%%%%%%%%%%%%%

\section{$\vE(\vJ)$ relation for anisotropic ``force-free" effects}
\label{s.CSM_ani}

The Double Critical State Model DCSM assumes two different limits for the critical current density. In the case when $\vJ$ is perpendicular to the local magnetic field $\vB$, the critical current density becomes $J_{c\perp}$. In the second case with $\vJ$ parallel to $\vB$, the parallel critical current density, $J_{c\parallel}$, applies. Further development of the DCSM results in Elliptic Critical State Model introduced ECSM by Badia and Lopez. The anisotropic $\vE(\vJ)$ power law \cite{badia15SST} based on ECSM is  
\begin{equation}
\vE(\vJ)=2m_0U_0\left[ \left( \frac{J_\parallel}{J_{c\parallel}}\right)^2 + \left( \frac{J_\perp}{J_{c\perp}}\right)^2\right]^{m_{0}-1}\cdot\left(\frac{J_\parallel}{J_{c\parallel}^2}{\ve}_\parallel
+\frac{J_\perp}{J_{c\perp}^2}{\ve}_\perp\right),
\label{e.ani_EJ}
\end{equation}
where ${m_0=(n+1)/2}$, ${U_0=E_cJ_{c\perp}/(n+1)}$, ${J_\parallel=\vJ\cdot\vB/|\vB|}$ and ${J_\perp=|\vJ\times\vB|/|\vB|}$. $\ve_\parallel$ and $\ve_\perp$ are unit vectors, where ${\ve_\parallel=\vB/|\vB|}$ and ${\ve_\perp=\vJ_\perp/|\vJ_\perp|}$. The modelling method is able to include any anisotropic $\vE(\vJ)$ relation, and hence it can model the electromagnetic response of samples with force-free effects.

\begin{figure}[tbp]
\centering
 \subfloat[][]
{\includegraphics[trim=0 0 -20 0,clip,width=4 cm]{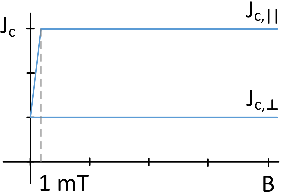}} 
\centering
 \subfloat[][]
{\includegraphics[trim=0 0 0 0,clip,width=4 cm]{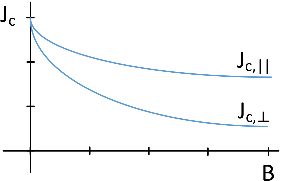}} 
\caption{Double critical state model with $J_c(B)$ dependence (a) with linear drop of $J_{c\parallel}$ to $J_{c\perp}$ and (b) including Kim model.}
\label{f.CSM_ani}
\end{figure}

The problem of the anisotropic power law is the undefined unit vector $\ve_\parallel$ when the local magnetic field is very low or zero. We suggest the following solution, in order to remove the uncertainty of the anisotropic $\vE(\vJ)$ relation. The assumption is that when the local magnetic field is below a certain value $B_{c0}$ (in our case we choose $B_{c0}$=1 mT), $J_{c\parallel}$ is linearly going to $J_{c\perp}$ as it is shown on figure \ref{f.CSM_ani}(a). The case with $B_{c0}\to 0$ exactly corresponds to the elliptic CSM.

Various electromagnetic modelling cases require $J_c(B)$ dependence, and hence we include Kim model for $J_{c\perp}(B)$, and $J_{c\parallel}(B)$ as it is on figure \ref{f.CSM_ani}(b).

%%%%%%%%%%%%%%%%%%%%%%%%%%%%%%%%%%%%%%%%%%%%%%%%%%%%%%%%%%%%%%%%%%%%%%%%%%%%%%%%%%%%%%%%%%%%%%%

\section{Applied vector potential}
\label{s.applied_Aa}

\begin{figure}[tbp]
\centering
 \subfloat[][]
{\includegraphics[trim=0 0 -30 0,clip,width=4 cm]{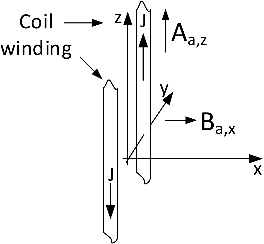}} 
\centering
 \subfloat[][]
{\includegraphics[trim=0 0 -30 0,clip,width=4 cm]{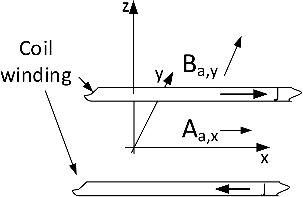}} 
\centering
 \subfloat[][]
{\includegraphics[trim=0 0 0 0,clip,width=3.5 cm]{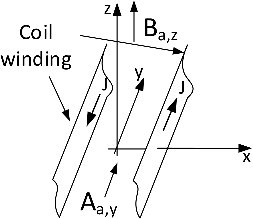}} 
\caption{The orientation of infinitely long external coils generating uniform applied magnetic field in the direction parallel to (a) $x$ axes, (b) $y$ axes and (c) $z$ axes.}
\label{f.Aa_coils}
\end{figure}

The MEMEP 3D method (section \ref{s.math_model}) requires the evaluation of $\vA$. Therefore, the applied vector potential, $\vA_a$, has to be defined well in the modelling tool. The applied vector potential is generated from an external source like a coil or a permanent magnet. The tool uses the interpretation that the applied vector potential is created by an infinitely long external coil, in order to generate the applied vector potential. The model uses only one external coil per component of the vector potential [figure \ref{f.Aa_coils}]. $\vA_a$ magnetizes the superconducting sample and induces screening current. 

We can assume only one component of the applied magnetic field $\vB_a=B_{az}\ve_z$, which caused by the applied vector potential $\vA_a$ by one external coil with infinite $Y$ direction [figure \ref{f.Aa_coils}(c)]. The vector potential is defined by Coulomb's gauge $\nabla\cdot\vA_a=0$, and hence it follows the direction of the current. The vector potential becomes  
\begin{equation}
\vA[\vJ](\vr)=\frac{\mu_0}{4\pi}\int_V\dif V'\frac{\vJ(\vr')}{|\vr -\vr'|}.
\label{}
\end{equation}
Then, if $\vJ(\vr)=J(\vr)\vu$, where $\vu$ is a constant unit vector, $\vA(\vr)=A(\vr)\vu$. 

In general, we can consider $\vA_a$ corresponding to $\vB_a$ in any direction, $\vA_a=A_{ax}\ve_x+A_{ay}\ve_y+A_{az}\ve_z$. According to the definition of vector potential, $\nabla\times \vA_a=\vB_a$  
\begin{eqnarray}
\vB_a=\left( \begin{array}{ccc}
\ve_x & \ve_y & \ve_z \\
\frac{\partial}{\partial x} & \frac{\partial}{\partial y} & \frac{\partial}{\partial z} \\
A_x & A_y & A_z
\end{array} \right)  \nonumber\\
\vB_a=\ve_x\left(\frac{\partial A_z}{\partial y}-\frac{\partial A_y}{\partial z}\right)+\ve_y\left(\frac{\partial A_x}{\partial z}-\frac{\partial A_z}{\partial x}\right)+\ve_z\left(\frac{\partial A_y}{\partial x}-\frac{\partial A_x}{\partial y}\right).
\label{B_components}
\end{eqnarray}
The second part of each magnetic field component $-\frac{\partial A_y}{\partial z}$, $-\frac{\partial A_z}{\partial x}$, $-\frac{\partial A_x}{\partial y}$ are set to zero, since they are generated by infinitely long coils in that direction. The magnetic field becomes $\vB_a=\frac{\partial A_z}{\partial y}\ve_x + \frac{\partial A_x}{\partial z}\ve_y + \frac{\partial A_y}{\partial x}\ve_z$, where the components can be rewritten as  
\begin{eqnarray}
B_{ax}=\frac{\partial A_z}{\partial y}\Rightarrow A_z=B_{ax}y, \nonumber\\
B_{ay}=\frac{\partial A_x}{\partial z}\Rightarrow A_x=B_{ay}z, \nonumber\\
B_{az}=\frac{\partial A_y}{\partial x}\Rightarrow A_y=B_{az}x. \nonumber\\
\label{B_components1}
\end{eqnarray}   
Then, the vector potential results in $\vA_a=B_{ay}z\ve_x+B_{az}x\ve_y+B_{ax}y\ve_z$, and hence 3 split coils infinite in the $x,y,z$ directions generate the vector potential in the $x,y,z$ direction, respectively [figure \ref{f.Aa_coils}].

\begin{figure}[tbp]
\centering
{\includegraphics[trim=0 0 0 0,clip,width=4.0 cm]{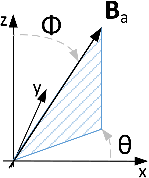}} 
\caption{The applied magnetic field direction is defined by $\phi,\theta$ in the spherical coordinate system.}
\label{f.Aa_angle}
\end{figure}

The applied magnetic field for sinusoidal waveform is defined by 
\begin{equation}
{\vB_a=B_{a\rm m}\sin\left(2\pi\omega f\right)\cdot\va},   
\label{}
\end{equation} 
where $B_{a\rm m}$ is the amplitude of the applied field, $f$ is the frequency and $\va$ is a unit vector in the direction of the applied field. The applied field direction is set by two angles $(\phi,\theta)$ in spherical coordinate system [figure \ref{f.Aa_angle}], where each component is calculated as follows  
\begin{eqnarray}
a_x=\sin{\phi}\cos{\theta}, \nonumber\\	%\alpha_x=\sin{\phi}\cos{\theta}, \nonumber\\
a_y=\sin{\phi}\sin{\theta}, \nonumber\\	%\alpha_y=\cos{\phi}, \nonumber\\
a_z=\cos{\phi}														%\alpha_z=\sin{\phi}\cos{\theta}. \nonumber\\
\label{B_components2}
\end{eqnarray}  
The applied field parameters like $B_{a\rm m},\phi,\theta,f$ are set in the input file (section \ref{c.input_file}), and hence the modelling tool can model any applied magnetic field.

%%%%%%%%%%%%%%%%%%%%%%%%%%%%%%%%%%%%%%%%%%%%%%%%%%%%%%%%%%%%%%%%%%%%%%%%%%%%%%%%%%%%%%%%%%%%%%%

\section{Eddy current problem}
\label{s.formulations}

The main goal of electro dynamic modelling is to find electromagnetic variables like $\vE$, $\vJ$, $\vB$, $\vA$ for any shape of chosen sample geometry with given initial conditions. Normal conductors, such as cooper or aluminium, are linear materials with constant conductivity. On the other hand, superconductors are highly non-linear, and hence to find the required electric variables is problematic. In addition, ferromagnetic materials with non-linear magnetic permeability could also be present. The properties of any material are defined by $\mu$, $\varepsilon$, $\rho$, where $\mu$ is the permeability, $\varepsilon$ is the permittivity and $\rho$ is the resistivity, so that $\vB=\mu(\vH)\vH$, $\vD=\varepsilon(\vE)\vE$, $\vE=\rho(\vJ)\vJ$. The magnetic field created by the superconductor always assumes permittivity of vacuum $\mu_0$, $\vB=\mu_0\vH$. 

Analytical solutions for arbitrary shapes and resistivities do not exist. However, there exist several formulations of Maxwell differential equations, which can find the electromagnetic response of any sample, such as those in the following sections. Several formulations are based on the Finite Element Method \cite{Lahtinen12SST}. Later, we also outline variational principles.      

%%%%%%%%%%%%%%%%%%%%%%%%%%%%%%%%%%%%%%%%%%%%%%%%%%%%%%%%%%%%%%%%%%%%%%%%%%%%%%%%%%%%%%%%%%%%%%%

\subsection{$\vA$-$\phi$-$\vJ$ formulation}
\label{s.A_phi_J}

The eddy current problem can be solved by many kinds of formulations. A common one is the $\vA$-$\phi$-$\vJ$ formulation. The formulation is based on $\vA$ vector potential, $\phi$ scalar potential and $\vJ$ current density. The two main differential equations are derived from Maxwell equations. The formulation starts with the magnetic field, $\vB$,    
\begin{equation}
{\vB = \nabla\times\vA},  
\label{e.rotA} 
\end{equation}
in order to derive the first equation. The Faraday's law is
\begin{equation}
{\nabla\times\vE = -\dot\vB}
\label{}
\end{equation}
and by substitution of magnetic field it becomes 
\begin{equation}
{\nabla\times\vE = -(\nabla\times\dot\vA)}.
\label{}
\end{equation}
The solution of the differential equation is 
\begin{equation}
{\vE = -\dot\vA - \nabla\phi},
\label{e.E}
\end{equation} 
which is the general equation of the electric field equation, $\vE$. This is the first equation of the formulation, where $\nabla\times(\nabla\phi)=0$ for any function $\phi$. 

The second equation definition starts from Ampere's law  
\begin{equation}
{\nabla\times\vH = \vJ + \dot\vD}.   
\label{e.Amper}
\end{equation}
The displacement current is neglected for quasi static model, resulting in 
\begin{equation}
{\nabla\times\vH = \vJ}.
\label{e.HJ}
\end{equation}

The magnetic field for non linear magnetic materials is 
\begin{equation}
{\vH = \mu^{-1}(\vB)\vB},
\label{e.muB}
\end{equation}
where $\mu$ is the anisotropic tensor of the magnetic material. The Ampere's law can be rewritten as 
\begin{equation}
{\vJ = \nabla\times[\mu^{-1}(\vB)\vB]}
\label{}
\end{equation}
and by substituting (\ref{e.rotA}) it becomes
\begin{equation}
{\vJ = \nabla\times\left((\nabla\times\vA)\mu^{-1}(\nabla\times\vA)\right)}.
\label{e.JA}
\end{equation}

The current density defined by the vector potential (\ref{e.JA}) is the second equation for the $\vA$-$\phi$-$\vJ$ formulation. For no magnetic materials and Coulomb's gauge, joining  equation (\ref{e.E}) and (\ref{e.JA}) becomes 
\begin{equation}
{\vE(-\nabla^2\vA)=-\dot\vA-\nabla\phi},
\label{}
\end{equation}
where $\vE(\vJ)$ follows the constitutive relation of the material. The transport current density is defined in the cross-section surface of each superconducting domain like   
\begin{equation}
{\int_{D_i} \vJ\cdot\dif\vS = I_i},
\label{}
\end{equation}
where $D_i$ is the domain of $i$ index and $\dif\vS$ is the differential area of the surface. The formulation has to discretize the air around the sample and solve it. However, the air is assumed as a non-conductive space. The main two differential equations (\ref{e.E}) and (\ref{e.JA}) are defined for each unknown variable $\vX$. The differential algebraic equation is combined to the single matrix equation 
\begin{equation}
{M\dot{\vX}={\bf{f}}(t,{\vX)}},
\label{e.M}
\end{equation}
where $M$ is the mass matrix of the problem at time $t$ and function $f$ depends on both $t$ and $\vX$.   

The solutions are the vector and scalar potentials. The matrix is solved in the time domain by a Partial Differential Equation (PDE) solver. These could be either in commercial software or self-programmed. 

%%%%%%%%%%%%%%%%%%%%%%%%%%%%%%%%%%%%%%%%%%%%%%%%%%%%%%%%%%%%%%%%%%%%%%%%%%%%%%%%%%%%%%%%%%%%%%%

\subsection{$\vT$-${\psi}$ formulation}
\label{s.T_psi}

Another type of eddy-current problem-formulation is $\vT$-${\psi}$, where $\vT$ is the vector current potential and $\psi$ is the scalar current potential. The formulation is based on the $\vT$ variable, which is defined as 
\begin{equation}
{\vJ = \nabla\times\vT}.   
\label{e.rotT}
\end{equation}
When we substitute (\ref{e.rotT}) into $\nabla\cdot\vJ=0$, then the function is always satisfied, since $\nabla\cdot(\nabla\times \vX)=0$ is valid for any function $\vX$. As a result of Ampere's law with neglected displacement current (\ref{e.HJ})and (\ref{e.rotT}),  
\begin{equation}
{\vH = \vT - \nabla\psi}.   
\label{e.T-p}
\end{equation}
These two equations (\ref{e.rotT}) and (\ref{e.T-p}) serve as base for solving the Maxwell equations. 

The Faraday's law with implemented $\vB=\mu\vH$ and $\vE=\rho\vJ$ relations results in 
\begin{equation}
{\nabla\times(\rho(\vJ)\vJ)=-\partial_{t} [\mu(\vH)\cdot\vH]}.
\label{e.1}
\end{equation}
The $\nabla\cdot\vB=0$ can be rewritten in the same way into 
\begin{equation}
{\nabla\cdot[\mu(\vH)\vH]=0}.   
\label{e.2}
\end{equation}
The substitution of (\ref{e.rotT}) and (\ref{e.T-p}) into (\ref{e.1}) and (\ref{e.2}) we get  
\begin{equation}
{\nabla\times[\rho(\nabla\times\vT)(\nabla\times\vT)]=-\partial_{t} [\mu(\vT-\nabla\psi)\cdot(\vT-\nabla\psi)]}   
\label{}
\end{equation}
and
\begin{equation}
{\nabla\cdot[\mu(\vT-\nabla\psi)(\vT-\nabla\psi)]=0}.   
\label{}
\end{equation} 
The previous two differential equations are the main formulation equations. The total current in a certain cross-section $\Omega$ is edge defined as 
\begin{equation}
{\int_{\partial\Omega} \vT\cdot \dif{l} = I},   
\label{}
\end{equation}
where $\dif{l}$ is the differential length of the edge. The formulation creates the matrix $M$ of all equations for chosen geometry in the space and time $t$. The similar matrix equation (\ref{e.M}) have to be solved by the solver in the Finite element method.

%%%%%%%%%%%%%%%%%%%%%%%%%%%%%%%%%%%%%%%%%%%%%%%%%%%%%%%%%%%%%%%%%%%%%%%%%%%%%%%%%%%%%%%%%%%%%%%

\subsection{$\vH$ formulation}
\label{s.H}

The next formulation is the $H$ formulation \cite{badia15SST}, which is based on a magnetic field. There is no gauge for vector and scalar potential. The $H$ formulation becomes popular and widely used, since commercial software like Comsol allows to enter differential equations directly into the PDE solver. 

The formulation is based on Ohm's law 
\begin{equation}
{\vE=\rho(\vJ)\vJ + E_c},   
\label{e.Ohm}
\end{equation}   
where $E_c$ is a critical voltage in the sub-domains. The substitution of Ampere's law with neglected displacement current rewrites Ohm's law into
\begin{equation}
{\vE=\rho(\nabla\times\vH)\nabla\times\vH+E_c}.
\label{}
\end{equation}  
After another substitution by Faraday's law and (\ref{e.muB}) the equation becomes
\begin{equation}
{ - \partial_{t}[\mu(\vH)\vH]=\nabla\times[\rho(\nabla \times\vH)\cdot\nabla\times\vH+E_c]}.
\label{}
\end{equation}
The previous equation is the core of the $H$ formulation. The current is defined in the cross-section of the $\Omega$ domain as
\begin{equation}
{\int_{\partial \Omega}\vH\cdot\dif\vl=I}.
\label{}
\end{equation}
The differential equations are defined in space for any geometry and solved by the solver via Finite Element Method FEM. 

%%%%%%%%%%%%%%%%%%%%%%%%%%%%%%%%%%%%%%%%%%%%%%%%%%%%%%%%%%%%%%%%%%%%%%%%%%%%%%%%%%%%%%%%%%%%%%%

\subsection{Variational principles}
\label{s.variational}

A completely different way of solving the eddy current problem is by the variational method. Any variational method is based on a certain functional. The Euler differential equations of the functional should correspond to the master equation of the electromagnetic quantity derived from Maxwell equations, such as the master equations of the $\vA-\phi-\vJ$, $\vT-\psi$ and $\vH$ above. There has to exist first and second functional derivatives of the functional. The first derivative obtains the differential equations at the extreme of the functional and the second derivative proves the uniqueness of the minimum, and hence uniqueness of the solution. If both conditions are satisfied, the solution of the functional is the same as the solution of the equivalent formulation by the differential equation. 

The functional is solved by a minimization algorithm, such as the maximum gradient method, Golden section search, Downhill simplex method or Powell's method \cite{Press88}. The variational method is written by a certain state variable; which can be defined around or only inside the sample, depending on the formulation and on the physical meaning of the functional. The minimization finds the solution of the unknown variable inside the sample in given boundaries and initial conditions. More details on Variational principles are in section \ref{s.minimi}.	       
 
%%%%%%%%%%%%%%%%%%%%%%%%%%%%%%%%%%%%%%%%%%%%%%%%%%%%%%%%%%%%%%%%%%%%%%%%%%%%%%%%%%%%%%%%%%%%%%%
%%%%%%%%%%%%%%%%%%%%%%%%%%%%%%%%%%%%%%%%%%%%%%%%%%%%%%%%%%%%%%%%%%%%%%%%%%%%%%%%%%%%%%%%%%%%%%%
%%%%%%%%%%%%%%%%%%%%%%%%%%%%%%%%%%%%%%%%%%%%%%%%%%%%%%%%%%%%%%%%%%%%%%%%%%%%%%%%%%%%%%%%%%%%%%%
%%%%%%%%%%%%%%%%%%%%%%%%%%%%%%%%%%%%%%%%%%%%%%%%%%%%%%%%%%%%%%%%%%%%%%%%%%%%%%%%%%%%%%%%%%%%%%%

\chapter{Model and Numerical method}
\label{s.method}

3D modelling tools are necessary, since 2D cross-sectional models cannot includes all finite size effects, and hence the models predictions are not accurate in difficult geometries of the sample. There are many 3D variational formulations like the $\vH$ one suggested by Bossavit \cite{bossavit94IEM}. Further development of the $\vH$ formulation was presented by Elliott \cite{elliott06JNA} and Kashima \cite{kashima08MNA}. A 2D-$\vJ$ formulation was introduced Prigozhin \cite{prigozhin96JCP,prigozhin97IES,prigozhin98JCP} either for infinitely long problems or thin films. Badia and Lopez introduced the Euler-Lagrange formalism for the $\vH$ formulation \cite{badia01PRL,badia12SST}.

The modelling method needs to handle a huge number of degrees of freedom, in order to solve the full 3D model with all finite size effects. Therefore, the calculation time has to be relatively fast even though there are a lot of elements in the mesh.      

Therefore, we focused on the development of a new modelling tool based on the variational method of the Minimum Electro-Magnetic Entropy Production in 3D (MEMEP 3D), which is a fast method. The MEMEP 3D method is based on a new formulation of $\vT$. It is proved by the Euler-equations of the 3D functional that the minimum of the functional is the solution of the Maxwell differential equations. The solution is a minimum and it is unique. This method is valid for any $\vE(\vJ)$ relation, including anisotropic force-free effects.

Any geometry of the sample is discretized and variables are set into the elements in the grid. The minimization solves the functional and finds the solution of the current modelling situation with initial conditions. Since the 3D object contains a huge number of degrees of freedom, the numerical method uses several strategies to speed up calculations, such as parallel computing, sectors and symmetry. Parallel computing uses OpenMP and BoostMPI protocols. The sector method decreases the computing time by reducing the minimized elements into sectors and solve them separately, which is faster than minimizing entire sample at once. Symmetry speeds up the calculation time of the solution, and hence only one quarter or one eight needs to be solved.     

%%%%%%%%%%%%%%%%%%%%%%%%%%%%%%%%%%%%%%%%%%%%%%%%%%%%%%%%%%%%%%%%%%%%%%%%%%%%%%%%%%%%%%%%%%%%%%%

\section{Mathematical model}
\label{s.math_model}

The Minimum Electro-Magnetic Entropy Production in 3D (MEMEP 3D) method is based on the variational method. The main core of the method is a 3D functional (\ref{e.funcdT}), which is solved by minimization. The minimum of the functional is the same as the solution of the general differential equation of the potential (\ref{e.pot}). 
The general electric field, $\vE$, equation is 
\begin{equation}
\vE(\vJ)+\dot\vA+\nabla\phi=0, 
\label{e.pot}
\end{equation}
where $\vA$ is the vector potential, $\phi$ is the scalar potential and $\vJ$ is the current density. The general potential equation can be rewritten by Maxwell equations to different forms, and hence there are many formulations such as $\vA-\phi-\vJ$, $\vT-\psi$ or $\vH$ (section \ref{s.formulations}). The MEMEP 3D uses Coulomb's gauge $\nabla\cdot\vA=0$. Since there are no magnetic materials, the magnetic field follows $\vB=\mu_0\vH$. Since $\nabla\times\vA=\vB$, the Ampere's law (\ref{e.Amper}) becomes
\begin{equation}
\nabla\times\frac{\nabla\times\vA}{\mu_0}=\vJ.
\label{}
\end{equation}
From vector calculus the equation with the double rotor of $\vA$ becomes
\begin{equation}
\frac{\nabla\left(\nabla\cdot\vA\right)-\nabla^2\vA}{\mu_0}=\vJ.
\label{}
\end{equation}
The first term vanishes with Coulomb's gauge $(\nabla\cdot\vA=0)$,
\begin{equation}
\frac{-\nabla^2\vA}{\mu_0}=\vJ.
\label{}
\end{equation}
The vector potential from the differential equation above can be found as the following volume integral of the current density  
\begin{equation}
\vA[\vJ](\vr)=\frac{\mu_0}{4\pi}\int_V\dif V'\frac{\vJ(\vr')}{|\vr -\vr'|}.
\label{}
\end{equation}
The general electric field equation can be rewritten by vector potential as 
\begin{equation}
\vE\left(\frac{-\nabla^2\vA}{\mu_0}\right)+\dot\vA+\nabla\phi=0. 
\label{}
\end{equation}
The time derivative of the vector potential is $\dot\vA\equiv\frac{\partial\vA}{\partial t}\approx\frac{\Delta\vA}{\Delta t}$. We assume that the electric field is time-independent between two time steps. The total vector potential is $\vA=\vA_0+\Delta\vA$, where $\vA_0$ is the vector potential at the previous time step and $\Delta\vA$ is the change between two time steps. The present time is $t=t_0+\Delta t$, where $t_0$ is the time at the previous time step and $\Delta t$ is the change in time between two time steps. The final form of the general electric field equation is 
\begin{equation}
\vE\left(\frac{-\nabla^2\left(\vA_0+\Delta\vA\right)}{\mu_0}\right)+\frac{\Delta\vA}{\Delta t}+\nabla\phi=0, 
\label{e.Egen}
\end{equation}
and Coulomb's gauge is 
\begin{equation}
\nabla\cdot\left(\vA_0+\Delta\vA\right)=0,
\label{e.Coulomb}
\end{equation}
which are differential equations. Equation (\ref{e.Egen}) corresponds to the general equation to the Eddy current problem in $\vA$ formulation (section \ref{s.A_phi_J}). 

The functional is simplified and later proofed that the minimization of the functional is the same as the solution of (\ref{e.Egen},\ref{e.Coulomb}). The functional is defined at each time step as  
\begin{eqnarray}
L[\Delta\vJ]& = & \int_{V}\dif V\left( \frac{1}{2}\Delta\vJ\cdot\frac{\vA[\Delta\vJ]}{\Delta t} \right . \nonumber\\
& + & \left . \Delta\vJ\cdot\frac{\Delta\vA_a}{\Delta t} + U(\vJ_0+\Delta\vJ)+\nabla\phi\cdot(\vJ_0+\Delta\vJ) \right) \nonumber\\
& = & \int_V\dif V\int_V\dif V' \frac{\mu_0}{8\pi\Delta t}\frac{\Delta\vJ\cdot\Delta\vJ'}{|\vr-\vr'|} \nonumber \\
& + & \int_V\dif V \left ( \Delta\vJ\cdot\frac{\Delta\vA_a}{\Delta t} + U(\vJ_0+\Delta\vJ)+\nabla\phi\cdot(\vJ_0+\Delta\vJ) \right),
\label{e.functional}
\end{eqnarray}
where $\vA_a$ is the applied vector potential, $\vJ_0$ is the current density at $t_0$, and the total current density at time $t=t_0+\Delta t$ is $\vJ=\vJ_0+\Delta\vJ$. The dissipation factor is defined as
\begin{equation}
U(\vJ)\equiv\int_0^{\bf J}\dif \vJ'\cdot\vE(\vJ'),
\label{e.UJ}
\end{equation}
and the solution for the isotropic power law of (\ref{e.EJ}) is
\begin{equation}
U(\vJ)=\frac{E_cJ_c}{n+1}\left(\frac{|\vJ|}{J_c}\right)^{n+1},
\label{}
\end{equation}
while for the force-free anisotropic $\vE(\vJ)$ relation of (\ref{e.ani_EJ}), the dissipation factor is 
\begin{equation}
U\left(\vJ,\vB\right)=U_0\left[\left(\frac{J_\parallel}{J_{c\parallel}}\right)^2+\left(\frac{J_\perp}{J_{c\perp}}\right)^2\right]^{m_0}.
\label{}
\end{equation}

The electric field created by the current density is well defined, since $\nabla_{\vJ}\times{\vE}(\vJ)=0$, and hence the line integral of (\ref{e.UJ}) does not depend on the integration path. Any physical $\vE(\vJ)$ relation follows $\nabla_{\vJ}\times{\vE}(\vJ)=0$ due to irreversible thermodynamical principles \cite{badia12SST}. According to the Onsager relations, the differential resistivity matrix should be symmetric, which causes $\nabla_{\vJ}\times\vE=0$. In addition $\nabla_{\vJ}U=\vE(\vJ)$.  

The extreme of the functional is when the first functional derivative is zero $\delta L[\Delta\vJ]=0$. The first variation is 
\begin{equation}
\delta L[\Delta{\bf J}]=\epsilon \left(  \frac{\dif}{\dif\epsilon} L[\Delta\vJ+\epsilon\vg ] \right)_{\epsilon=0},
\label{}
\end{equation}
where $\epsilon$ is an arbitrary small parameter and $\vg$ is an arbitrary function with continuous second derivatives except at the sample surface and equal zero outside the sample. The first variation of the functional becomes (\ref{e.dvi})
\begin{eqnarray}
\delta L[\Delta\vJ] & = & \epsilon \int_V\dif V \vg\cdot \int_V \dif V' \frac{\mu_0}{4\pi\Delta t} \frac{\Delta\vJ'}{|\vr-\vr'|} \nonumber \\
& + & \epsilon\int_V\dif V \vg\cdot \left( { \frac{\Delta \vA_a}{\Delta t} + \vE(\vJ_0+\Delta\vJ) + \nabla\phi } \right ) \nonumber \\
& = & \epsilon \int_V\dif V\vg\cdot \left( { \frac{\vA[\Delta\vJ]+\Delta \vA_a}{\Delta t} + \vE(\vJ_0+\Delta\vJ) + \nabla\phi } \right).
\label{}
\end{eqnarray}
The Euler equation (appendix \ref{s.appendix_B}) applies on the functional. Then, 
\begin{equation}
\vE(\vJ_0+\Delta J)+\frac{(\vA[\Delta\vJ]+\Delta\vA_a)}{\Delta t}+\nabla\phi=0,
\end{equation}
which corresponds to the extreme $\delta L[\Delta\vJ]$=0. The functional is the same as the general potential equation (\ref{e.Egen}). Therefore, the minimization of (\ref{e.functional}) is the same as solving the differential equation (\ref{e.Egen}). The extreme of the functional is minimum and the minimum is unique when the second derivative is always positive, $\delta^2 L[\Delta\vJ]>0$. The second variation of the functional is defined as 
\begin{equation}
\delta^2 L\equiv \half \epsilon^2 \left ( \frac{\dif^2}{\dif\epsilon^2} L[\Delta\vJ+\epsilon\vg ] \right )_{\epsilon=0},
\label{}
\end{equation}
and following the formulas in appendix \ref{s.appendix_B} it becomes 
\begin{eqnarray}
\delta^2L[\Delta\vJ] & = & \half\epsilon^2\int_V\dif V\int_V \dif V' \frac{\mu_0}{4\pi\Delta t}\frac{\vg(\vr)\cdot\vg(\vr')}{|\vr-\vr'|} \nonumber \\
& + & \half\epsilon^2\int_V\dif V\vg(\vr)\overline{\overline{\rho}}(\vJ_0+\Delta\vJ)\vg(\vr).
\label{}
\end{eqnarray}
The first term is the magnetic interaction energy, which is always positive, and $\bar{\bar\rho}$ is the differential resistivity matrix, which is always positive definite due to thermodynamical principles. A matrix is $\bar{\bar M}$ positive definite if $\upsilon^T\bar{\bar{M}}\upsilon$ is always positive, where $\upsilon$ is a vector. Then, $\delta^2L>0$ always.

The functional contains the scalar potential, which is unknown during the minimization of $L[\Delta\vJ]$. The possible solution is to use a second functional of the scalar potential 
and proof the extreme of it. The functional is 
\begin{equation}
L[\Delta\vJ]= \int_V\dif V \nabla\phi\cdot(\vJ_0+\Delta\vJ).
\label{Lphi}
\end{equation}
The first functional derivative of the scalar potential functional is  
\begin{equation}
\delta L[\phi]=\epsilon \frac{\dif}{\dif\epsilon}L[\phi+\epsilon g]=\int_V\dif Vg \nabla\cdot(\vJ_0+\Delta\vJ).
\end{equation}
The Euler equation of this functional is $\nabla\cdot\vJ=0$. Then, the extreme of the functional imposes current conservation. The problem of the scalar potential functional is that the second derivative is zero, and hence the extreme of functional is not a minimum. Therefore, the functional cannot be split and solved by minimization separately.  

Another approach to solve this problem is to use a different formulation. $\vT$ is defined as the effective magnetization, so that 
\begin{equation}
\vJ=\nabla\times\vT.
\end{equation}
Since there are no surface currents, the tangential components of $\vT$ on the surface needs to be continuous. Since the effective magnetization vanishes outside the sample, the tangential component of $\vT$ vanishes on the surface. 
The total current density inside the sample is the current from magnetization and transport current density $\vJ_t$,
\begin{equation}
\vJ=\nabla\times\vT+\vJ_t.
\label{}
\end{equation}
The total current crossing the outer surface of the sample is due to transport current, since the effective magnetization creates no net current 
\begin{equation}
I=\int_S\dsur\cdot\vJ=\int_S\dsur\cdot\vJ_t.
\label{	}
\end{equation}
The functional with $\vT$ formulation becomes
\begin{eqnarray}
L[\Delta\vT] & = & \int_V\dif V\left (  \half \rotDT\cdot\frac{\vA[\rotDT]}{\Delta t} + \rotDT\cdot\frac{(\Delta\vA_a+\Delta\vA_t)}{\Delta t} \right . \nonumber\\
& + & U(\vJ_0+\Delta\vJ_t+\rotDT)+\nabla\phi\cdot(\vJ_0+\Delta\vJ_t+\rotDT) \Bigg),
\label{}
\end{eqnarray}
where $\vA_t$ is the vector potential created by $\Delta\vJ_t$. The last term contains the scalar potential, which according to vector calculus follows 
\begin{eqnarray}
&& \int_V\dif V \nabla\phi\cdot(\vJ_0+\Delta\vJ_t+\rotDT)=\int_V\dif V \nabla\phi\cdot(\vJ_t+\rotT) \nonumber \\
&& =\int_{S_i}\dsur\cdot(\phi\vJ_t) + \int_{S_o}\dsur\cdot(\phi\vJ_t).
\label{}
\end{eqnarray}
The integral domains are $S_i,S_o$, which are the surfaces of input and output of $\vJ_t$. This term does not depend on the effective magnetization, and hence it can be dropped out from the functional. If we choose $S_i$ and $S_o$ as equipotentials, this integral turns into $\Delta\phi I_t$, where $\Delta\phi$ is the voltage drop between the sample ends. In conclusion, the complete functional with $\vT$ formulation is 
\begin{eqnarray}
L[\Delta\vT] & = & \int_V\dif V \Bigg( \rotDT\cdot\frac{\vA[\rotDT]}{2\Delta t} \nonumber \\
& + & \rotDT\cdot\frac{(\Delta\vA_a+\Delta\vA_t)}{\Delta t} + U(\vJ_0+\Delta\vJ_t+\rotDT) \Bigg) \nonumber \\
& = & \int_V\dif V \Bigg( \rotDT\cdot\frac{(\Delta\vA_a+\Delta\vA_t)}{\Delta t} + U(\vJ_0+\Delta\vJ_t+\rotDT) \Bigg) \nonumber\\
& + & \int_V\dif V\int_V\dif V' \frac{\mu_0}{8\pi\Delta t} \frac{(\rotDT)\cdot(\rotDTp)}{|\vr-\vr'|},
\label{e.funcdT}
\end{eqnarray}
which is the main core of the entire MEMEP 3D method with $\vT$ as unknown variable. The minimum of the functional corresponds to the physical entropy production. Then, the minimum of the functional corresponds to the minimum of the entropy production. This gives the name of the method, as Minimum Electro-Magnetic Entropy Production \cite{badia01PRL,pardo15SST}. 

%%%%%%%%%%%%%%%%%%%%%%%%%%%%%%%%%%%%%%%%%%%%%%%%%%%%%%%%%%%%%%%%%%%%%%%%%%%%%%%%%%%%%%%%%%%%%%

\section{Discretization}
\label{s.discretization}

\begin{figure}[tbp]
\centering
 \subfloat[][]
{\includegraphics[trim=0 0 -20 0,clip,width=4.0 cm]{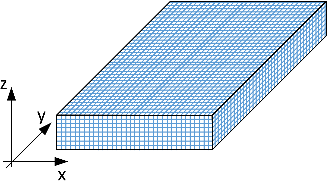}} 
 \subfloat[][]
{\includegraphics[trim=0 0 0 0,clip,width=4.0 cm]{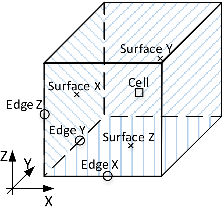}} 
\caption{(a) The mesh inside the modelling sample after discretization. (b) The cell with all elements such as surfaces and edges of type $(X,Y,Z)$. $J_x$, $J_y$ and $J_z$ are assumed constant in surface $X,Y,Z$, respectively. The same applies for $T_x$, $T_y$ and $T_z$ for edges $X,Y,Z$.}
\label{f.disc_geometry}
\end{figure}

\begin{figure}[tbp]
\centering
 \subfloat[][]
{\includegraphics[trim=0 0 0 0,clip,width=5.0 cm]{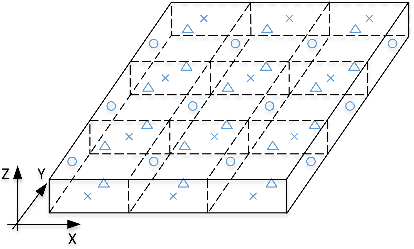}} 
 \subfloat[][]
{\includegraphics[trim=0 0 0 0,clip,width=5.0 cm]{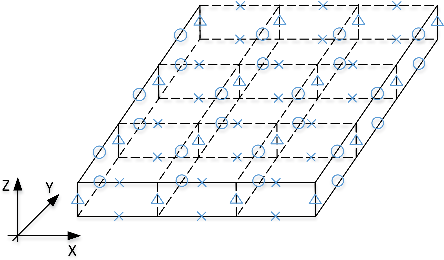}} 
\caption{The mesh elements inside the sample volume. The dicretization creates (a) 3 kinds of surfaces $(X,Y,Z)$ and (b) 3 kinds of the edges $(X,Y,Z)$.}
\label{f.disc_elements}
\end{figure}

The modelling tool needs to create mesh according to the sample size specifications in the input file. The discretization process creates an orthogonal mesh in the $x,y,z$ directions and saves the variables into the data structures. In this thesis, we consider only uniform mesh but the program is also prepared for non-uniform mesh. The MEMEP 3D method avoids taking variables in the air around the modelling sample, and hence the discretization creates a mesh only inside the sample. The functional (section \ref{s.math_model}) contains a lot of variables, which have to be dedicated in the correct positions inside the mesh, in order to evaluate properly $\nabla\times\vT$ and other quantities. The sample is split into a lot of small elements [figure \ref{f.disc_geometry}(a)]. The current state of the modelling tool uses rectangular or square prisms, called cells [figure \ref{f.disc_geometry}(b)]. The code uses many types of elements such as cells, three kinds of surfaces $(X,Y,Z)$ and three kinds of edges $(X,Y,Z)$ [figure \ref{f.disc_geometry}(b)]. The normal of each kind of surface is parallel to the axis of its label ($X$ surfaces are perpendicular to the $x$ axis, $Y$ to $y$, and $Z$ to $z$) [figure \ref{f.disc_elements}(a)]. Each kind of edge is parallel to the axis of its label ($X$ edges are parallel to $x$ axis, $Y$ to $y$, and $Z$ to $z$) [figure \ref{f.disc_elements}(b)]. 

Each type of element contains many parameters and variables saved in the RAM memory during the calculation time. Apart from geometrical variables, the cells contain the dissipation factor $U$, power law value $n$, critical current densities $J_c, J_{c\perp}, J_{c\parallel}$, magnetic field $\vB$, electric field $\vE$, and interpolated current density $\vJ_{\rm{int}}$ and effective magnetization $\vT_{\rm{int}}$. Each surface contains the address of the 4 neighbour edges, addresses of 8 equivalent surfaces for symmetry, $(J_0, \Delta J, J)$, and vector potential perpendicular to the surface $(A_0,\Delta A,A)$. Each type of edge contains the address of the 4 neighbour surfaces, address of 4 neighbour cells, address of 8 equivalent edges for symmetry, and effective magnetization along the edge $(T_0,\Delta T, T)$. All variables are listed in table \ref{t.disc}. 

\begin{table}[tpb]
\begin{center}
\begin{tabular}{ll}
\hline
\hline
{\bf Elements} & {\bf Variables}\\
\hline
Cell & $\vr_c$, $V$, address $(i,j,k)$,$(a,b,c)$, $U$, $n$, $J_{c\perp}$, $J_{c\parallel}$, $\vB$, $\vE$, $\vJ_{\rm{int}}$, $\vT_{\rm{int}}$ \\
Surface & $\vr_c$, address $(i,j,k)$, $(a,b)$,$S$, $J$, $\Delta J$, $J_0$, $V_i$, $A$, $\Delta A$, $A_0$ \\
Edge & $\vr_c$, $l$, address $(i,j,k)$, $T$, $\Delta T$, $T_0$\\
\hline
\hline
\end{tabular}
\caption{Variables in the elements stored in the memory during the minimization process, where $\vr_c$ is the central position vector, $V$ is the volume, $(i,j,k)$ is the address in the mesh, $(a,b,c)$ are the sizes of the edges, $U$ is the dissipation factor, $n$ is the power law exponent, $J_{c\perp}$ is the perpendicular critical current density, $J_{c\parallel}$ is the parallel current density, $\vB$ is the magnetic field, $\vE$ is the electric field, $\vJ_{\rm{int}}$ interpolated current density from surfaces, $\vT_{\rm{int}}$ interpolated effective magnetization from the edges, $S$ is the surface area, $J$, $J_0$ are the current density at the time $t$, $t_0$, respectively, $\Delta J=J-J_0$, $V_i$ is the volume of influence, $A$, $A_0$ are the vector potentials at appropriate time, $\Delta A=A-A_0$, $l$ is the length of edge, $T$, $T_0$ are the effective magnetization at appropriate time, and $\Delta T=T-T_0$.}
\label{t.disc}
\end{center}
\end{table}

\begin{figure}[tbp]
\centering
 \subfloat[][]
{\includegraphics[trim=0 0 0 0,clip,width=5.0 cm]{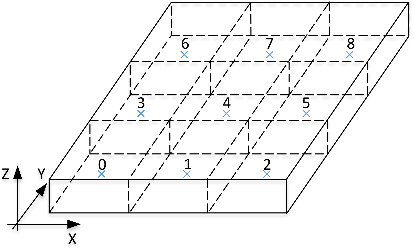}} 
 \subfloat[][]
{\includegraphics[trim=0 0 0 0,clip,width=5.0 cm]{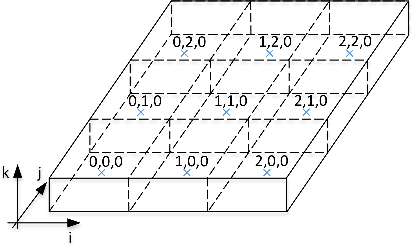}} 
\caption{The cells address by (a) sequence order and by (b) coordinate system with $(i,j,k)$ indexes.}
\label{f.disc_indexes}
\end{figure}

There are two types of addresses for each type of elements, in order to access the variable values faster and easier. The first type of address is on figure \ref{f.disc_indexes} (a); where the elements are cells. This cell address is by sequence order. The second type of the address is according the $(i,j,k)$ coordinate system figure \ref{f.disc_indexes}(b). The sequence address is suitable for code loops with all elements of one type. The coordinate address is suitable in the case of filtering any elements by any restrictions, as well as to easily determine neighbour elements. 

The functional in integral form (\ref{e.funcdT}) is changed by the discretization to the following form with $\vJ$ as variable 
\begin{eqnarray}
L[\Delta\vJ] & = & \frac{1}{2\Delta t}\sum_{s\in\{x,y,z\}}\sum_{i,j=1}^{n_s}V_{si}V_{sj}\Delta J_{si}\Delta J_{sj}a_{sij} \nonumber \\
& + & \sum_{s\in\{x,y,z\}}\sum_{i=1}^{n_s}V_{si}\Delta J_{si}\Delta A_{a,si} + \sum_{c=1}^{n_c}V_{c}U_c ,
\label{e.sumJ}
\end{eqnarray}
where $s$ is the type of the surface, $i,j$ are the surface indexes, $c$ is the cell, $n_c$ the total number of cells, $a_{sij}$ is the average vector potential and $A_{a,si}$ is the applied vector potential at surface type $s$ and index $i$. We can set all components of the average vector potential, $a_{sij}$, as the interaction matrix.

%%%%%%%%%%%%%%%%%%%%%%%%%%%%%%%%%%%%%%%%%%%%%%%%%%%%%%%%%%%%%%%%%%%%%%%%%%%%%%%%%%%%%%%%%%%%%%

\section{Thin film approximation and stacks of many tapes}
\label{s.film_approx}

Second generation HTS superconducting tapes are produced as thin films. In addition, thin films such as REBCO are used in RF cavities and electronics, in order to drastically improve their properties. Therefore, a thin film model is very useful and necessary to explain all effects. The thin film model contains only one layer of cells in the thickness, and hence the current can flow only in the $z$ plane with $J_x$ and $J_y$ components. The missing $J_z$ component and the fact that no current crosses the sample surface reduces $\vT=(T_x,T_y,T_z)$ to only the $T_z$ component, since the tangential component of $\vT$ at the sample boundary vanishes. The reduced number of state variables speeds up the calculation time.

The stack of tapes is a potential alternative to superconducting bulks. The screening current path is still not clear, since a non-insulated stack could contain coupling current in the metal stabilization or solder connecting tapes. For stacks of insulated tape, we can assume the approximation that the model contains only $J_x$ and $J_y$ components like the uncoupled case, which is a homogeneous bulk approximation. The homogeneous bulk approximation contains only the $T_z$ component, and hence the calculation time is again reduced by reduction of the unknown variables.

%%%%%%%%%%%%%%%%%%%%%%%%%%%%%%%%%%%%%%%%%%%%%%%%%%%%%%%%%%%%%%%%%%%%%%%%%%%%%%%%%%%%%%%%%%%%%%

\section{Current lines}
\label{s.current_lines}

The current lines are important, in order to see the current direction in any geometry of the sample. The current lines by definition, follow the direction of $\vJ$ and the separation between lines is inversely proportional to $|\vJ|$. The colour maps in this thesis contain current lines calculated by two ways: for 2D and 3D case.

The 2D model is the thin film approximation, and hence it contains only a $T_z$ non-zero component (section \ref{s.film_approx}). The effective magnetization is defined as $\nabla\times\vT=\vJ$ and for thin film case it becomes $\vJ=\frac{\partial T_z}{\partial y}\ve_x-\frac{\partial T_z}{\partial x}\ve_y$. 

In thin films, the level curves of $T_z$, are the current lines. This can be seen as follows. $\nabla T_z$ is perpendicular to the level curves of $T_z$. The $\nabla T_z\times \ve_z$ is tangent to the level curves of $T_z$ and it follows the level curve line. Since $T_z\times\ve_z=\vJ$ the level curves of $T_z$ follow the current density direction. The separation between level curves of $T_z$ is inversely proportional to $|\nabla T_z|$. Since $|\nabla T_z|=|\vJ|$, the separation between level  curves is inversely proportional to $|\vJ|$. 

The 3D case cannot assume level curves of $\vT$, since $\vT$ contains all three components in the 3D space, and hence the numerical method is used. The 3D numerical method for plotting current lines set the initial point in the space and calculates the unit vector of the current density at that point. Other points follows the current direction separated by a small distance and make a line until they close the loop. This method is more general, but requires a small separation between points.The reasons is that a large separation can create a current line loop that does not close due to imprecise unit vector.

%%%%%%%%%%%%%%%%%%%%%%%%%%%%%%%%%%%%%%%%%%%%%%%%%%%%%%%%%%%%%%%%%%%%%%%%%%%%%%%%%%%%%%%%%%%%%%

\section{Minimization}
\label{s.minimi}

Minimization is the main part of the solver. The minimization algorithm is developed especially for this kind of problem. The minimization solves the functional (section \ref{s.math_model}) by looking for its minimum value through the variables in the mesh (section \ref{s.discretization}). 

Before the minimization starts, other sub-routines calculate the applied vector potential $\vA_a$ (section \ref{s.applied_Aa}) on the surfaces according the applied magnetic field $B_a$, instant time $t$ and time step $i_t$. The change in one time step of the variables inside the mesh, such as effective magnetization $\Delta\vT$ at edges, current density $\Delta\vJ$ and vector potential $\Delta\vA$ at the surfaces, are initialized to zero. 

\subsection{Evaluation of $\vJ$ and $\vA$}

\begin{figure}[tbp]
\centering
{\includegraphics[trim=0 0 0 0,clip,width=6.0 cm]{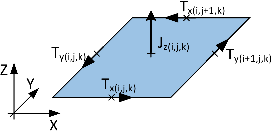}} 
\caption{A $Z$-surface with uniform current density component $J_z$ with index address $(i,j,k)$ and neighbour edges with appropriate index addresses. $T$ along the edge is assumed uniform.}
\label{f.mini_Jz}
\end{figure}

\begin{figure}[tbp]
\centering
{\includegraphics[trim=0 0 0 0,clip,width=6.0 cm]{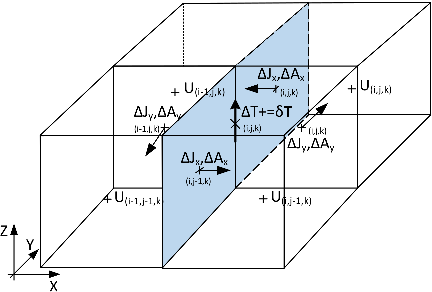}} 
\caption{The change of variable $\Delta T$ in $Z$-edge Z$_{(i,j,k)}$ creates a change in the variables $\Delta \vJ, U, \Delta \vA$ in the shown neighbour surfaces and cells.}
\label{f.minimization}
\end{figure}

Once $\Delta T$ is known at all edges, we evaluate the current density $\vJ$, vector potential $\vA$ in the surfaces and dissipation factor $U$ in the cells (figure \ref{f.minimization}). The current density is calculated via 
\begin{equation}
{\Delta J_z=\oint \frac{\Delta\vT\cdot\dif\vl}{S_z}},    
\label{}
\end{equation} 
where the integral is done on the edge of the elementary surface where $\Delta J_z$ is evaluated. The sum of the close loop integral is then 
\begin{eqnarray}
\ S_{z,i,j,k}J_{z,(i,j,k)} & = & l_{x,(i,j,k)}T_{x,(i,j,k)} + l_{y,(i+1,j,k)}T_{y,(i+1,j,k)}  \nonumber \\ 
           & - & l_{x,(i,j+1,k)}T_{x,(i,j+1,k)} - l_{y,(i,j,k)}T_{y,(i,j,k)},   
\label{} 
\end{eqnarray} 
where $i,j,k$ are the indexes of the elements [figure \ref{f.mini_Jz}]. The components $\Delta J_x$ and $\Delta J_y$ are found in the same way.

The vector potential is calculated by            
\begin{equation}
{\Delta \vA(\vr)=\frac{\mu_0}{4\pi}\int\frac{\Delta \vJ(\vr')}{|\vr-\vr'|}d\vr}.    
\label{}
\end{equation}
Then, one component of the vector potential is 
\begin{equation}
{\Delta A_{xi}(\vr)=\frac{\mu_0}{4\pi}\int\frac{\Delta J_{xi}(\vr')}{|\vr-\vr'|}d\vr},    
\label{}
\end{equation} 
The numerical evaluation of the $s$ component of the vector potential due to $\Delta J_s$ is   
\begin{eqnarray}
{\Delta A_{si}=\int \dif V A_s(\vr)h_{si}(\vr)=\sum_{j=1}^{n_s}V_{sj}\Delta J_{sj}\Delta a_{sij}}.
\end{eqnarray}

The average vector potential for the interaction matrix is [see section \ref{s.e_cells} and equation (\ref{e.sumJ})] 
\begin{eqnarray}
a_{sij}=\frac{\mu_{0}}{4\pi V_{si}V_{sj}}\int_{V}d^3r\int_{V}d^3r'\frac{h_{si}(r)h_{sj}(r')}{|\vr-\vr'|},
\label{e.asij}
\end{eqnarray} 
where $V_{si}$,$V_{sj}$ is volume of influence of surface $s$ with index $i$ and $j$ [equation (\ref{e.Vh})], $\vr$ and $\vr'$ are vector positions of the surfaces $i$,$j$, and $h_{si}(\vr)$,$h_{sj}(\vr)$ are the interpolation functions of figure \ref{f.e_cells}. 

\subsection{Minimization algorithm}

\begin{figure}[tbp]
\centering
{\includegraphics[trim=0 0 0 0,clip,width=7.0 cm]{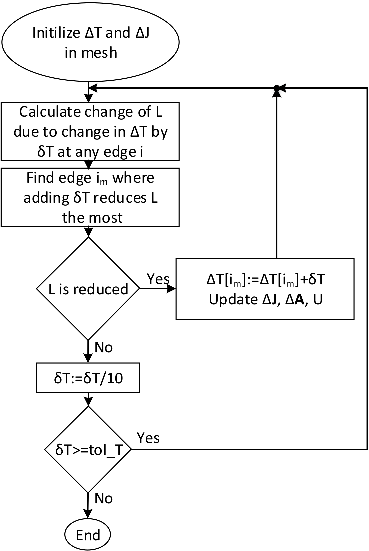}} 
\caption{The simplified minimization block diagram for a thin film with only $T_z$ component. For a bulk as a full 3D object, the minimization diagram is the same with all $\vT$ components. The minimization routine is developed for MEMEP 3D modelling tool.}
\label{f.min_diagram}
\end{figure}

The following minimization algorithm explains the whole method on the 2D film case, which contains only the $T_z$ component. The minimization block diagram is on figure \ref{f.min_diagram}. However, the same algorithm applies for all 3 components $(T_x,T_y,T_z)$ in the 3D case. 

The program calculates the change in the functional due to a change of $\delta T$ and $-\delta T$ at the first edge of type $Z$, $Z_1$. At the beginning, we start with a $\delta T$ value according to the current density $J_c$ and the dimensions of the elements         
\begin{equation}
{\delta T=J_c\frac{S_x}{l_z}},   
\label{}
\end{equation} 
where $S_x$ is the surface area of the surface $X$ and $l_z$ is the length of the edge $Z$. For uniform mesh, $\delta T$ creates current density equal the $J_c$ value. The other variables are not updated due to change of $\delta T$ in the edge $Z_1$. The sub-routine takes the second edge $Z_2$ and finds the value of the functional for it. The algorithm continues until all edges $Z$ are tested. Next, the program chooses the edge $Z_m$ that minimizes the functional the most. The algorithm sets the new value at the edge $Z_m$ and updates all variables inside the whole mesh. The minimization restarts again looking for the other edge $Z_m$ in the mesh, which minimizes the most the functional. The procedure repeats until there is no edge $Z_m$ that reduces the value of the functional. 

The solution of $\vT$ distribution in the sample is coarse. For this reason, $\delta T$ is divided by factor 10, in order to find a finer $\vT$ solution. The minimization algorithm starts again looking for all edges $Z$, which minimizes the functional. The sub-sequent decrease of $\delta T$ and minimization is repeated up to the tolerance of $J$, which is set in the input file (section \ref{c.input_file}). 

The final solution of $\vT$ distribution and all variables are saved into the output files for the present time step $i_t$ and variables $\Delta\vT, \Delta\vJ, \Delta\vA$ are added to the values of the ones at the previous time step, $\vT_0,\vJ_0,\vA_0$. Then, the values $\Delta\vT, \Delta\vJ, \Delta\vA$ are set to zero and the applied vector potential $\vA_a$ is recalculated to the values according to the next time step $i_t$. The minimization looks for a solution for each time step $i_t$ one by one. The minimization algorithm uses the same sub-routines for the mesh splits into sectors (section \ref{s.sectors}). However, each sector contains a relatively small number of elements, and hence the minimization is very fast.                   

%%%%%%%%%%%%%%%%%%%%%%%%%%%%%%%%%%%%%%%%%%%%%%%%%%%%%%%%%%%%%%%%%%%%%%%%%%%%%%%%%%%%%%%%%%%%%%

\section{Sectors}
\label{s.sectors}

\begin{figure}[tbp]
\centering
{\includegraphics[trim=0 0 0 0,clip,width=8.0 cm]{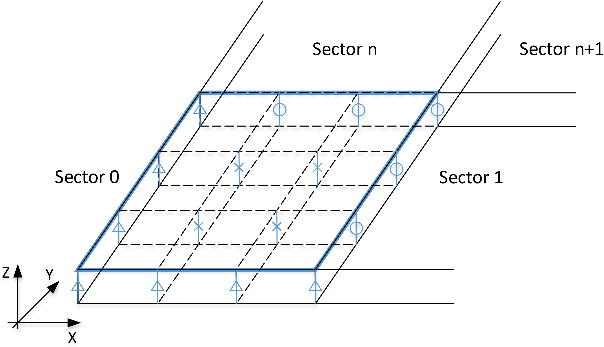}} 
\caption{The boundary conditions at the edges inside a sector on the sample boundary. The edges on the sample surface are with triangles, edges at the sector border are with circles, and edges inside the sector without any restrictions are with crosses.}
\label{f.sector_set}
\end{figure}

\begin{figure}[tbp]
\centering
 \subfloat[][]
{\includegraphics[trim=0 0 0 0,clip,width=4.0 cm]{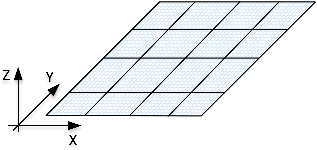}} 
 \subfloat[][]
{\includegraphics[trim=0 0 0 0,clip,width=4.0 cm]{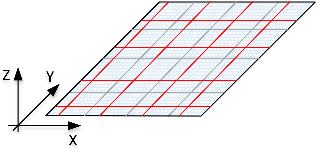}} 
 \subfloat[][]
{\includegraphics[trim=0 0 0 0,clip,width=4.0 cm]{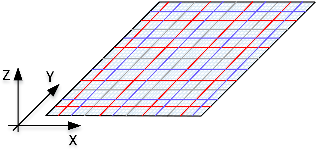}} 
\caption{The sectors distribution in the sample. (a) The first set of sectors (black lines). (b) The second set of sectors (red lines) is with  $1/3$ diagonal shift compared to Sector set 1. (c) The third set of sectors (blue lines) is with $2/3$ diagonal shift compared to the Sector set 1. The sets of sectors drastically reduce the minimization time.}
\label{f.sectors}
\end{figure}

The 3D modelling tool has to handle a huge number of elements, and hence a fast calculation method is necessary. The minimization with sectors \cite{Pardo16SST,Kapolka18IESa} speeds up the computation time (see the end of this section). The computing time for minimization increases with the second power of the total numbers elements (section \ref{s.cal_time}). 

The entire sample is split into many smaller sectors [figure \ref{f.sectors}(a)] that contain a relatively small number of elements, and hence the minimization of each sector is very fast. The minimization is looking for $\vT$ values at the cell's edges (section \ref{s.minimi}). Therefore, the boundaries between sectors are set at the edges. One sector with all boundary conditions is on figure \ref{f.sector_set}. The thin film model is with one cell in thickness and hence only $T_z$ component is with non-zero value. The sector in this example contains $3\times 3$ cells and $4\times 4$ edges. However, we found that the optimum value for the calculations is $12\times 12\times 1$ cells in each sector in the 2D case and $9\times 9\times 9$ cells in the 3D case. There are three conditions for $T_z$ edges in blue on figure \ref{f.sector_set}. The current density outside the sample is zero and there is no surface current density, and therefore $T_z$ at the edges on the sample surface is zero. The boundary edges on the sample surface are with triangles. The edges at the border between the sectors are with circles. The minimization does not minimize the sector border's edges and keeps them with the same values as before minimization. The edges with a cross change values during the minimization. There are 3 sets of sectors, in order to solve the edges at all the borders, which the minimization sub-routine keeps as constant. The second set of sectors contains sectors with the same boundary conditions as the first set,  but the sectors position is shifted by $1/3$ of the sector size [figure \ref{f.sectors}(b)]. The third set of sectors shift all sector positions by $2/3$ of sector size [figure \ref{f.sectors}(c)]. Therefore, the border edges from the first set of sectors are solved in the second or third set of sectors, and opposite. 

The three sets of sectors speed up minimization time compared to the minimization without the sectors (section \ref{s.cal_time}). The total computing time of one iteration without any speed up method is $t=a\cdot n_{cx}^6$, where $a$ is certain constant factor and $n_{cx}$ is the number of cells along one side of the 3D cube sample. Then, the total computing time $t_s$ of one iteration by sectors is $t_s=a\cdot n_{cx}^6\cdot 3/n_s^3$, where $n_s$ is the number of sectors per one edge of the cube and multiplied by factor 3, since the mesh contains 3 sets of the sectors. The reduction factor can be defined as $t/t_s$ and is equal to $n_s^3/3$. Therefore, the total theoretical speed up factor by sectors is $S/3$, where $S$ is the total number of sectors $n_s^3$.

One disadvantage of the sector method is that each set of sectors increase the usage of the memory RAM, since the minimization requires all interaction matrices. However the total memory usage is small, a few Gbs, and hence an increase of 3 times is still low compared to the available RAM memory (64 GB) in the computer cluster. The required RAM memory is low thanks to the uniform mesh. For this case, the independent entries of the interaction matrix can be drastically reduced, since it only depends on the difference of the addresses between two surfaces $(i_1-i_2,j_1-j_2,k_1-k_2)$, where $(i_1,j_1,k_1)$ and $(i_2,j_2,k_2)$ are the addresses of two arbitrary surfaces. 

%%%%%%%%%%%%%%%%%%%%%%%%%%%%%%%%%%%%%%%%%%%%%%%%%%%%%%%%%%%%%%%%%%%%%%%%%%%%%%%%%%%%%%%%%%%%%%

\section{Symmetry}
\label{s.symmetry}

\begin{figure}[tbp]
\centering
{\includegraphics[trim=30 0 10 0,clip,width=6.0 cm]{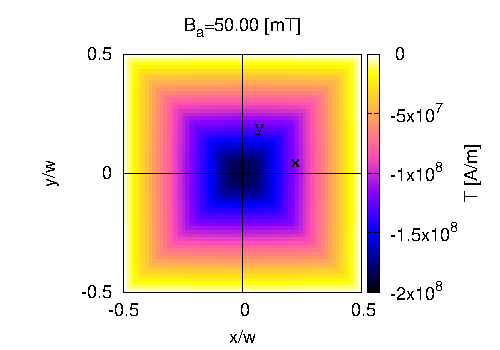}} 
\caption{The $\vT$ distribution in the thin film with 50 mT perpendicular applied field and critical current density $3\cdot 10^{10}$ A/m$^{2}$ and $n$ value 30. The $x-y$ mirror lines are with origin at $0,0$. Ideally, symmetry reduces the minimization time to 1/4.}
\label{f.symmetry_T}
\end{figure}

\begin{figure}[tbp]
\centering
 \subfloat[][]
{\includegraphics[trim=0 0 0 0,clip,width=6.0 cm]{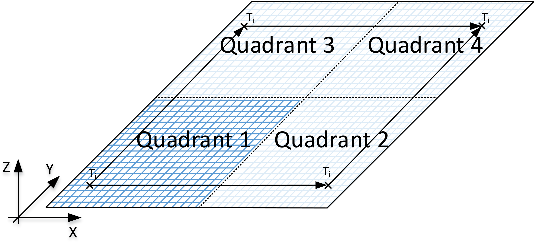}} 
 \subfloat[][]
{\includegraphics[trim=0 0 0 0,clip,width=5.0 cm]{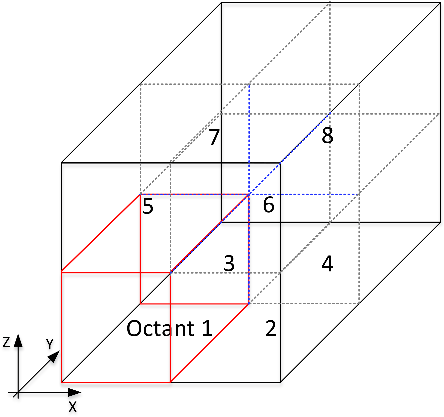}} 
\caption{The symmetry in (a) thin film and (b) cube reduces sectors into the first quadrant in 2D case and octant in the 3D case.}
\label{f.symmetry_sketch}
\end{figure}

The symmetry method reduces calculation time further. The simple modelling example of $\vT$ in magnetizing a thin film is on figure \ref{f.symmetry_T}. The modelling case contains homogeneous distribution of critical current density $J_c$ and power law $n$ value, and hence there exist symmetry in the $\vT$ solution. The solution presents mirror symmetry with respect to the $x$ and $y$ axis with origin at the sample center. The independent fourth of the sample is set as the quadrant at the left-bottom corner [figure \ref{f.symmetry_sketch}(a)]. The solution at the other quadrants is found from the independent quadrant by mirror symmetry. The same method applies for the 3D case. The cube mesh is split into 8 octants and the independent one mirrors the $\vT$ values into the remaining octants according to the $x,y,z$ axes lines [figure \ref{f.symmetry_sketch}(b)]. The symmetry can be exploited only when the applied magnetic field is perpendicular to any of sample surfaces. 

\begin{figure}[tbp]
\centering
{\includegraphics[trim=0 0 0 0,clip,width=6.0 cm]{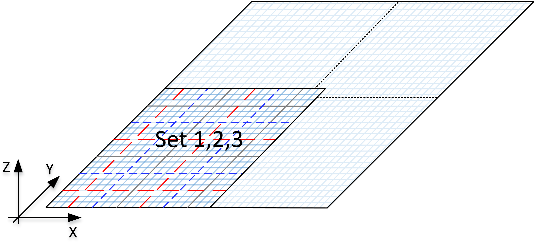}} 
\caption{The three Sets of sectors in the first quadrant overlap the mirror lines $x,y$, and hence it does not reach ideal sped up time.}
\label{f.symmetry_2D}
\end{figure}

The code implements symmetry only in case of an applied field angle of $\theta=0\degree$. The sub-routines in the code minimize $\vT$ in the sectors of the independent quadrant and the symmetry reduces the number of sectors 4 times in 2D case and 8 times in 3D case. The sectors in the independent quadrant have to be with full size, and hence sectors in various sets overlap the mirror planes (figure \ref{f.symmetry_2D}). Therefore, the number of sectors is not reduced as much as factor 4 (2D case) or 8 (3D case), and hence the minimization time speeds up only 3-3.5 times (2D) and 6-7.5 times (3D). The speed up factor increases with the number of sectors. The $\vT$ solution after each iteration of the independent quadrant is copied into the other quadrants. The sub-routines are in the recalculation block in the code structure diagram (figure \ref{f.block_diagram}), in order to evaluate $\vJ,U,\vA,\vB$ in the entire sample (block``Recalc. $\vJ$,$U$,$\vA$,$\vB$" in figure \ref{f.block_diagram}). The evaluation of current density and dissipation factor is fast. The evaluation of $\vA$ and $\vB$ is time consuming, scaling as the square of the total number of cells. However, constant $J_c$ requires the evaluation of $\vB$ only at the end of each time step $i_t$. The minimization requires evaluation of $\vA$ in the independent quadrant only, and hence further speed up of the calculation time is possible. The code contains all possible combinations of input options (section \ref{c.input_file}).            

%%%%%%%%%%%%%%%%%%%%%%%%%%%%%%%%%%%%%%%%%%%%%%%%%%%%%%%%%%%%%%%%%%%%%%%%%%%%%%%%%%%%%%%%%%%%%%

\section{Parallel programming}
\label{s.programming}

The 3D modelling tool is dealing with a huge number of degrees of freedom, and hence calculation time is an important feature for any calculation method. The calculation tool of any physical case with computing time of a several weeks is not feasible. The modelling case often needs to adapt according to the requirements with a slight change in an input parameters, and hence the final results will take several months if not optimized. The tool can be sped up numerically as it is explained in previous sections \ref{s.minimi}, \ref{s.sectors}, \ref{s.symmetry} or by parallel programming in the code itself. The modelling tool is written in C++\cite{C++} and parallel programming is implemented by OpenMP\cite{OpenMP} (on a single computer CPU) and BoostMPI\cite{BoostMPI} (on cluster nodes) standards.       

%%%%%%%%%%%%%%%%%%%%%%%%%%%%%%%%%%%%%%%%%%%%%%%%%%%%%%%%%%%%%%%%%%%%%%%%%%%%%%%%%%%%%%%%%%%%%%

\subsection{Structure of the code}
\label{s.code}

\begin{figure}[tbp]
\centering
{\includegraphics[trim=0 0 0 0,clip,height=20.0 cm]{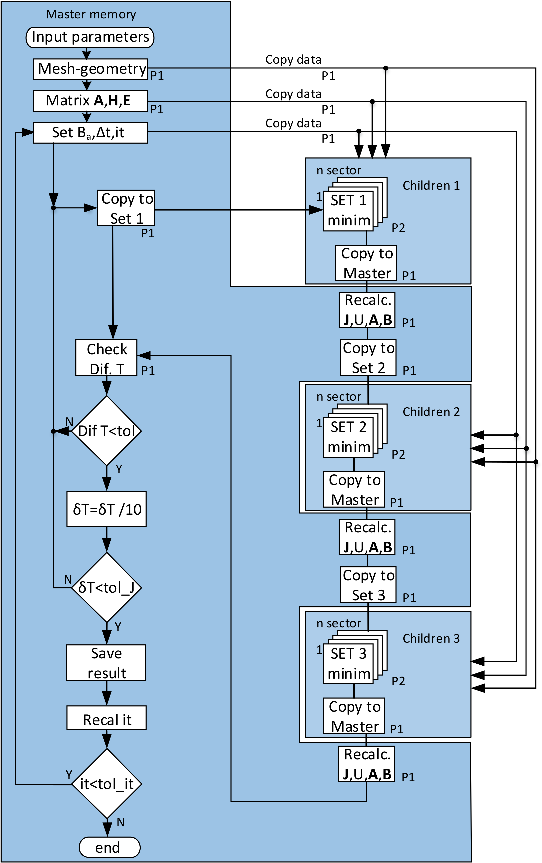}} 
\caption{The simplified block diagram of the code structure. The highly parallelized code structure exploits the C++ object-based properties to create ``children" object from a ``parent". $P_1$ and $P_2$ is the type of parallel calculation (section \ref{s.OpenMP}) and $n$ is the number of sector.}
\label{f.block_diagram}
\end{figure}

The MEMEP 3D modelling tool has a lot of input options, reflecting the complex code structure. This section explains the calculation steps written on around 7000 lines in C++ code. The big advantage of C++ is the possibility to organize the code by classes and objects, which are used in the program. The code is split into several main blocks. The simplified block diagram is on figure \ref{f.block_diagram}. 

The user friendly input file contains all input parameters (chapter \ref{c.input_file}). The program loads the input file, calculates the mesh and creates all variables for cells, surfaces and edges. Then, it also sets the values of critical current density $J_c$ and power law exponent $n$ in the cells. 

The most optimization complex calculation are the interaction matrices for the vector potential $\vA$, and magnetic field $\vH$. The non-uniform mesh with total number of cells 70000 requires full interaction matrix of size around 117 GB RAM memory. The interaction matrix for uniform mesh depends on the relative distance, and hence the matrix can be reduced to a few MB of RAM. 

The initialization sub-routine sets the applied magnetic field with angle $\theta$, the current time $t$, time step $\Delta t$ and number of time steps. All the previous data is saved in the memory in a ``parent" object. Later, the program copies all mesh and parameter data to ``children" objects, where the minimization is done. The flow of data is shown on figure \ref{f.block_diagram}. Each ``children" object contains one of the 3 sets of sectors (section \ref{s.sectors}). Next, follows the main minimization part of the code. $\vT$ is solved in the first set of sectors by minimizing the functional in each sector in parallel. Then, the solution of $\vT$ is copied to the ``parent" object, where all side variables are re-calculated, such as the dissipation factor $U$, the vector potential $\vA$, the magnetic field $\vB$, the current density $\vJ$ or the critical current density $J_c$; according to the chosen dependence in the input file. 

Afterwards, the data moves into the second ``children" object, where minimization finds the new solution for the second set of sectors. The sub-routines again copy the solution into the ``parent" object and recalculate the side variables. Next, the third set of sectors receives the data from the ``parent" and finds the solution for the last time. The data moves into the ``parent" and recalculates all necessary variables. 

The solution from all 3 sets is compared with the solution of the previous iteration step by finding the maximum difference in $\vT$ at each edge. If the difference is higher than the chosen tolerance, the entire iteration procedure is repeated. If the difference is lower than $\delta T$ (explained in section \ref{s.minimi}), $\delta T$ is divided by factor 10 and the iteration starts again. The iterations stop only if $\delta T$ is below the certain input current density of tolerance $d_J$. The final solution of the present time is saved into output files. Finally, a new time step and applied field is set. The new time step starts the entire minimization process in order to find a new solution. The code calculates all time steps and saves them into the output files. 

%%%%%%%%%%%%%%%%%%%%%%%%%%%%%%%%%%%%%%%%%%%%%%%%%%%%%%%%%%%%%%%%%%%%%%%%%%%%%%%%%%%%%%%%%%%%%

\subsection{Parallel programming in one computer by OpenMP}
\label{s.OpenMP}

Parallel programming by OpenMP speeds up the calculation time by using all cores or threads in the computer. Nowadays, computers have several threads (for instance, 8 threads in affordable Intel i7 4000 series processors), reaching 40-60 or more threads in powerful workstations. Then, parallel programming could be speed up as many times as the number of threads. 

There are two ways of parallel computing, which are used in the code. The first way can be used in each loop ``for" with a lot of iterations. The simple loop example is 
\begin{verbatim}
#pragma omp parallel for private(n) num-threads(i)
for(n=0;n<x;n++) {a[n]=n;}
\end{verbatim}
puts the $n$ value to the vector $a$ of size $x$. The OpenMP commands starts with ``$\#$pragma omp parallel for", which is dedicated for loop ``for". The private variables means that each thread will create its own variable and it does not override the value between threads. Num-threads sets the number of required threads. The previous simple example will split a single master thread into many threads. The single master thread is performing calculation in series and many threads will perform the calculation in parallel and faster than with only one thread. The second way is used in the short loop with complex calculation such as minimization of a sector. The code example is 

\begin{verbatim}
#pragma omp parallel for private(n) num-threads(i)
for(n=0;n<set1;n++) {minimization-sector(n);}.
\end{verbatim}
However, all calculation in sub-routines [below minimization-sector(n)] cannot contain any pragma commands. Variables cannot depend on an other sector, since the master thread is already split into many threads, each tread minimizing its own sector. If the thread finishes, the calculation on the current sector will continue with another free sector, which is not calculated. 

The simplified block diagram of the code (figure \ref{f.block_diagram}) is marked with ``P1" and ``P2". P1 is parallelization by the first way and P2 by the second way. The sub-routines in P2 blocks are parallelized in top level, and hence the efficiency of the parallel programming is very high. The high efficiency is because the master thread is split into many threads only once, and hence each thread is calculating its own part of the code for a relatively long time. This avoids time-consuming creation of new threads. In the code, there still exist parts calculated in series, and hence parallel programming efficiency decreases a bit. However, the code reaches very high parallel efficiency, more than 90\%. 

%%%%%%%%%%%%%%%%%%%%%%%%%%%%%%%%%%%%%%%%%%%%%%%%%%%%%%%%%%%%%%%%%%%%%%%%%%%%%%%%%%%%%%%%%%%%%%

\subsection{Parallel programming on a cluster by BoostMPI}
\label{s.BoostMPI}

\begin{figure}[tbp]
\centering
{\includegraphics[trim=0 0 0 0,clip,height=18.0 cm]{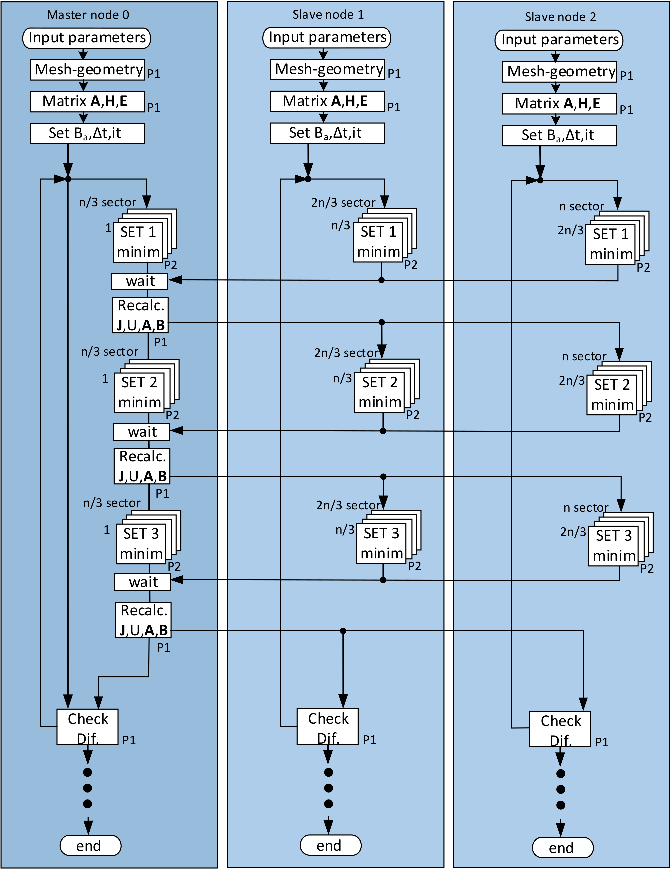}} 
\caption{The simplified block diagram of the code structure with BoostMPI hierarchy for 3 nodes, where $n$ is the number of sectors. Time consuming data transfer between nodes is minimized, and hence  parallel code efficiency is high.}
\label{f.block_diagramMPI}
\end{figure}

\begin{figure}[tbp]
\centering
{\includegraphics[trim=0 0 0 0,clip,height=5.5 cm]{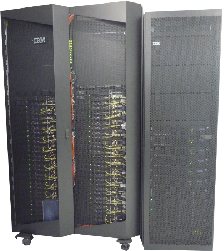}} 
\caption{Slovak Academy of Science - Institute of informatics HPC Cluster with 52 x nodes of IBM dx360 M3.}
\label{p.cluster}
\end{figure}

Further speeding up the modelling tool is possible by computing on a computer cluster. Parallel programming on a computer cluster is more complex than on a single computer. The used a computer cluster in the Slovak Academy of Sciences (picture \ref{p.cluster}) that contains 52 nodes of IBM dx 360 M3 and each node contains 12 cores (12 threads). Parallel programming in clusters using C++ language is only supported by  BoostMPI protocol. The program is written with both OpenMP and BoostMPI protocols, and hence it can use all nodes in the cluster and all cores available per node.

The simplified block diagram of the code for three nodes is on figure \ref{f.block_diagramMPI}. The compiler file contains all input parameters for the cluster, such as number of nodes, number of cores per node and the code file. The cluster terminal loads the code into the nodes and each node performs the same code. The structure of the code is similar to that on figure \ref{f.block_diagram}. Each node loads an identical input file and calculates the mesh and interaction matrices, in order to avoid unnecessary data exchange between nodes, which reduces parallel computing efficiency. One node is dedicated as ``master" node and the remaining nodes are ``slaves". The calculation is performed by all cores in each node separately. The parallel computing by OpenMP commands is marked by ``P1" and ``P2" as it is explained in previous section \ref{s.OpenMP}. Each node contains the appropriate $1/n_n$ number of sectors for each set of sectors, where $n_n$ is the number of nodes. The minimization sub-routines of the sectors calculate for each node in parallel and separately. The result of $\vT$ is copied into the master node by BoostMPI commands (send, received). The master node waits until all data is received from all nodes and then it evaluates variables $\vJ,U,\vA,\vB$. The evaluated variables in the master node are sent back to the other nodes, and then minimization of the second set of sectors continues. The same process of the exchange of data between master node and other nodes continues until all 3 sets are solved. The master node collects, evaluates, and finally broadcasts all data for the last time. Each node decides to either continue with another iteration or not as it is explained in section \ref{s.code}

The example on figure \ref{f.block_diagramMPI} uses 3 nodes, and hence the theoretical speed up of the calculation time is 3 times. The process of the exchange of data decreases parallel efficiency of the code, because the nodes are not calculating until all data are completely received or sent. The symmetric distribution of calculation tasks on each node is very important, in order to secure the synchronicity of the redistribution of the data and decrease time of not calculating nodes. However, the parallel programming efficiency is very high, around 90\%. The block diagram (figure \ref{f.block_diagramMPI}) is simplified. Actually, the evaluation of variables such as $\vB$ and $\vA$ is calculated by all nodes, in order to speed up calculation time further. The entire code contains around 7000 lines and more than 250 sub-routines.       

%%%%%%%%%%%%%%%%%%%%%%%%%%%%%%%%%%%%%%%%%%%%%%%%%%%%%%%%%%%%%%%%%%%%%%%%%%%%%%%%%%%%%%%%%%%%%%

\subsection{Computation time}
\label{s.cal_time}

\begin{figure}[tbp]
\centering
{\includegraphics[trim=0 0 0 0,clip,height=5.5 cm]{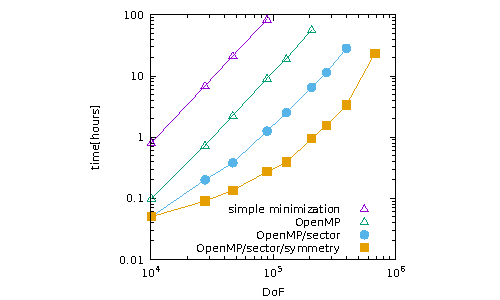}} 
\caption{The calculation time of the MEMEP 3D modelling tool with increasing number of degrees of freedom and various sped up methods on a single standard computer. The model case is a cube with perpendicular applied field parallel to $z$ axis and 10 time steps, in order to reach the peak of the applied field of 200 mT. The DoF for symmetry corresponds to the full sample.}
\label{f.speed}
\end{figure}

In conclusion, the modelling tool reaches fast calculation time. All the methods to speed up the calculation time according to the previous sections are implemented in the C++ code. The example case with various number of degrees of freedom shows the calculation time dependence of the modelling tool. The modelling situation is a simple cube magnetization with the applied magnetic field of 200 mT parallel to the $z$ axes and critical current density $1\cdot 10^8$ A/m$^2$. The current density tolerance is $1\cdot 10^{-3}J_c$. The MEMEP 3D method calculates 10 time steps until the model reaches the peak of applied field. 

We study the effect on the computing time by increasing the total number of cells after each code improvement (\ref{f.speed}). The mesh starts with $15\times 15\times 15$ (3375) cells and raises up to $61 \times 61\times 61$ (226981) cells, which is around 680000 Degrees of Freedom DoF. The calculations are performed on one node of the computer cluster, which is similar to a desk-top computer with Intel core i7-4771 CPU@8x3.5Ghz. The pure minimization without any speed up method can calculate 90000 degrees of freedom (30000) cells in less than 100 hours. The parallel programming by OpenMP protocol speeds up the calculation time around 10 times, since the cluster node contains 12 cores and parallel efficiency is around 90\%. The other method improvement is by using sectors (section \ref{s.sectors}). The sectors speed up the calculation time by almost 8.5 times. The sectors efficiency increase with the number of sectors and with the number of elements. The other method to decrease the calculation time is symmetry (section \ref{s.symmetry}). The symmetry reduces computation time 7.5 times and its efficiency increases with the number of elements. For symmetry, the ideal theoretical speed factor for the 3D case is 8 times. In conclusion, we are able to reduce the entire computing time by two and three orders of magnitude on one computer for the routines without and with symmetry, respectively. The calculation time of $31 \times 31 \times 31$ (29791) cells (90000 DoF) reduces from 90 hours to less than 20 minutes with symmetry. 

\begin{figure}[tbp]
\centering
{\includegraphics[trim=0 0 0 0,clip,height=5.5 cm]{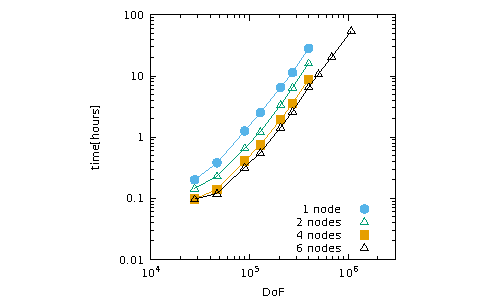}} 
\caption{The computing time of MEMEP 3D method on the computer cluster. The reduction of the computing time increases with number of nodes. However, the efficiency of parallelization decreases by many nodes.}
\label{f.speed1}
\end{figure}

Parallel programming by BoostMPI protocol on the computer cluster reduces the calculation time further. The previous methods speed up the calculation time many times, and hence the computing tasks for the cluster are not heavy and parallel programming efficiency on the cluster is low. In order to use the cluster effectively, we did not use symmetry, which cannot be used for all modelling cases. The same modelling example as for one computer is used for the study of the computing time on a cluster. The calculation time as a function of the degrees of freedom is on figure \ref{f.speed1}. The increase of the total number of elements (or DoF) increases the efficiency of the parallel computing. For very low total number of cells, the dominant calculation time is data transfer between nodes, which is negligible for high numbers of cells. The calculation time speeds up on 2 nodes by factor 1.75, on 4 nodes by factor 3.23 and be 6 nodes 4.31 times for 397953 DoF. However, the 6 nodes of the cluster allows us to model the sample with more than 1 million of DoF in less than 55 hours (figure \ref{f.speed1}). The reason for low efficiency with increased number of the nodes is that the computing load of each node is small and data transfer between nodes starts to be dominant. The efficiency can be improved by further increase of the total number of cells. The efficiency, of the parallel computing on the cluster is more than 80\% for the present range of the number of cells in the mesh. Further advanced modelling by MEMEP 3D method will require higher number of cells in the mesh, and hence the parallel computing efficiency will increase.         

%%%%%%%%%%%%%%%%%%%%%%%%%%%%%%%%%%%%%%%%%%%%%%%%%%%%%%%%%%%%%%%%%%%%%%%%%%%%%%%%%%%%%%%%%%%%%%

\section{Elongated cells}
\label{s.e_cells}

Elongated cells in the mesh enables to model long samples without any further increase in the total number of elements or reduction of the number of elements. The reduction of the number of elements speeds up the computation time of the modelling tool. The modelling sample geometry with elongated cells plays an important role for long thin films (2D) and long bars (3D).

\begin{figure}[tbp]
\centering
 \subfloat[][]
{\includegraphics[trim=0 0 -30 0,clip,width=6.0 cm]{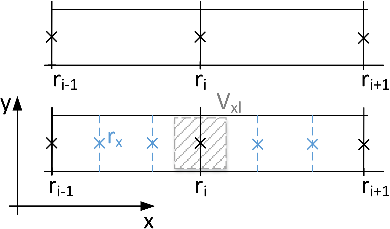}} 
 \subfloat[][]
{\includegraphics[trim=0 0 0 0,clip,width=6.0 cm]{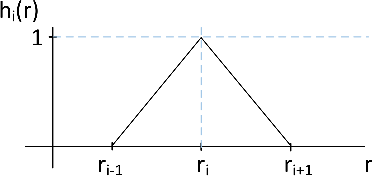}}
\caption{(a) Elongated cells (top) with surface positions $r_{i-1},r_i,r_{i+1}$. The elongated cells are divided into sub-elements (bottom). (b) $h_i(r)$ is a linear interpolation function with value 1 at surface $r_i$ and 0 at the surrounding surfaces $r_{i-1},r_{i+1}$.}
\label{f.e_cells}
\end{figure}

The interaction matrix for non-elongated cells is calculated by two formulas. The vector potential is evaluated in the surface elements (section \ref{s.discretization}). Therefore, the criterion to choose between formulas is according to the surface positions. The general definition of vector-potential interaction matrix is 
\begin{eqnarray}
a_{sij}=\frac{\mu_{0}}{4\pi V_{si}V_{sj}}\int_{V}d^3r\int_{V}d^3r'\frac{h_{si}(r)h_{sj}(r')}{|\vr-\vr'|},    
\label{e.1asij}
\end{eqnarray} 
where $s$ is the type of surface $s\in(x,y,z)$; $i,j$ are surface indexes; $V_{si},V_{sj}$ are volumes of influence; $h_{si}(r)$, $h_{sj}(r')$ are interpolation functions; and $r,r'$ are vector positions of the surfaces. The interaction matrix element of (\ref{e.1asij}) is the average vector potential created by the $s$-surface $j$ on surface $i$ per unit current density in the $j$ surface. The volume of influence is calculated as 
\begin{eqnarray}
V_{si}\equiv\int_{V_{si}}d^3rh_{si}(r),    
\label{e.Vh}
\end{eqnarray} 
where the linear interpolation function of first-order is on figure \ref{f.e_cells}(b). The self-interaction of two surfaces $i=j$ uses the following approximation, which is analytical 
\begin{eqnarray}
a_{sii}=\frac{\mu_{0}}{4\pi V^2_{si}}\int_{V_{si}}d^3r\int_{V_{si}}d^3r'\frac{1}{|\vr-\vr'|}.    
\label{e.as1}
\end{eqnarray}
The full length of the analytical expressions is in appendix \ref{s.appendix_A}. For cubic or 2D square cells and $i\ne j$ we can use the approximation  
\begin{eqnarray}
a_{sij}\approx\frac{\mu_{0}}{4\pi|\vr-\vr'|}.    
\label{e.ap1}
\end{eqnarray}

For elongated cells, we need to use sub-elements, as it is shown on figure \ref{f.e_cells}(a-bottom). The number of sub-elements increases until the sub-elements are as square as possible, in order to increase the accuracy of the numerical calculation of the average vector potential. The general equation for two sub-elements $l,m$ of surfaces $i,j$ is     
\begin{eqnarray}
a_{sijlm}=\frac{\mu_{0}}{4\pi V_{sl}V_{sm}}\int_{V_{sl}}d^3r\int_{V_{sm}}d^3r'\frac{h_{si}(r_{sl})h_{sj}(r_{sm}')}{|\vr_{sl}-\vr_{sm}'|}.    
\label{}
\end{eqnarray} 
where $V_{sl}$, $V_{sm}$ are the volume of influence of sub-elements $l$ and $m$, as defined in figure \ref{f.e_cells} (a-bottom). The case of overlapping sub-elements $l=m$ uses the analytical formula 
\begin{eqnarray}
a_{sijll}=\frac{\mu_{0}h_{si}(r_{sl})h_{sj}(r_{sl})}{4\pi V^2_{sl}}\int_{V_{sl}}d^3r\int_{V_{sl}}d^3r'\frac{1}{|\vr_{sl}-\vr_{sl}'|}
\label{e.as2}
\end{eqnarray}
with full expression in appendix \ref{s.appendix_A}. The sub-elements with position $l\ne m$ use the approximated formula 
\begin{eqnarray}
a_{sijlm}\approx\frac{\mu_{0}h_{si}(r_{sl})h_{sj}(r_{sm})}{4\pi |\vr_{sl}-\vr_{sm}'|}.
\label{e.ap2}
\end{eqnarray}

%%%%%%%%%%%%%%%%%%%%%%%%%%%%%%%%%%%%%%%%%%%%%%%%%%%%%%%%%%%%%%%%%%%%%%%%%%%%%%%%%%%%%%%%%%%%%%%
%%%%%%%%%%%%%%%%%%%%%%%%%%%%%%%%%%%%%%%%%%%%%%%%%%%%%%%%%%%%%%%%%%%%%%%%%%%%%%%%%%%%%%%%%%%%%%%
%%%%%%%%%%%%%%%%%%%%%%%%%%%%%%%%%%%%%%%%%%%%%%%%%%%%%%%%%%%%%%%%%%%%%%%%%%%%%%%%%%%%%%%%%%%%%%%
%%%%%%%%%%%%%%%%%%%%%%%%%%%%%%%%%%%%%%%%%%%%%%%%%%%%%%%%%%%%%%%%%%%%%%%%%%%%%%%%%%%%%%%%%%%%%%%

\chapter{Experimental method and samples}
\label{c.experiments}

The experimental measurements of AC loss further verify the modelling method. The discussion of comparison between measurements and model is in section \ref{s.film_2filament}. In addition, the measurements are important by themselves, since they provide information on the real samples. Here, the measurement set-up is introduced with a following part about preparation of samples.  

The AC loss is an important parameter in any superconducting tape or wire. Since AC loss increases with the tape width, the AC loss in HTS tapes under perpendicular applied field is large, because of the relatively large tape width. The striations \cite{Grilli16SST} the reduce hysteresis loss, since the screening current closes in narrower loops in the thinner filaments. These striations cut the superconducting layer into thinner tapes to the substrate, with or without resistive coupling. The coupled case introduces AC coupling loss, because of the present resistive material between the filaments. The coupling AC loss can be reduced by increasing the coupling resistivity up to the uncoupled case. However, the fully uncoupled case reduces the thermal stabilization of the tape, since the filaments do not share current in case of a fault in one filament. A modelling tool that can predict the AC loss with high accuracy is very demanding. The following section is dedicated to the very convenient case of AC loss measurement of two superconducting tapes soldered together and comparison with modelling results of the MEMEP 3D method.  

The cross-field demagnetization of a superconducting bulk sample with a cube geometry is still not well understood. Therefore, an accurate 3D model with all finite size effects is required. We performed the same experiments as the calculation in section \ref{s.cube_demagnetization} with collaboration of the University of Cambridge. The measurements validate the whole method and confirms new findings of MEMEP 3D method. The test sample is prepared by the colleagues in the Superconducting Bulk Group in the University of Cambridge and I performed the measurement of the sample during my stay in Cambridge.

%%%%%%%%%%%%%%%%%%%%%%%%%%%%%%%%%%%%%%%%%%%%%%%%%%%%%%%%%%%%%%%%%%%%%%%%%%%%%%%%%%%%%%%%%%%%%%

\section{Sample of two tapes joined by normal metal for coupling effects measurements}
\label{s.couple_sample}

\begin{figure}[tbp]
\centering
 \subfloat[][]
{\includegraphics[trim=0 0 0 0,clip,width=12.0 cm]{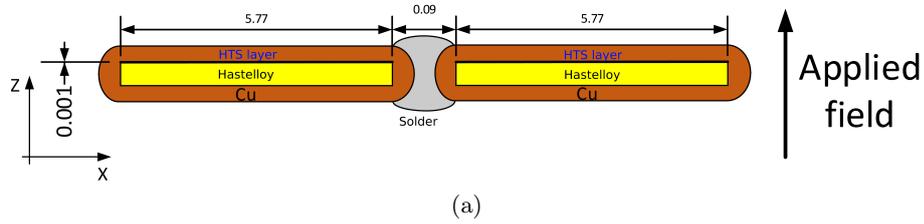}} 
\caption{Sketch of the cross-section of the soldered sample. Dimensions are in mm.}
\label{p.couple_sample}
\end{figure}

The experimental sample consists on two superconducting tapes soldered together by indium. The superconducting commercial tape is 6 mm wide with 20 $\mu$m cooper stabilization from SuperOx \cite{Chepikov17SST}. The tapes are soldered side by side (figure \ref{p.couple_sample}). First, we prepared a sample of 50 mm length, and after measuring, we cut it to 22 mm length. The resistance between tapes in the sample is measured at the liquid nitrogen temperature (77 K) by electric measurements. The resistance of the 50 mm sample is 710 n$\ohm$, while the 22 mm sample has the same conductance per unit tape length. In order to compare to the uncoupled situation, we made a second sample made of two superconducting tapes of 50 mm length placed side by side and insulated by Kapton tape. 

%%%%%%%%%%%%%%%%%%%%%%%%%%%%%%%%%%%%%%%%%%%%%%%%%%%%%%%%%%%%%%%%%%%%%%%%%%%%%%%%%%%%%%%%%%%%%%

\section{Calibration free method for AC loss measurement}
\label{s.measurements_coupling}

The AC loss is measured by the calibration-free method \cite{souc05SST}. The calibration free method is a direct technique to measure AC loss. The method is based on AC power evaluation of the system by measuring voltage. The measurement method is based on series connection of two identical field coils, where the fist coil is the measurement coil for inserting the sample and induce the magnetic field. The second coil is a compensation coil. The pick-up coils are identical and wound parallel and isolated within both field coil windings. The anti-series connection of pick-up coils result in no voltage signal without inserted sample. The measured voltage by pick-up coils is completely due to the sample flux $V=-\frac{d\psi}{dt}$, where $\psi$ is the sample magnetic flux. Since the pick-up coils follow the field coils winding, the measured flux in the pick-up coil is the same as in the field coil. The power provided to the field coils, in order to compensate the sample dissipation is $P=-I\frac{d\psi}{dt}$. By energy conservation, the power integrated over one cycle is the power loss of the sample.     

\begin{figure}[tbp]
\centering
{\includegraphics[trim=0 0 0 0,clip,width=9.0 cm]{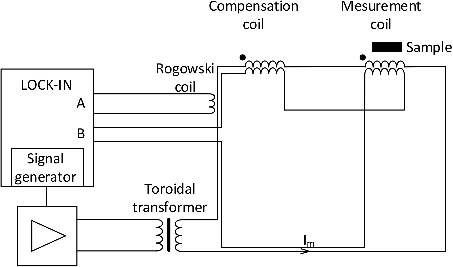}} 
\caption{The electric scheme of the measurement set-up of the calibration free method\cite{souc05SST}.}
\label{p.ACloss_scheme}
\end{figure}

The electric scheme is on figure \ref{p.ACloss_scheme}. The frequency generator creates the AC signal, which is amplified in the AC amplifier and galvanically separated from the AC magnets by a toroidal transformer. The lock-in measures the voltage on the Rogowski coil induced by the magnet current $I$, $V_{rog}=-\omega k_{rog}\sqrt{2}I_{rms}\sin{\omega t}$, where $k_{rog}$ is the Rogowsky coil constant, $I_m=\sqrt{2}I_{rms}\cos{\omega t}$ (channel A) and the voltage generated by the sample $V=\sqrt{2}V_{R,rms}\cos\omega t+\sqrt{2}V_{I,rms}\sin\omega t$ (channel B), and hence the AC power loss of one cycle is $Q=I_{rms}V_{I,rms}f$. 

\begin{figure}[tbp]
\centering
{\includegraphics[trim=0 0 0 0,clip,width=11.0 cm]{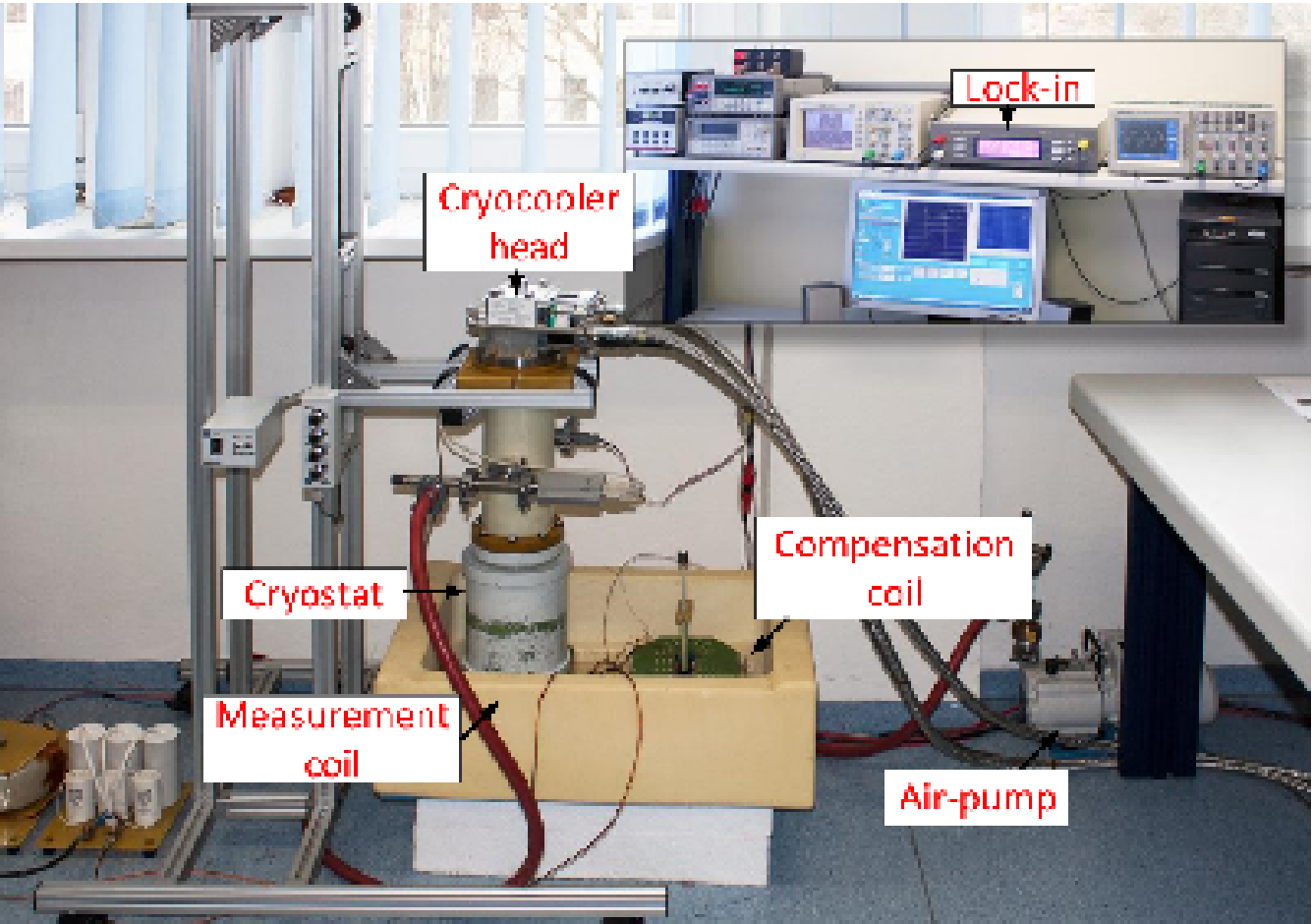}} 
\caption{Picture of the calibration-free measurement set-up \cite{kovac12PhC,kovacj13PhC}.}
\label{p.ACloss_setup}
\end{figure}

The coils are race-track coils with size of bore 140$\times$22 mm and height 23 mm. The measurement set-up picture is on figure \ref{p.ACloss_setup}. There is an unbalance between coils, since they are not completely identical. The unbalance is corrected by small pick up coils. Compensation is done by position adjustment of wire until different voltage signal of both coils is zero without measurement sample \cite{kovacj13PhC}.

The set-up provides applied field in the range of 0.1-100 mT amplitude for frequencies 72 and 144 Hz and up to 18 mT for a wider frequency range, between 2.3-1152 Hz. A double stage cryocooler allows to reach temperatures below 20 K. 

%%%%%%%%%%%%%%%%%%%%%%%%%%%%%%%%%%%%%%%%%%%%%%%%%%%%%%%%%%%%%%%%%%%%%%%%%%%%%%%%%%%%%%%%%%%%%%

\section{Cube sample for demagnetization by cross-field}
\label{s.demag_sample}

\begin{figure}[tbp]
\centering
{\includegraphics[trim=0 0 -20 0,clip,height=6.5 cm]{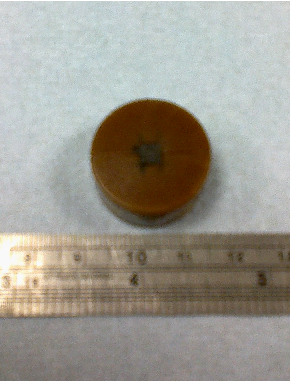}} 
\caption{The GdBCO cube sample inside a cylindrical holder.}
\label{p.demag_sample}
\end{figure}

A superconducting sample pellet with diameter of 12 mm and thickness 10 mm was grown by the Top Seeded Melted Grown TSGM method \cite{Namburi16JECS}. The GdBCO pellet contains 10wt\% of Ag. A cube was cut down from the center of the pellet, in order to have more homogeneous sample without any cracks or inhomogeneities coming from the growth method close to the edges \cite{Namburi16JECS}. The cube size is $6.08\times 6.04\times 5.98$ mm$^3$ (figure \ref{p.demag_sample}, the cube sample is inside the cylinder holder). 

%%%%%%%%%%%%%%%%%%%%%%%%%%%%%%%%%%%%%%%%%%%%%%%%%%%%%%%%%%%%%%%%%%%%%%%%%%%%%%%%%%%%%%%%%%%%%%

\section{Measurement of bulk demagnetization by cross-fields}
\label{s.measurements_demag}

The measurement contains two operations. The first operation is to magnetize the sample and the second is to apply an alternating cross-field and measure demagnetization. 

\begin{figure}[tbp]
\centering
{\includegraphics[trim=0 0 0 0,clip,width=12.0 cm]{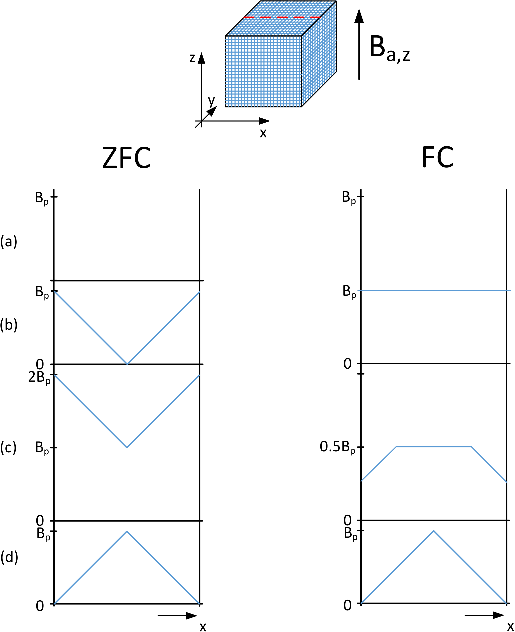}} 
\caption{The Zero Field Cool (ZFC) and Field Cool (FC) method to magnetize a superconducting bulk sample in three steps from top to down with three applied fields of value $B_p$, $2B_p$ and $0B_p$ for ZFC and $B_p$, $0.5B_p$ and $0B_p$ for FC. $B_p$ is the penetration field of the sample.}
\label{p.demag_FCM}
\end{figure}

The cube is magnetized by the Field Cool method FC instead of the Zero Field Cool \cite{Ainslie15SST} or pulse method \cite{Srpcic18IES}. The magnetization process with resulting trapped field by ZFC (left column) and FC (right column) is on figure \ref{p.demag_FCM}. The applied field is along the $z$ axes. The ZFC method starts the magnetization process with already cooled superconducting sample and zero applied field [figure \ref{p.demag_FCM}(a)]. The applied field of $B_p$ partially [figure \ref{p.demag_FCM}(b)] and with $2B_p$ fully saturates the sample [figure \ref{p.demag_FCM}(c)]. At the end of the magnetization process, ramping down the field to zero, there is trapped field [figure \ref{p.demag_FCM}(d)]. The FC method starts with the sample at room temperature and zero applied field [figure \ref{p.demag_FCM}(a)]. The applied field is ramped up to the penetration field [figure \ref{p.demag_FCM}(b)], which is followed by cooling down the sample. The sample is fully magnetized after ramping down the applied field, as shown at the profiles for $0.5B_p$ [figure \ref{p.demag_FCM}(c)] and for $0B_p$ [figure \ref{p.demag_FCM}(d)]. 

\begin{figure}[tbp]
\centering
{\includegraphics[trim=0 0 -10 0,clip,height=7.0 cm]{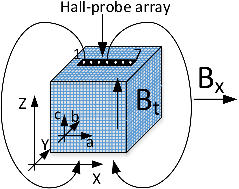}} 
\caption{The skecth of the cross-field orientation to the sample and to the Hall probe array.}
\label{p.demag_measur}
\end{figure}

For our case, the magnetization operation contains the following steps:
\begin{enumerate}
 \item Ramp up the split coil electromagnet up to 1.3 T in 10 s with already inserted the bulk sample at room temperature.
 \item Cool down the sample by liquid nitrogen over 15 minutes.
 \item Ramp down the electromagnet with ramp rate 13 mT/s  
\end{enumerate}  
The sample is fully magnetized along the $c$ plane and the demagnetizing operation starts after 900 seconds of the relaxation time, with these steps: 
\begin{enumerate}
 \item Move the sample into the second coil with transverse applied field $B_{ax,\rm{max}}$ up to 170 mT.
 \item Apply the transverse field of various amplitudes $B_{ax,\rm{max}}=B_t/2$, $B_t/4$, $B_t/8$ and frequencies 0.1, 1.0 Hz, where $B_t$ is trapped field 100 $\mu$m above the sample surface [figure \ref{p.demag_measur}(a)]. 
 \item Measure reduction of the trapped field by transverse field in the following 10 minutes by Hall-probe sensors. 
\end{enumerate}

\begin{figure}[tbp]
\centering
 \subfloat[][]
{\includegraphics[trim=0 0 -10 0,clip,width=7.0 cm]{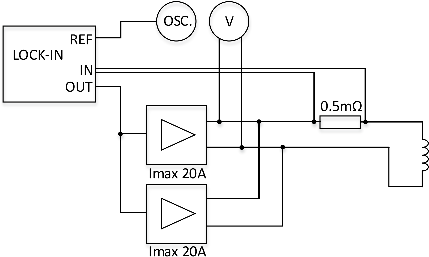}} 
 \subfloat[][]
{\includegraphics[trim=0 0 0 0,clip,height=7.5 cm]{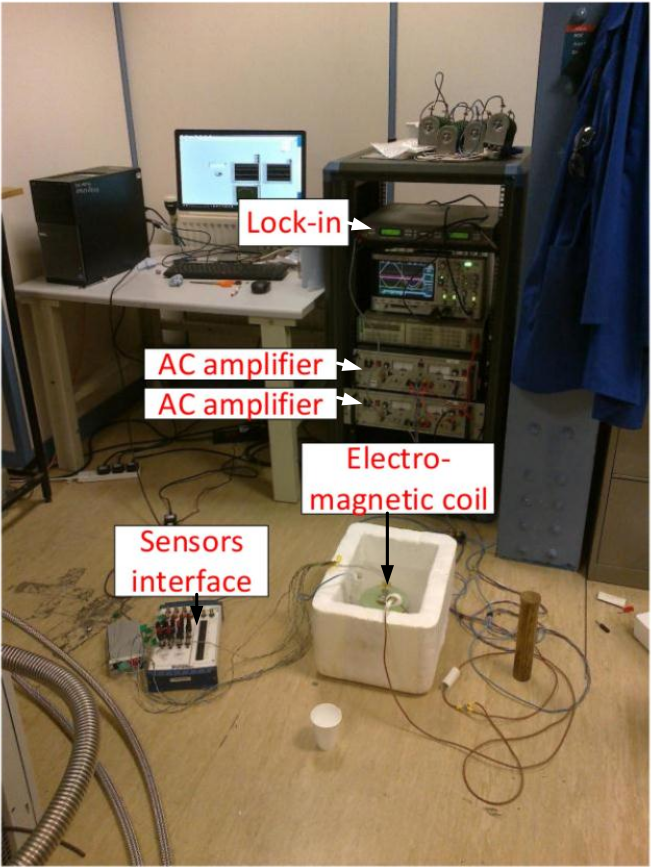}} 
\caption{(a) The electric scheme of demagnetization measurements by cross-field. (b) The set-up measurement with electromagnetic coil.}
\label{p.demag_schem}
\end{figure}

The measuring set up is based on a lock-in amplifier that generates AC signal and measures the voltage drop on a 0.5 m$\ohm$ shunt. The lock-in output AC signal is amplified by two parallel amplifiers, to generate AC current for the coil and also the required AC magnetic field. The set up is on figure \ref{p.demag_schem}(b) and the electrical scheme is on figure \ref{p.demag_schem}(a). The Hall probe array with 7 sensors Multi-7U \cite{Arepoc} measure the magnetic field 100 $\mu$m above the sample along the $x$ axes (figure \ref{p.demag_measur}). The sensor array is 3.5 mm long only while the sample side is around 6 mm. Therefore, the magnetic field is not measured close to the edges of the sample.

%%%%%%%%%%%%%%%%%%%%%%%%%%%%%%%%%%%%%%%%%%%%%%%%%%%%%%%%%%%%%%%%%%%%%%%%%%%%%%%%%%%%%%%%%%%%%%%
%%%%%%%%%%%%%%%%%%%%%%%%%%%%%%%%%%%%%%%%%%%%%%%%%%%%%%%%%%%%%%%%%%%%%%%%%%%%%%%%%%%%%%%%%%%%%%%

\chapter{Model tests and verification}
\label{c.test}

Superconductors are non-linear materials. This complicates the calculation of non-linear eddy currents, resulting in high computing times (section \ref{s.cal_time}). Therefore, new faster methods of calculation are required. Any new theoretical way of calculation must be validated by other methods or measurements in order to be reliable. There have been many studies of thin film models. The solution of infinite prisms by the CSM is in \cite{chen89JAP}. Analytical thin film studies of the CSM are in \cite{norris70JPD,halse70JPD,brandt95PRL,brandt95PRB,clem94PRB} and with isotropic power law in \cite{prigozhin98JCP,navau08JAP}. Thin film sample with a hole and $\vE(\vJ)$ power law is \cite{barrett12SST}.  

The following sections contain a validation of the MEMEP 3D method. The model tests use a ``thin film" geometry like a an infinitely thin sheet approximation (section \ref{s.film_test}) and a thin disk. We also qualitatively compare our results for a square thin film with those in \cite{brandt95PRL}. The AC coupling loss of filaments joined by a normal metal are compared with measurements in the section \ref{s.film_2filament}. 

%%%%%%%%%%%%%%%%%%%%%%%%%%%%%%%%%%%%%%%%%%%%%%%%%%%%%%%%%%%%%%%%%%%%%%%%%%%%%%%%%%%%%%%%%%%%%%%

\section{Comparison of thin film model with analytical formulas.}
\label{s.film_test}

The Critical State Model (CSM) is well defined for simple geometries like infinitely long thin strips and thin disks. Therefore, we model both situations in order to validate our numerical method. We have found a very good agreement for both cases, and hence we have successfully validated the method. 

\begin{figure}[tbp]
\centering
\subfloat[][]
{\includegraphics[trim=-10 0 -10 0,clip,width=4.5 cm]{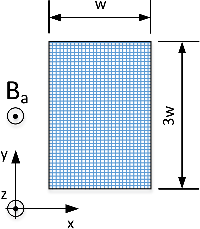}}
\subfloat[][]
{\includegraphics[trim=-10 0 -10 0,clip,width=4.5 cm]{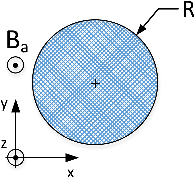}}
\subfloat[][]
{\includegraphics[trim=-10 0 -10 0,clip,width=4.5 cm]{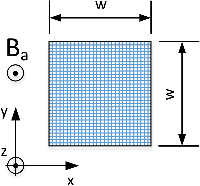}}
\caption{Sketch of the thin film modelling geometry (a) thin strip (b) thin disk (c) square film. The results obtained with these geometries agree with calculations with previous models.}
\label{f.film_geometry}
\end{figure}

The first case of the long thin strip is with dimensions $4\times 12\times 0.001$ mm and sinusoidal perpendicular applied magnetic field with frequency 50 Hz. The sketch of the geometry is on the figure \ref{f.film_geometry}(a). The mesh is created by square elements with $107\times 321\times 1$ cells, which results in more than 34000 degrees of the freedom. The critical current density $J_c$ is $2.72\cdot10^{10}$ A/m$^2$.

\begin{figure}[tbp]
\centering
{\includegraphics[trim=90 0 90 0,clip,width=5.5 cm]{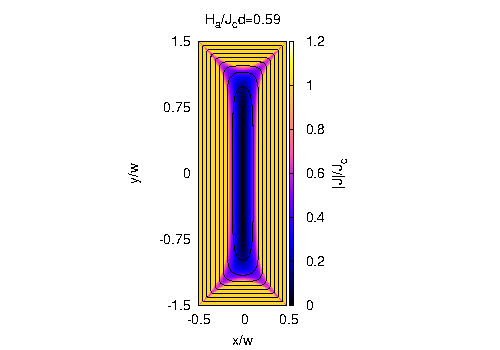}}
\caption{The modulus of the current density and current lines at the peak of the AC applied magnetic field 20 mT. The current lines are parallel to $y$ axis except close to the ends of the tape.}
\label{f.strip}
\end{figure}

\begin{figure}[tbp]
\centering
{\includegraphics[trim=0 0 0 0,clip,width=8.5 cm]{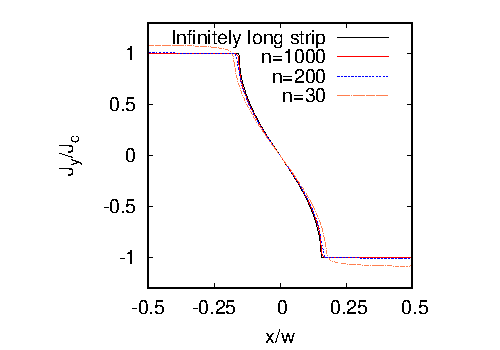}}
\caption{The $J_y$ component at the mid cross-section plane $y=0$ with various power-law exponents $n$ 1000,200,30. The model with $n$ 1000 agrees with the thin strip formula \ref{e.Brandt}.}
\label{f.strip1}
\end{figure}

The gradual penetration of the current density is shown on figure \ref{f.strip} with current lines at the peak of the applied field of 20 mT. The current path is changing only close to the ends of the tape, otherwise the $J_y$ component of the current density is parallel to the $y$ axis. The mid cross-section of the strip at the plane $y$=0 mm is on figure \ref{f.strip1}. The model assumes various power-law exponents like $n$=1000,200,30, in order to get results comparable with the CSM and real values of the superconducting tapes. The case of $n$=1000 is basically the same as the CSM, thanks to the high power-law exponent. The $J_y$ component with $n$=1000 agrees very well with the thin strip analytical formula of \cite{brandt93PRBa,halse70JPD}, being at the initial magnetization curve

\begin{eqnarray}
\ J_{y}(x) & = & \frac{2J_{c}}{\pi}\arctan{ \frac{cx}{\sqrt{(b^{2}-x^{2})}}},  |x|<b,  \nonumber \\ 
           & = & J_{c}\frac{x}{|x|},   b<|x|<w/2,   
\label{e.Brandt} \nonumber \\
\end{eqnarray} 
where 
\begin{equation}
{b=\frac{w/2}{\cosh{\frac{H_a}{H_c}}}},   
\label{} \\
\end{equation}  
\begin{equation}
{c=\tanh{\frac{H_{a}}{H_{c}}}},   
\label{} \\
\end{equation}  
and
\begin{equation}
{H_{c}=\frac{J_{c}d}{\pi}}.   
\label{} \\
\end{equation}
Above, $w$ is the width of the the tape, $H_a$ is the instantaneous applied field and $d$ is the thickness of the sample. 

Real superconductors present lower $n$ factors. The $n$-factor 30 is a realistic value of REBCO superconductors. In the Critical State Model, the current density never exceeds $J_c$ but the real superconductor with $n$=30 allows higher $|\vJ|$, around 10\% above $J_c$ in our case, which figure \ref{f.strip1} confirms. 

\begin{figure}[tbp]
\centering
{\includegraphics[trim=0 0 0 0,clip,width=8.5 cm]{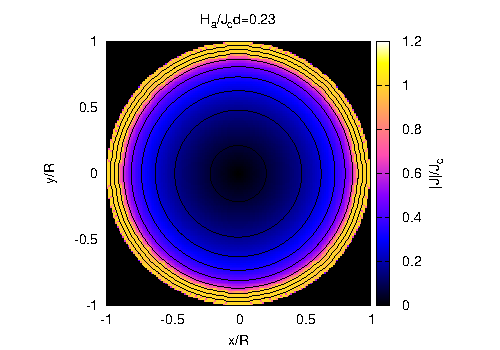}}
\caption{The modulus of the current density at applied field 7.8 mT for a thin disk. The current lines form circular loops, while the model does not assume any cylindrical symmetry.}
\label{f.disk}
\end{figure}

\begin{figure}[tbp]
\centering
{\includegraphics[trim=40 0 25 0,clip,width=8.5 cm]{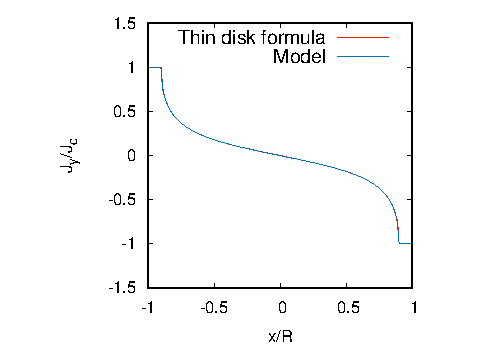}}
\caption{The $J_y$ component at the mid cross-section plane $y=0$ of a thin disk with a power-law exponent $n=$1000 at applied field of 7.8 mT. The model agrees very well with the thin disk formula \cite{clem94PRB}.}
\label{f.disk1}
\end{figure}

\begin{figure}[tbp]
\centering
{\includegraphics[trim=40 0 25 0,clip,width=8.5 cm]{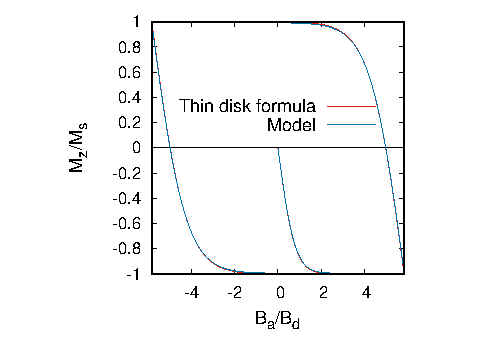}}
\caption{The thin disk magnetization loop agrees with the analytical formulas \ref{e.Clem1}.}
\label{f.disk2}
\end{figure}

The second simple geometry is a thin disk with radius $R=$6 mm and thickness $d=$1 $\mu$m (sketch \ref{f.film_geometry}(b)). The critical current density is $2.72\cdot10^{10}$ A/m$^2$ and sinusoidal applied magnetic field of 50 Hz. The model takes $n$ factor 1000. The penetration of the current density at 7.8 mT is on figure \ref{f.disk}. The current path is circular according the outer shape of the sample, in spite of the fact that the applied vector potential, $\vA_a$, follows the $y$ direction (section \ref{s.applied_Aa}). The $J_y$ component at the mid cross-section (figure \ref{f.disk1}) agrees very well with the analytical equation (\ref{e.Clem}). The formula in \cite{clem94PRB} for thin disk is

\begin{eqnarray}
\ J_{y}(r) & = & -\frac{2J_{c}}{\pi}\arctan{ \frac{\frac{r}{R}\sqrt{(R^2-a^2)}}{\sqrt{(a^2-r^{2})}}},  r<R,  \nonumber \\ 
           & = & -J_{c},   a\leq r<R,   
\label{e.Clem} \nonumber \\
\end{eqnarray} 
where 
\begin{equation}
{a=\frac{R}{\cosh{\frac{H_{a}}{H_{d}}}}},   
\label{} \\
\end{equation}  
and
\begin{equation}
{H_{d}=\frac{J_{c}d}{2}}.   
\label{} \\
\end{equation}
The magnetization formula is split into 3 branches. The initial magnetization is 
\begin{equation}
{M_{zi}(H_a)= -\chi_0H_aS\left(\frac{H_a}{H_d}\right)},
\label{e.Clem1}
\end{equation}
where
\begin{equation}
{\chi_0=\frac{8R}{3\pi d}}
\label{}
\end{equation}
and
\begin{equation}
{S(x)=\frac{1}{2x}\left[\arccos{\frac{1}{\cosh x}}+\frac{\sinh x}{\cosh^2x}\right]}.
\label{}
\end{equation}
Then, the magnetization from positive to negative peak of the applied magnetic field and vice versa is 
\begin{equation}
{M_{z\downarrow}=M_{zi}(H_m)-2M_{zi}\left(\frac{H_m-H_a}{2}\right)} 
\label{e.Clem2}
\end{equation}
and
\begin{equation}
{M_{z\uparrow}=-M_{zi}(H_m)+2M_{zi}\left(\frac{H_a-H_m}{2}\right)}, 
\label{e.Clem3}
\end{equation}
where $H_m$ is the maximum applied field. The relation between $M_{zi}$ and $M_{z\downarrow}$, $M_{z\uparrow}$ applies also for other shapes with high symmetry, such as thin strips slabs and cylinders. The hysteresis loop of the thin disk on figure \ref{f.disk2} agrees with equations (\ref{e.Clem1},\ref{e.Clem2},\ref{e.Clem3}). 

Both simple situations confirm that the MEMEP 3D method can describe the Critical State Model and agrees with it. This is important validation of the model. 

%%%%%%%%%%%%%%%%%%%%%%%%%%%%%%%%%%%%%%%%%%%%%%%%%%%%%%%%%%%%%%%%%%%%%%%%%%%%%%%%%%%%%%%%%%%%%%%

\section{Finite-length superconducting thin film with constant ${J_{c}}$}
\label{s.film_Jc}

A more advanced validation of the method is by modelling of finite size rectangular films with all finite size effects. The qualitative comparison is by current loops inside a square film. This is a common shape for investigation of characteristic parameters of superconducting films like $J_c$ or AC loss. Therefore, the validation by square film is desirable. Further, it serves as a basic validation step for more complex modelling as it is anisotropy with perpendicular and parallel applied fields. The following modelling case is the usual electromagnetic response of a superconducting square film on perpendicular applied magnetic fields to the surface with constant $J_c$. The size of the sample is $12\times 12\times 0.001$ mm and the sketch of the square case is on figure \ref{f.film_geometry}(c). The applied magnetic field is 50 mT with frequency 50 Hz. The number of the elements in the mesh is $60\times 60\times 1$. The critical current density is $3\cdot 10^{10}$ A/m$^2$ and we take a realistic $n$ factor of 30. The model calculated 160 time steps per cycle.  

\begin{figure}[tbp]
\centering
 \subfloat[][]
{\includegraphics[trim=30 0 50 0,clip,height=4.5 cm]{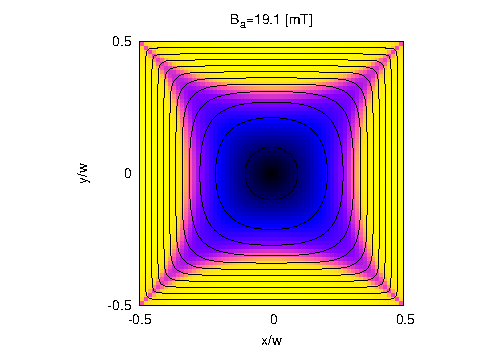}} 
 \subfloat[][]
{\includegraphics[trim=75 0 75 0,clip,height=4.5 cm]{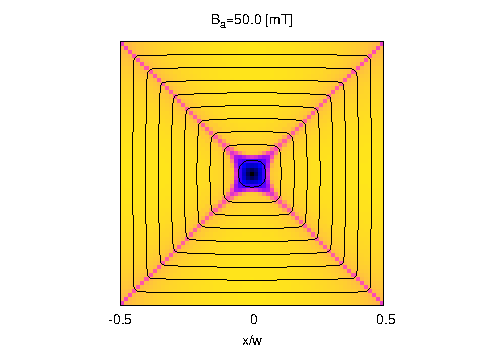}} 
 \subfloat[][]
{\includegraphics[trim=50 0 25 0,clip,height=4.5 cm]{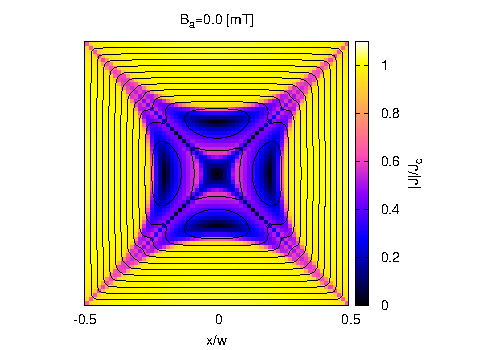}}
\caption{The gradual penetration of the current density into a superconducting square thin film with constant $J_c$ dependence and perpendicular applied magnetic field to the surface with amplitude 50 mT.}  
\label{f.Jc}
\end{figure}

The small applied magnetic field of only 19.1 mT already causes penetration of the screening current density into the square sample [figure \ref{f.Jc} (a)]. The penetration depth of the critical current density is symmetric along both $x$ and $y$ axes. The sample is almost fully saturated at the peak of the applied field [figure \ref{f.Jc}(b)]. When the applied magnetic field decreases to zero, the region with $|\vJ|=J_c$ penetrates again from the edges to the center of the sample with the opposite sign, and hence the penetration front erases the previous screening current. 

The applied field must be two times higher in order to achieve a $J_c$ penetration to the same depth. Therefore, at the remanent state with zero applied field the sample is penetrated only partially (figure \ref{f.Jc}(c)). The sample becomes completely penetrated at the minus peak of the applied magnetic field, $B_a$= -50 mT, achieving the same current density as figure \ref{f.Jc}(b) but with opposite sign.

The measurements of thin films are based on inversion. The inversion method measures the magnetic field close to the surface sample and then calculates $\vJ$ from $B_z$ \cite{solovyov09IES,Yoo08PCS,jooss02RPP}, which qualitatively agree with our results.

\begin{figure}[tbp]
\centering
 \subfloat[][]
{\includegraphics[trim=60 0 60 0,clip,width=6.5 cm]{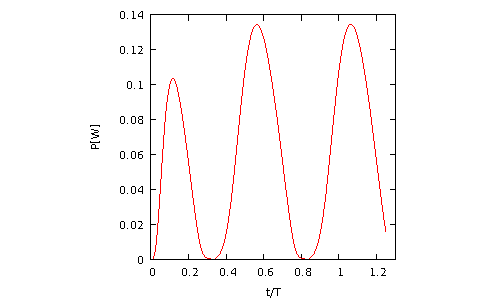}}
 \subfloat[][]
{\includegraphics[trim=60 0 60 0,clip,width=6.5 cm]{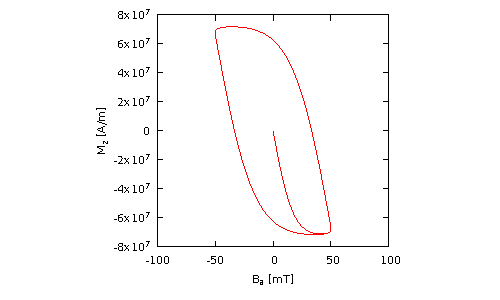}}
\caption{(a) AC power loss and (b) hysteresis loop of the superconducting thin film with constant $J_c$ and power-law exponent n=30. The applied magnetic field amplitude is 50 mT and the frequency is 50 Hz.} 
\label{f.lossJc}
\end{figure}

The MEMEP 3D method is able to calculate the electric field, and hence also instantaneous power loss. The AC loss is on figure \ref{f.lossJc}(a) and the hysteresis loop is on figure \ref{f.lossJc}(b). These results qualitatively agree with previous calculations on thin films \cite{brandt93PRBa,prigozhin98JCP,navau08JAP}. Therefore, these calculations further validate the results. 

%%%%%%%%%%%%%%%%%%%%%%%%%%%%%%%%%%%%%%%%%%%%%%%%%%%%%%%%%%%%%%%%%%%%%%%%%%%%%%%%%%%%%%%%%%%%%%%

\section{Finite superconducting thin film with ${J_c(B)}$ dependence}
\label{s.film_JcB}

The next test includes $J_c(B)$ dependence in the model, since the superconductor reduces $J_c$ by local magnetic field even by self-field, and hence the modelling tool becomes more realistic and complex. The $J_c(B)$ dependence is a further step to model force-free anisotropy inside superconductors and increases the accuracy of the AC loss calculation \cite{jiangZ14SST,souc09SST,pardo12SSTa}. The calculations with $J_c(B)$ dependence have the same input parameters like the previous ones with constant $J_c$ in section \ref{s.film_Jc}, except that the method assumes Kim model as $J_c(B)$ dependence [equation (\ref{e.Kim})] with $m$=0.5 and $B_0$=20 mT. With these parameters, we choose a value of $J_{c0}$ so that the transport critical current $I_c$ on an infinitely long tape is the same as that for constant $J_c=3\cdot 10^{10}$ A/m$^2$, resulting in $J_{c0}=3.615\cdot 10^{10}$ A/m$^2$.  

\begin{figure}[tbp]
\centering
 \subfloat[][]
{\includegraphics[trim=50 0 25 0,clip,height=6.0 cm]{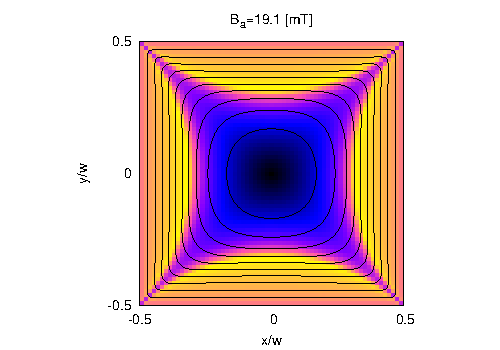}} 
 \subfloat[][]
{\includegraphics[trim=50 0 25 0,clip,height=6.0 cm]{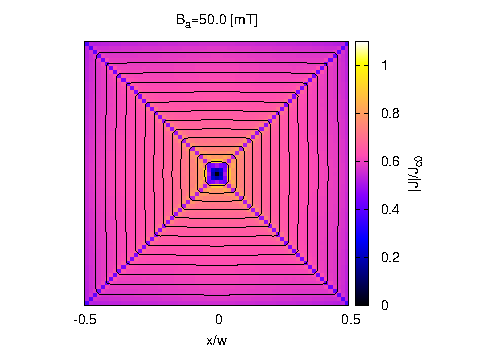}}\\ 
 \subfloat[][]
{\includegraphics[trim=50 0 25 0,clip,height=6.0 cm]{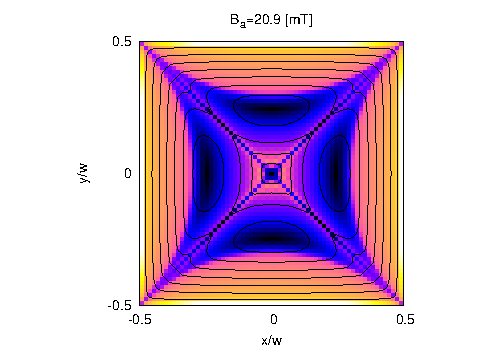}}
 \subfloat[][]
{\includegraphics[trim=50 0 25 0,clip,height=6.0 cm]{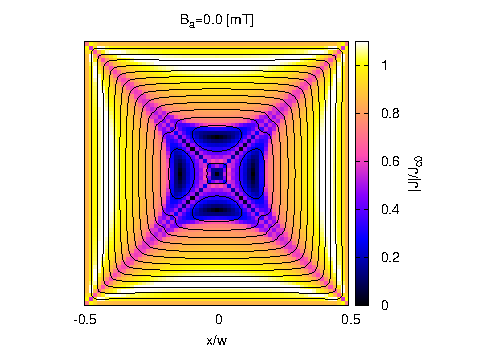}}
\caption{The gradual penetration of the current density into the superconducting square thin film with $J_c(B)$ dependence and perpendicular applied magnetic field to surface of 50 mT amplitude.}
\label{f.JcB}
\end{figure}

The applied magnetic field magnetizes the square film, and hence the current density penetrates the sample [figure \ref{f.JcB}(a)]. The penetration depth is roughly the same as for $J_c$ constant [figure \ref{f.Jc}(a)]. The current density is symmetric regarding penetration along both $x$ and $y$ axis and $|\vJ|$ at the critical region front is around $J_{c0}$. The local magnetic field decreases the critical current density, and so $|\vJ|$ does between the sample border and the moving front, where $B_z\approx 0$. The square film is almost fully saturated at the peak of the applied field with current density magnitude around $0.5J_c$ at the edges [figure\ref{f.JcB}(b)]. When the applied magnetic field decreases back to zero, the new screening current penetrates from the edges to the center of the sample. The applied field of around 20 mT creates zones with zero local magnetic field close to the edges, where the screening current is around $J_{c0}$ [figure\ref{f.JcB}(c)]. The penetration front and zero local field zones move further into the center by continual decreasing the applied field to the remnant state [figure \ref{f.JcB}(d)] and beyond. The penetration depth is higher compared to the case of constant $J_c$ (figure \ref{f.Jc}). Since $J_c$ is reduced by the local magnetic field, the screening current needs bigger penetrated zone to shield the applied field.

\begin{figure}[tbp]
\centering
 \subfloat[][]
{\includegraphics[trim=60 0 60 0,clip,width=6.5 cm]{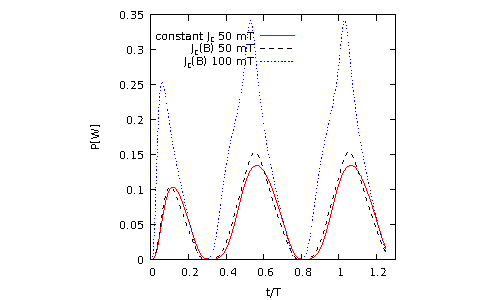}}
 \subfloat[][]
{\includegraphics[trim=70 0 70 0,clip,width=6.0 cm]{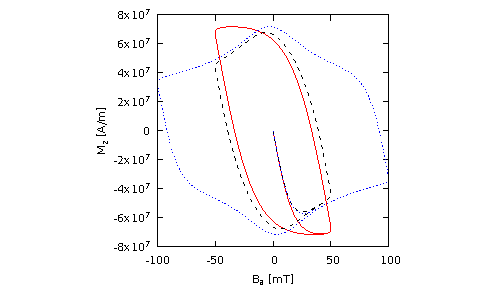}}
\caption{Superconducting thin film (a) AC power loss and (b) hysteresys loop of the square with ${J_{c}(B)}$ dependence (Kim model) and sinusoidal applied magnetic field ($B_a=B_{a,\rm{max}}\sin\omega t$) with amplitude of 50 mT and 50 Hz frequency.} 
\label{f.lossJcB}
\end{figure}

The magnetization loop with $J_c(B)$ dependence showed lower magnetization closer to the applied field peaks compare to the constant $J_c$. Since the $J_c$ is reduced by the local magnetic field. The magnetization curve is higher close to the remanent state, because of higher penetration depth [\ref{f.lossJcB}(b)]. The instantaneous AC loss between $B_a=+B_{a,\rm{max}}$, being $B_{a,\rm{max}}$ the peak, and the remanent is higher for $J_c(B)$ [figure \ref{f.lossJcB}(a)]. The reason is the following. For $J_c(B)$, the magnetic field decreases the local $J_c$ between $B_a=+B_{a,\rm{max}}$ and the remanent. This causes a larger penetration of the critical zone, increasing the instantaneous AC loss. The second calculation of the AC loss with $J_c(B)$ dependence and applied field of 100 mT shows even higher instantaneous AC loss, since the time derivative of the applied field is larger. The reduced current density at the peaks of the AC field decreases as well the magnetization [figure \ref{f.lossJcB}(b)]. The area of the magnetization loop increases [figure \ref{f.lossJcB}(b) point line], therefore the AC loss per cycle is higher. The total AC loss per cycle is in table \ref{t.film_ACloss}. The cycle is from the first positive peak up to the second positive peak of the applied field. The total AC loss is similar for both $J_c$ constant and $J_c(B)$ dependence with applied field 50 mT. The cause is that the local magnetic field is of the order of self-field. The two times higher applied field amplitude of 100 mT shows roughly two times higher AC loss [table \ref{t.film_ACloss}(3)].

\begin{table}[tpb]
\begin{center}
\begin{tabular}{llll}
\hline
\hline
{\bf{Number}} & {\bf{Dependence}} & {B$_{a,\rm{max}}$[mT]} & {{\bf AC loss}[mJ] } \\
\hline
1 & $J_c$ & 50 &1.22503	\\
2 & $J_c(B)$ & 50 & 1.26545	\\
3 & $J_c(B)$ & 100 & 2.70117	\\
\hline
\hline
\end{tabular}
\caption{Calculated AC loss per cycle for different cases of the thin film.}
\label{t.film_ACloss}
\end{center}
\end{table}

%%%%%%%%%%%%%%%%%%%%%%%%%%%%%%%%%%%%%%%%%%%%%%%%%%%%%%%%%%%%%%%%%%%%%%%%%%%%%%%%%%%%%%%%%%%%%%%

\section{Coupling effects in multi-filamentary tapes}

Modelling the coupling effects is important, in order to fully understand the AC loss in the multi-filamentary tapes or coupled tapes. The coupling currents causes the dominant part of the AC loss in low fields.

%%%%%%%%%%%%%%%%%%%%%%%%%%%%%%%%%%%%%%%%%%%%%%%%%%%%%%%%%%%%%%%%%%%%%%%%%%%%%%%%%%%%%%%%%%%%%%%

\subsection{Coupling effects in two-tape conductor compared with FEM model}
\label{s.film_filament}

\begin{figure}[tbp]
\centering
 \subfloat[][]
{\includegraphics[trim=0 0 0 0,clip,width=5.5 cm]{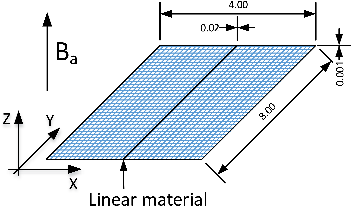}}
 \subfloat[][]
{\includegraphics[trim=0 0 0 0,clip,width=5.3 cm]{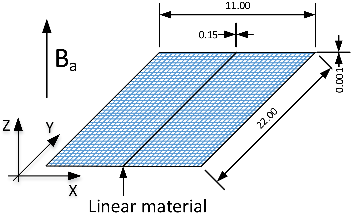}}
\caption{The geometry of the long strip with filaments. The superconductors are electromagnetically coupled by a linear material. Dimensions are in mm. (a)(b) Two-filament tape with different dimensions.} 
\label{f.filament_geometry}
\end{figure}

The last results from the thin film series is a model of coupling effects like coupling AC loss and coupling current. The modelling situation is two superconducting tapes coupled by a normal metal. The sinusoidal applied magnetic field is perpendicular to the surface, with an amplitude of 20 mT and frequency of the power network (50 Hz). The sample size is $4\times 8\times 0.001$ mm$^3$ and the sketch of the geometry is on figure \ref{f.filament_geometry}(a). The normal conducting material is of 20 $\mu$m width with the effective resistivity $\rho_{eff}$ of $2.4\cdot 10^{-11}$ $\Omega$m. The effective resistivity is defined from measurement. The conductivity $\sigma_m$ is measured along the tapes width and the resistive gap. The resistive width $w_{r,m}$ is estimated from the sample dimension. The model assumes any width of the resistive gap $w_{r,cal}$, and hence $\sigma_{cal}=w_{r,m}\sigma_m/w_{r,call}$. From where the effective resistivity is $\rho_{eff}=1/\sigma_{cal}$. The critical current density is $3\cdot 10^{10}$ A/m$^2$. The mesh is dedicated to 40 cells per each superconducting filament and 1 cell for a metallic joint along the $x$ axis. The total number of elements is $(80+1)\times 160\times 1$.      

\begin{figure}[tbp]
\centering
 \subfloat[][]
{\includegraphics[trim=80 0 90 22,clip,height=6.0 cm]{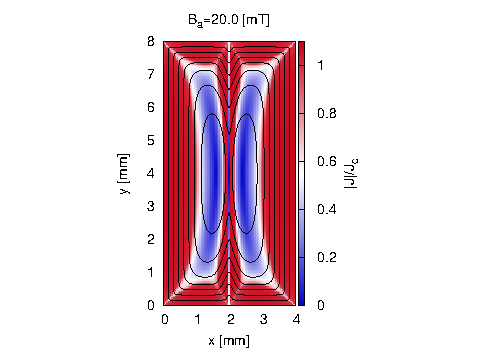}}
 \subfloat[][]
{\includegraphics[trim=30 0 -30 0,clip,height=5.95 cm]{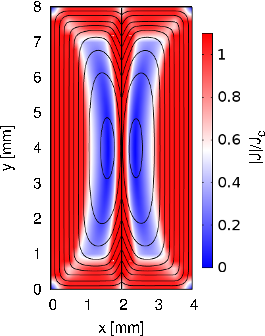}} 
\caption{Distribution of the current density in a two-filament tape at the applied field of 20 mT with constant $J_c$. The results are of two methods: (a) MEMEP 3D, (b) FEM in $\vH$ formulation.}
\label{f.filament_comparison}
\end{figure}

The usual gradual penetration of the current density at the peak of the applied field is on the figure \ref{f.filament_comparison}(a). The result is compared with the same case calculated by Finite Element Method (FEM) in the $\vH$ formulation by Francesco Grilli \ref{f.filament_comparison}(b). Both methods show the same current path and current density \cite{Kapolka18IES}.

%%%%%%%%%%%%%%%%%%%%%%%%%%%%%%%%%%%%%%%%%%%%%%%%%%%%%%%%%%%%%%%%%%%%%%%%%%%%%%%%%%%%%%%%%%%%%%%

\subsection{Coupling effects in two-filament tape and measurements}
\label{s.film_2filament}

The similar coupling case with different sample size is calculated and later compared with the measurements (details in section \ref{s.measurements_coupling}). The sample size is $11\times 22\times 0.001$ mm with 150 $\mu$m width of the coupling linear material [figure \ref{f.filament_geometry}(b)]. The critical current density of superconductor is $2.72\cdot 10^{10}$ A/m$^2$, which is the average value of measured tape ($I_c=160$ A of 6 mm width tape) and the effective resistivity (defined in section \ref{s.film_filament}) of linear material is $39.4\cdot 10^{-10}$ $\Omega$m. The applied field frequency is 144 Hz and the relatively low amplitude of 20 mT ensures not fully saturated sample. The mesh contains $(70+1)\times 141\times 1$ elements. 

\begin{figure}[tbp]
\centering
 \subfloat[][]
{\includegraphics[trim=80 0 108 0,clip,height=7.5 cm]{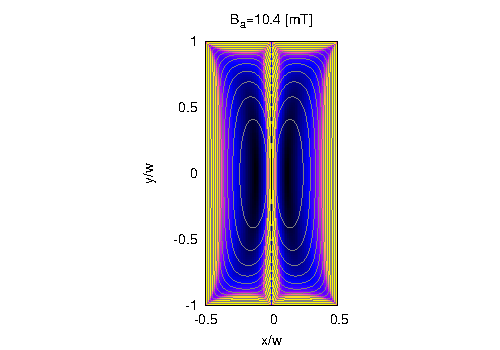}} 
 \subfloat[][]
{\includegraphics[trim=125 0 120 0,clip,height=7.5 cm]{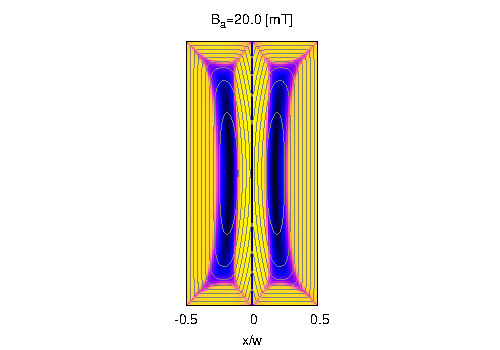}} 
 \subfloat[][]
{\includegraphics[trim=100 0 95 0,clip,height=7.5 cm]{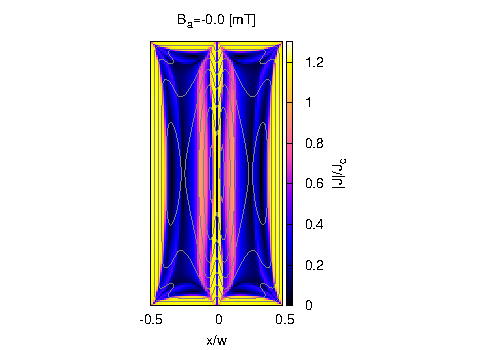}} 
\caption{Gradual penetration of the current density into the two-filament tape with constant $J_c$ coupled by a normal material due to perpendicular applied magnetic field to the surface (50 mT amplitude).} 
\label{f.filament_2}
\end{figure}

The screening current penetrates into the sample already under the small applied field of 10.4 mT [figure \ref{f.filament_2}(a)]. The current lines are parallel to the $x$ and $y$ axis except the metallic joint. The metallic joint conductivity is lower than superconductor and therefore the current line loop bends around the metallic joint and passes through it further at the center of the metallic line [figure \ref{f.filament_2}(a)]. The superconductor around the metallic joint is fully saturated with $J_c$. There are two paths of the current loops. The first one follows the sample outline and passes through the metallic material. The second one closes the loop in his own superconducting filament [figure \ref{f.filament_2}(b)]. The first type of the current loop is the coupling current. The remnant state presents a complex current path of the new penetration front coming from the edges. Now, there appear current lines that cross the normal joint but they close in the superconducting region next to the joint, while at the peak of the AC field current lines crossing the joint close to the outer edges of the superconductor [figure \ref{f.filament_2}(c)]. 

\begin{figure}[tbp]
\centering
{\includegraphics[trim=0 0 0 0,clip,height=7.5 cm]{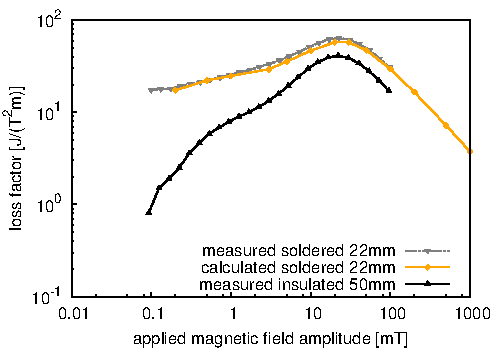}}
\caption{The AC loss factor, defined as the loss per cycle and length divided by the square of the applied field amplitude. Comparison of the measurements and model with two-tape conductor and frequency of the perpendicular applied field of 144 Hz. The calculations for the coupled case agree with the measurements.}
\label{f.filament_ACloss}
\end{figure}

The comparison of the AC loss measurement with the sample of the same parameters is on figure \ref{f.filament_ACloss}. The samples are measured in the applied field range from 0.1 to 100 mT amplitude (section \ref{s.measurements_coupling}). The calculation of the AC loss is extended to 1T. The AC loss measurements agree very well with the calculation. The total AC loss contains two contributions: the hysteresis loss from superconducting part and the coupling loss from metallic joint. The maximum 4\% deviation of the AC loss is around 20 mT, being around the value of the self-field of the tape. The calculation assumes constant $J_c$, and hence the modelling accuracy can be even more precise by including the $J_c(B)$ dependence. The measurement of the electrically uncoupled 50 mm long tapes is in the same graph with the black curve on figure \ref{f.filament_ACloss}. For this case, the AC loss contains only hysteresis loss created in the superconductors. Since it is much smaller than for the coupled tapes, the coupling loss is the dominant part of the total AC loss in the soldered sample. Understanding the effects of the linear material is key for the reduction of AC loss in such structures, since the coupling current in the metal joint creates increases the loss by creating coupling loss and increasing superconductor loss. The coupling loss is a main part of the AC loss in low applied fields, while superconducting loss dominates at high applied fields. 

%%%%%%%%%%%%%%%%%%%%%%%%%%%%%%%%%%%%%%%%%%%%%%%%%%%%%%%%%%%%%%%%%%%%%%%%%%%%%%%%%%%%%%%%%%%%%%%
%%%%%%%%%%%%%%%%%%%%%%%%%%%%%%%%%%%%%%%%%%%%%%%%%%%%%%%%%%%%%%%%%%%%%%%%%%%%%%%%%%%%%%%%%%%%%%%
%%%%%%%%%%%%%%%%%%%%%%%%%%%%%%%%%%%%%%%%%%%%%%%%%%%%%%%%%%%%%%%%%%%%%%%%%%%%%%%%%%%%%%%%%%%%%%%
%%%%%%%%%%%%%%%%%%%%%%%%%%%%%%%%%%%%%%%%%%%%%%%%%%%%%%%%%%%%%%%%%%%%%%%%%%%%%%%%%%%%%%%%%%%%%%%

\chapter{Results and discussion}
\label{c.results}

Superconductors are still relatively expensive conductors. Therefore, it is an advantage to have modelling predictions of any superconducting machine before it is built and measured. Superconductors are anisotropic materials with critical current density affected by the magnetic field magnitude and its orientation. The critical current density also depends on the direction of the current density relative to the magnetic field, appearing force-free effects when $\vJ$ has a parallel component to $\vB$. Therefore, models with high accuracy and with $J_c(B)$ and anisotropy dependence are very required. 3D models with high calculation speed are needed to model properly finite superconducting samples and superconductor machines. 

The AC loss in superconductors is important, since it can inflict losses the cryogenic system, causing malfunction of the machine when the heat generation is higher than the cryogenic system can pump out. Striations reduce AC loss. Therefore, AC loss prediction in multi-filament tape is important. Bulk samples are an alternative to permanent magnets, but cubic bulks are still not well understood. Therefore, a detailed study of screening current is very interesting. The magnetization of the prism with decreasing thickness shows the intermediate state between infinitely thin sheet approximation and infinite bar. The tilted applied field angle in bulks and stacks of tapes shows the fundamental difference between them. Cross-field demagnetization predicts how long the magnetization in bulks withstands under cross-fields, which is the situation in motors. The anisotropy study of thin films and prisms shows the real response of REBCO tapes as highly anisotropic material.

Therefore, the MEMEP 3D results are focused on the previously explained cases. The following sections contain original results obtained by the MEMEP 3D method. It is the most important part of thesis, apart from the method itself (section \ref{s.math_model}). The results study fundamental cases focusing on the screening current, magnetization, demagnetization and anisotropy in the samples. The results are important for further improvement of the superconducting power applications.
 
%%%%%%%%%%%%%%%%%%%%%%%%%%%%%%%%%%%%%%%%%%%%%%%%%%%%%%%%%%%%%%%%%%%%%%%%%%%%%%%%%%%%%%%%%%%%%%%

\section{Coupling loss in multi-filament tape}
\label{s.film_6filament}

The following AC loss study is based on infinitely thin film approximation (section \ref{s.film_approx}). Other thin film and coupling situations are in section \ref{c.test}. Although other authors modelled electrically coupled striated tapes, the results here provide higher accuracy, and hence these calculations can be consider as an original results. 

The following studied coupling situation is a tape with 6 filaments. The size of the sample is $12\times 24 \times 0.001$ mm (figure \ref{f.filament_6g}) with the applied field of 20 mT and frequency 144 Hz. The gap between superconductors is 20 $\mu$m with resistivity 39.4$\cdot 10^{-10}$ $\ohm$m. 

The coupling current lines show the same behaviour like the case of the two filaments tape (see figure \ref{f.filament_2}). These current lines pass through the metallic joints at instantaneous applied fields of already 10.4 mT [{figure \ref{f.filament_6}(a)]. The current density reaches the value of $J_c$ close to the edges of the sample and around the metallic joints [figure \ref{f.filament_6}(b)]. The remnant state shows the mixed state of the new penetration current at the edges and the previous screening current at the center of the sample [figure \ref{f.filament_6}(c)]. 

A transient state appears in the model since it contains metallic parts. At remanent, there also appear closed current loops around the metal joints. The superconducting and normal magnetization currents present a phase shift from each other. Both the superconducting and normal magnetization currents present a phase shift with the applied field. This phase shift is larger than that of the two-filament tape (section \ref{s.film_2filament}), and hence the phase shift increases with the number of filaments. The phase shift is the most visible in the remnant state. The penetration front with the opposite sign penetrates less than half of the previous penetration depth. The method allows to model any number of filaments, as long as the mesh contains sufficiently high number of the elements for each filament. The current state of the MEMEP 3D code can accurately model up to 10 filaments. 

\begin{figure}[tbp]
\centering
{\includegraphics[trim=0 0 0 0,clip,width=6.5 cm]{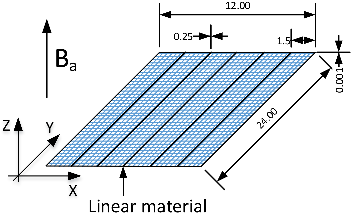}}
\caption{The geometry of the long strip with 6 filaments. Dimensions are in mm.} 
\label{f.filament_6g}
\end{figure}

\begin{figure}[tbp]
\centering
 \subfloat[][]
{\includegraphics[trim=80 0 108 0,clip,height=7.5 cm]{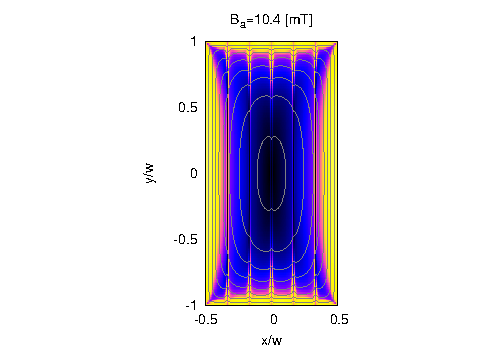}} 
 \subfloat[][]	
{\includegraphics[trim=125 0 120 0,clip,height=7.5 cm]{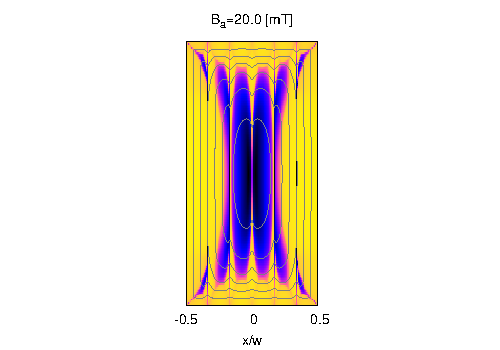}} 
 \subfloat[][]
{\includegraphics[trim=100 0 95 0,clip,height=7.5 cm]{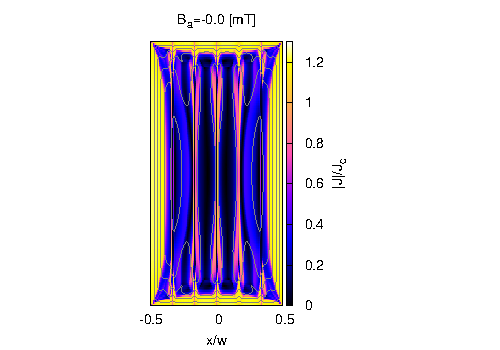}} 
\caption{The gradual penetration of the current density into the six filament tape coupled by a linear material. Due to a perpendicular applied magnetic field to the surface with amplitude 20 mT and frequency 144 Hz. The superconductor has constant $J_c$.} 
\label{f.filament_6}
\end{figure}

The six-filament tape at the peak of the applied field (20 mT) shows that the coupling current path passes through all the metallic joints [figure \ref{f.filament_6}(b)]. The largest AC loss density is at the linear material [figures \ref{f_filament_Ploss}], where the maximum loss density is 60 times higher than that in the superconductor. However, when analysing the total power loss, the superconductor is responsible of most of the dissipation, as discussed below.

The AC loss is split into coupling and hysteresis loss on figures \ref{f.filament_ACloss_f}. The applied field of 10 mT creates comparable AC losses in superconducting and normal material [figure \ref{f.filament_ACloss_f}(a)]. However, at the applied field of 20 mT the peak instantaneous AC loss in the superconductor is 6 times higher than at the normal metal [figure \ref{f.filament_ACloss_f}(b)]. Therefore, at very low field (below 10 mT in our case) the dominant part of entire AC loss is coupling loss. 

The AC loss factor dependence on various applied field frequencies 1.4, 144 and 1400 Hz with the superconducting gaps of 90 $\mu$m is on figure \ref{f.6filament_ACloss}. The model shows the highest AC loss at 144 Hz. The peaks of the AC loss factor are at the same amplitude, except small shift at 1.4 kHz frequency. 

\begin{figure}[tbp]
\centering
{\includegraphics[trim=70 0 70 0,clip,width=5.5 cm]{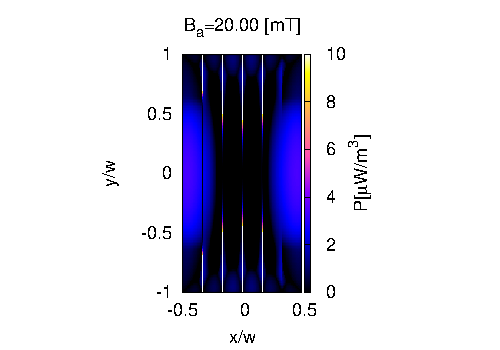}}
\caption{Instantaneous power loss at the peak of the applied field for the same situation as figure \ref{f.filament_6}.}
\label{f_filament_Ploss}
\end{figure}

\begin{figure}[tbp]
\centering
 \subfloat[][]
{\includegraphics[trim=45 0 55 0,clip,width=6.5 cm]{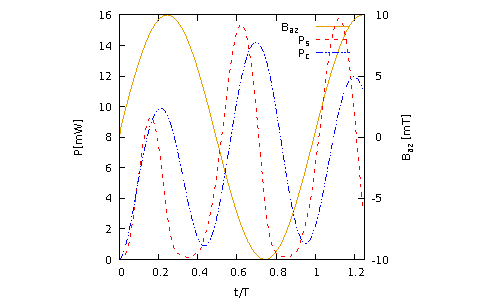}}
 \subfloat[][]
{\includegraphics[trim=45 0 55 0,clip,width=6.5 cm]{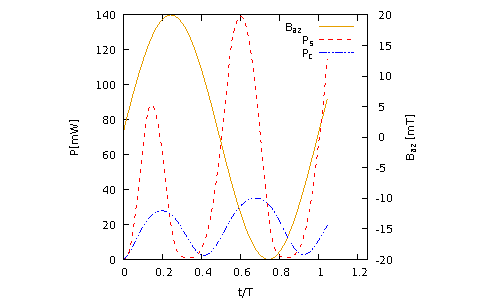}}
\caption{The AC loss in 6 filament tape split into superconductor loss ($P_s$) and coupling current loss ($P_c$). Cases with applied magnetic field amplitude (a) 10 mT and (b) 20 mT with frequency 144 Hz. 
$B_{az}$ is the instantaneous applied magnetic field. The coupling loss increases with lower applied field.}
\label{f.filament_ACloss_f}
\end{figure}

\begin{figure}[tbp]
\centering
{\includegraphics[trim=0 0 10 0,clip,height=7.5 cm]{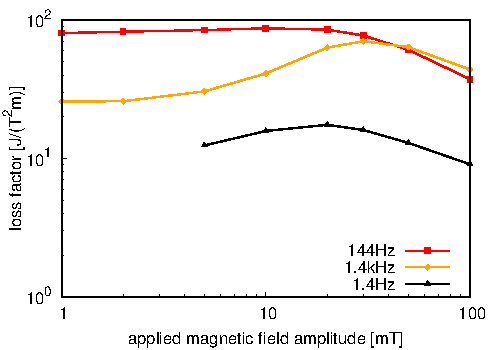}}
\caption{The AC loss factor is defined per cycle, divided by square of applied field and length of the sample. Calculation of the six filament tape model. Dominant part of the AC loss at low fields is coupling loss.}
\label{f.6filament_ACloss}
\end{figure}

%%%%%%%%%%%%%%%%%%%%%%%%%%%%%%%%%%%%%%%%%%%%%%%%%%%%%%%%%%%%%%%%%%%%%%%%%%%%%%%%%%%%%%%%%%%%%%%

\section{Magnetization of isotropic rectangular prisms}
\label{s.iso_bulk}

Before this thesis, the geometry like a cube or prism is still not completely solved and current path inside it is not well understood in spite of it simplicity, therefore a full model with all finite size effects is necessary. The isotropic bulk or prism model is a full 3D model with many layers of cells in the thickness. The MEMEP 3D method allows current to flow in all 3 dimensions $(x,y,z)$ and hence there is $\vJ$ with all 3 components $(J_x,J_y,J_z)$, as well as $\vT=(T_x,T_y,T_z)$. The full 3D model is demanding on the calculation speed, due to the high number of variables. 

%%%%%%%%%%%%%%%%%%%%%%%%%%%%%%%%%%%%%%%%%%%%%%%%%%%%%%%%%%%%%%%%%%%%%%%%%%%%%%%%%%%%%%%%%%%%%%%

\subsection{3D Magnetization currents with constant ${J_c}$}

\subsubsection{Screening currents}
\label{s.cube_Jc}

The magnetization model of a superconducting cube shows the real 3D current path. We take that the cube edge is 10 mm and the applied field amplitude is 200 mT with frequency of 50 Hz is parallel to the $z$ axis [figure \ref{f.cube_geometry}(a)]. The critical current density is $1\cdot 10^{8}$ A/m$^{2}$ with constant $J_c$. We assume an isotropic power law with $n=$100. The mesh contains $41\times 41 \times 41$ elements, which represents more than 200000 degrees of the freedom.

\begin{figure}[tbp]
\centering
 \subfloat[][]
{\includegraphics[trim=0 0 -10 0,clip,width=5.5 cm]{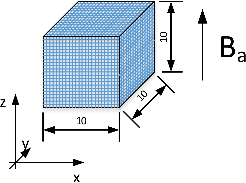}}
 \subfloat[][]
{\includegraphics[trim=-10 0 0 0,clip,width=5.0 cm]{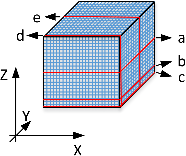}}
\caption{The geometry of the cube sample. Dimensions are in mm. (a) Cube and direction of the applied field. (b) The positions of the cross-sectional planes in the cube are ``a", z/d=0; ``b", z/d=-0.39; ``c", z/d=-0.49; ``d", y/d=-0.49; and ``e", y/d=0.} 
\label{f.cube_geometry}
\end{figure}

The cube is not fully saturated with screening currents, because the applied field is lower than the saturation field [figure \ref{f.cubeJcJy}]. The mid plane at $y/w=0$ shows the $J_y$ component on figure \ref{f.cubeJcJy}. The current distribution in this plane is similar to 2D cross-sectional models for infinitely long samples along the $y$ axis \cite{prigozhin96JCP}. Close to $z=0$, the penetration depth of $|\vJ|\approx J_c$ into the cube is only around $x/w=$0.2 and the penetration front is almost flat at the center. The screening current penetrates only in a thin area from the top and bottom. The $J_x$ component at plane $x/w=0$ is symmetric to $J_y$ because of the cube symmetry. 

\begin{figure}[tbp]
\centering
{\includegraphics[trim=0 0 0 0,clip,height=5.5 cm]{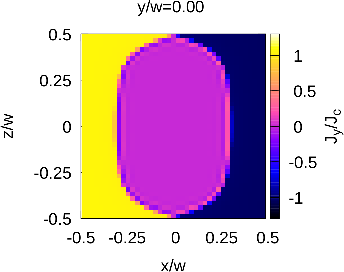}}
\caption{The $J_y$ component at the mid cross-section plane $y/d=0$ of the cube in figure \ref{f.cube_geometry} at the applied field amplitude of 200 mT. The power law exponent is n=100.}
\label{f.cubeJcJy}
\end{figure}

The colour maps in three different heights of the cross-sectional planes show the same penetration depth of the current modulus from all sides [figure \ref{f.cube_geometry}(b)]. At the mid plane $(z/d=0)$, the screening current penetrates $x/d\approx$0.2 from each edge of the cube and the current lines are square, following the outer shape of the cube [figure \ref{f.cubeJc}(a)]. The penetration depth is higher at a lower plane $z/d=-0.39$ and the current lines closer to the center starts to be rounded [figure \ref{f.cubeJc}(b)]. The last plane, $z/d=-0.49$, shows full penetration and fully rounded current lines at the center of the plane [figure \ref{f.cubeJc}(c)]. 

The MEMEP 3D model also finds the $J_z$ component of the current density inside the cube. The maximum magnitude of $J_z$ is around $0.3J_c$, being the highest around the diagonal lines in the plane $z/d=-0.39$ [figure \ref{f.cubeJc}(e)]. This is the reason why we chose the $z/d=-0.39$ plane for the maps. The lateral surface ${y/d=-0.49}$ shows symmetrical distribution of $J_z$ [figure \ref{f.cubeJc}(d)]. $J_z$ slightly bends the main screening current in the $z$ direction. The full 3D current path is on figure \ref{f.cube3D}. The current path goes up and down in the places where $J_z$ has the highest value. $J_z$ at the mid plane $z/d=0$ is zero, because of cube symmetry. The main reason of the existence of $J_z$ is a shape of the sample. The cylinder does not have $J_z$ component, because the circular screening current loops shield the magnetic field completely. However, the square current loops cannot create uniform self-field in the corners of the sample in order to completely shield the magnetic field. The self-field in the corners is higher than along the rest of the current path, and hence the $J_z$ currents are necessary to completely shield penetrating field at the region with $|\vJ|< J_c$. The $J_z$ current exists only in the case of not fully saturated sample \cite{badia05APL}. Previous works from Badia \cite{badia05APL,chen89JAP} justified that the current loops always need to be rectangular, also below full penetration. However, the argumentation assumes that $|\vJ|$ can only be 0 or $J_c$, while the Critical-State Model allows any $|\vJ|<J_c$. 

The same model study is on the cube with the similar input parameters except the power law exponent $n$=30. The calculation shown the $J_z$ (figure \ref{f.cubeJcJz1}) is qualitatively the same as case with $n=$100. The result confirmed that the $J_z$ is independent on the power law exponent.  

\begin{figure}[tbp]
\centering
{\includegraphics[trim=0 -10 0 0,clip,height=4.2 cm]{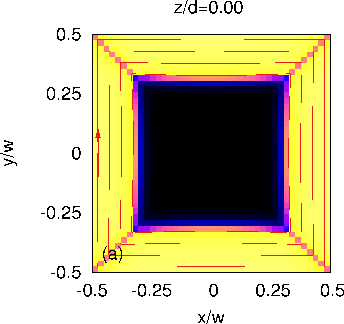}}
{\includegraphics[trim=0 -10 0 0,clip,height=4.2 cm]{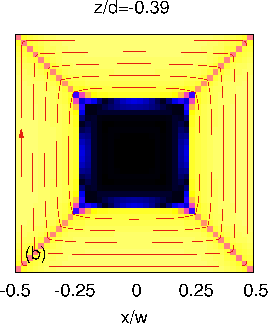}}  
{\includegraphics[trim=0 -10 0 0,clip,height=4.2 cm]{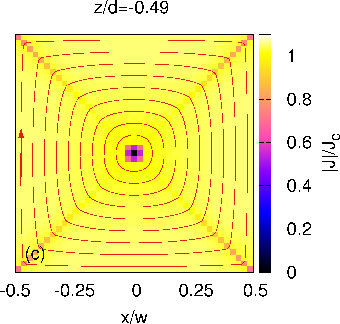}} \\
{\includegraphics[trim=0 0 0 0,clip,height=4.2 cm]{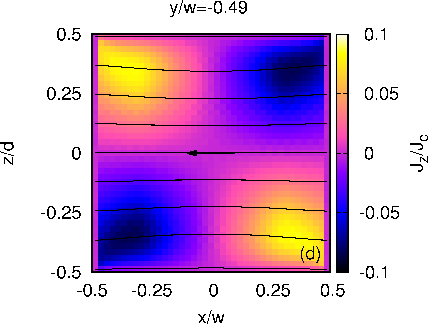}}
{\includegraphics[trim=0 0 0 0,clip,height=4.2 cm]{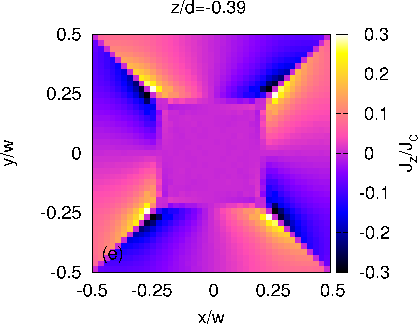}}
\caption{The penetration of the current density modulus into the cube with power law $n=$100 at the peak of the applied field of 200 mT amplitude. The cross-sectional planes are at (a) $z/d=0$, (b) $z/d=-0.39$, (c) $z/d=-0.49$, (d) $y/w=-0.49$ and (e) $z/d=-0.39$.}
\label{f.cubeJc}
\end{figure}

\begin{figure}[tbp]
\centering
{\includegraphics[trim=0 0 0 0,clip,height=5.5 cm]{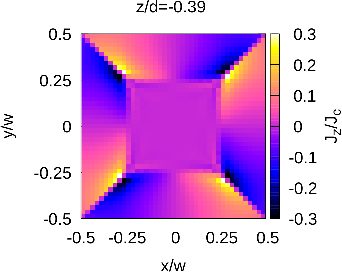}}
\caption{The $J_z$ component at the mid cross-section plane $z/d=-0.39$ at the applied field amplitude of 200 mT. The power law exponent $n$=30 proved independence of the $J_z$ on the CSM.}
\label{f.cubeJcJz1}
\end{figure}

\begin{figure}[tbp]
\centering
{\includegraphics[trim=10 20 25 20,clip,width=8.0 cm]{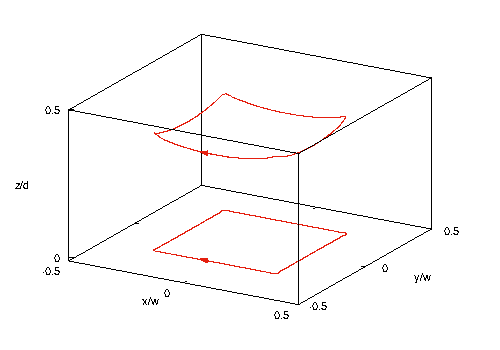}}
\caption{The upper half of the cube sample with real 3D current lines in two positions. The current lines are bended from $z$ plane by $J_z$ component, which improve shielding of the applied fields in corners.}
\label{f.cube3D}
\end{figure}

The second magnetization calculation is with applied field 1 T. The applied field is higher than the saturation field, which is $B_s=$413 mT. The current density $|\vJ|$ is practically equal to $J_c$ because of the high power law $n$=100. The current line loops are all square [figure \ref{f.cubeJcJ1}(a),(b),(c)]. Since the sample is fully saturated $|\vJ|=J_c$ everywhere, and hence current loops need to be square due to current conservation \cite{badia05APL}. The model shows zero value of $J_z$ at the planes $y/d=-0.49$ and $z/d=-0.39$ on figures \ref{f.cubeJcJ1}(d)(e), and hence the model confirms the Critical State prediction for long bars.   

\begin{figure}[tbp]
\centering
{\includegraphics[trim=0 -10 0 0,clip,height=4.2 cm]{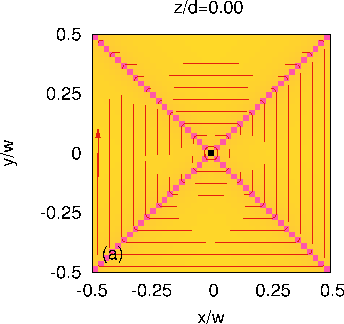}}
{\includegraphics[trim=0 -10 0 0,clip,height=4.2 cm]{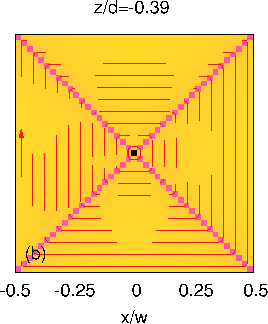}}  
{\includegraphics[trim=0 -10 0 0,clip,height=4.2 cm]{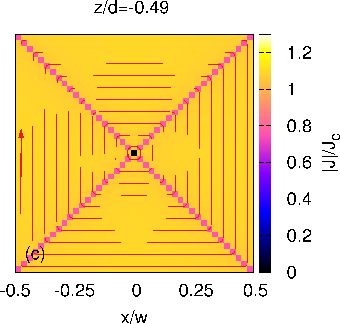}} \\
{\includegraphics[trim=0 0 0 0,clip,height=4.2 cm]{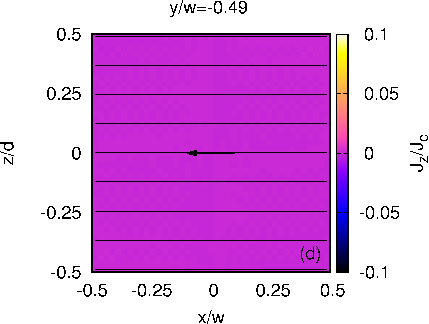}} 
{\includegraphics[trim=0 0 0 0,clip,height=4.2 cm]{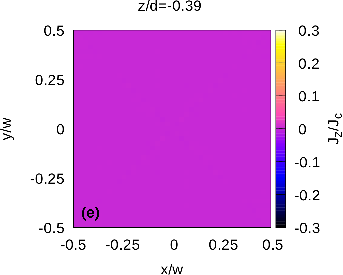}}
\caption{The penetration of the current density modulus into the cube with power law $n=$100 at the instantaneous applied field of 418 mT. The cross-sectional planes are at (a) $z/d=0$, (b) $z/d=-0.39$, (c) $z/d=-0.49$, (d) $y/w=-0.49$ and (e) $z/d=-0.39$.}
\label{f.cubeJcJ1}
\end{figure}

%%%%%%%%%%%%%%%%%%%%%%%%%%%%%%%%%%%%%%%%%%%%%%%%%%%%%%%%%%%%%%%%%%%%%%%%%%%%%%%%%%%%%%%%%%%%%%%

\subsubsection{Effect of aspect ratio of rectangular prisms}
\label{s.prism_Jc}

A prism with any intermediate thickness is more realistic than any slab, cylinder or thin sheet approximation, and hence the model includes all finite size effects. This model is more accurate and reveals new effects. 

The next modelling situation is the magnetization of prisms with different aspect ratio $c=d/w$, where $d$ is the thickness of the sample and $w$ is its width. The cross-section of the prisms is $12\times 12$ mm and their thicknesses are $d=20, 10, 5, 2, 1$ mm (figure \ref{f.prism_geometry}). The critical current density is $J_c=1\cdot 10^{8}$ A/m$^{2}$ for all aspect ratios $c=2,1,0.5,0.2,0.1$ and the $n$ factor is 100. The applied magnetic field is parallel to the $z$ axis with frequency 50 Hz. 

The mesh of various $c$ cases are created by different number of cells such as $c=0.1$ $(101\times 101\times 11)$, $c=0.2$ $(81\times 81\times 21)$, $c=0.5$ $(61\times 61\times 31)$ and $c=1$ $(41\times 41\times 41)$. Therefore, the colour maps are finer in the $z$ plane for thinner samples. The applied magnetic field for $c=1$ is 200 mT, which is 0.484 of the penetration field, $B_p$. We assume the same ratio for the applied field, $B_a=0.484B_p$, and hence the applied field for the prism with various aspect ratios are $B_a=200, 153.02, 87.65, 51.81$ mT. We calculate and discuss the penetration field in section \ref{s.prism_magnetization}.      

\begin{figure}[tbp]
\centering
{\includegraphics[trim=0 0 -20 0,clip,width=6.5 cm]{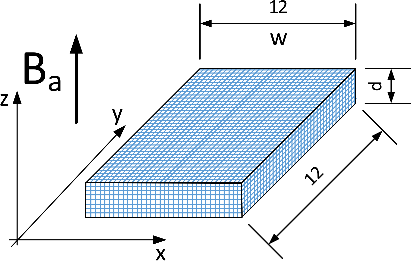}}
\caption{The geometry and dimensions of the prism (in mm) with the direction of the applied field, where $d$ is the thickness of the prism and $w$ is its width. The aspect ration is defined as $c\equiv d/w$.} 
\label{f.prism_geometry}
\end{figure}

The $J_y$ component at the mid-cross section plane $y/w=0$ for each case of $c$ is on figure \ref{f.prismJyc}. Close to the center, the $J_c$ penetration front for $c=1$ is around $x/w=$0.2 from both sides of the prism and is almost flat along the $z$ axes. There is an additional close to the center small penetration from top and bottom [figure \ref{f.prismJyc}(a)]. The thinner prism with $c=0.5$ shows shorter flat part of the $J_y$ [figure \ref{f.prismJyc}(b)]. There is no flat part of $J_y$ along the $z$ axis in the case of $c=0.2$ and the not penetrated zone is roughly elliptic [figure \ref{f.prismJyc}(c)]. The thinner prism with aspect ratio 0.1 presents again the same penetration depth and not saturated central zone, which is basically a thinner ellipse [figure \ref{f.prismJyc}(d)]. The $J_y$ distribution shows that even in the thinnest prism, the current penetration depth is similar to the bulk one [figure \ref{f.prismJyc}(a)].   

\begin{figure}[tbp]
\centering
{\includegraphics[trim=0 30 0 0,clip,width=6.5 cm]{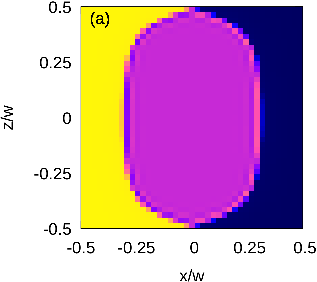}}\\
{\includegraphics[trim=0 43 0 0,clip,width=6.5 cm]{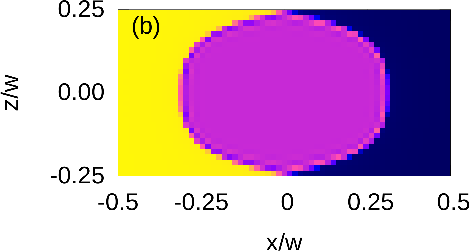}}\\
{\includegraphics[trim=0 43 0 0,clip,width=6.5 cm]{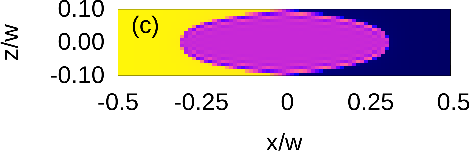}}\\
{\includegraphics[trim=0 -10 0 0,clip,width=6.5 cm]{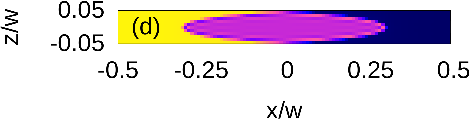}}\\
{\includegraphics[trim=0 0 0 58,clip,width=6.5 cm]{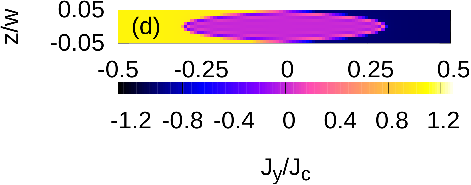}}
\caption{The $J_y$ component of the current density in the mid plane $y/w=0$ for various aspect ratios $c=w/d$ of the prism with $n$=100 and the applied field of the same ratio $B_a=0.484B_s$, where $B_s$ is the saturation field defined in section \ref{s.prism_magnetization}. (a) c=1, (b) c=0.5, (c) c=0.2, (d) c=0.1. The penetration depth of $J_y$ is the same for each case of $c$.}
\label{f.prismJyc}
\end{figure}

The next colour maps contain the $J_z$ distribution in different planes $z/d$ (figure \ref{f.prismJzc}). These are the planes that show the highest value of $J_z$ for each $c$. The cubic bulk shows the highest $J_z$ along the diagonal lines in plane $z/d=-0.39$ [figure \ref{f.prismJzc}(a)]. The prism with $c=0.5$ has the same $J_z$ distribution in relatively lower plane $z/d=-0.32$ [figure \ref{f.prismJzc}(b)]. The penetration depth in a thinner prism with $c=0.2$ is slightly smaller and $J_z$ has smooth and round penetration front [figure \ref{f.prismJzc}(c)] instead of the flat parts along the $x$ and $y$ axis like in the cube [figure \ref{f.prismJzc}(a)]. The round penetration front comes from non-square current loops in the $z$ plane projection, which are explained it the next section \ref{s.prism_c0.1}. The last prism, with $c=0.1$, still has $J_z$ around $0.3J_c$ along the diagonal lines but in plane $z/d=-0.27$ [figure \ref{f.prismJzc}(d)]. The $z/d$ plane with the highest $J_z$ is moving to the center of the prism exactly according the additional $J_y$ penetration in figure \ref{f.prismJyc} from the bottom and the top of the prism. The plane of maximum $J_z$ corresponds roughly to the center of this zone with additional penetration. The cause is that $J_z$ is caused by the self-field, being the highest close to the sample ends. Moreover, $J_z$ approaches zero at exactly the sample end. The model shows that the $J_z$ component is not vanishing with decreasing the aspect ratio and that the penetration front starts to be rounded. 

\begin{figure}[tbp]
\centering
{\includegraphics[trim=0 26 0 0,clip,width=6.5 cm]{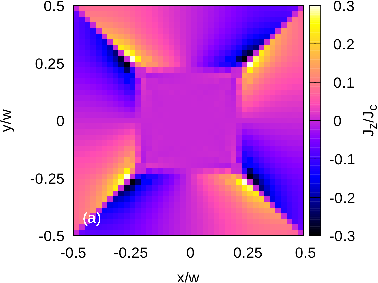}}\\
{\includegraphics[trim=0 26 0 0,clip,width=6.5 cm]{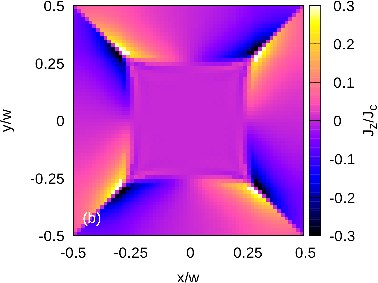}}\\
{\includegraphics[trim=0 26 0 0,clip,width=6.5 cm]{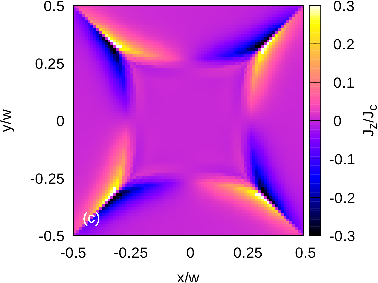}}\\
{\includegraphics[trim=0 0 0 0,clip,width=6.5 cm]{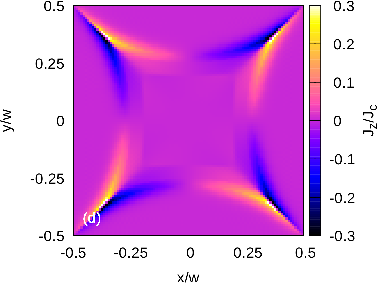}}
\caption{The $J_z$ component of the current density in the prism with various aspect ratios $c$ in the $z$ plane with the highest value. (a) $c$=1, $z/d$=-0.39 (b) $c$=0.5, $z/d$=-0.32 
(c) $c$=0.2, $z/d$=-0.29 (d) c=0.1, $z/d$=-0.27.}
\label{f.prismJzc}
\end{figure}

%%%%%%%%%%%%%%%%%%%%%%%%%%%%%%%%%%%%%%%%%%%%%%%%%%%%%%%%%%%%%%%%%%%%%%%%%%%%%%%%%%%%%%%%%%%%%%%

\subsubsection{Thin rectangular prism (aspect ratio $c=0.1$)}
\label{s.prism_c0.1}

This section is focused on the 3D current path in the thinnest prism, with aspect ratio $c=0.1$. The modulus of the current density is on figure \ref{f.prismJ}. The figure also shows the current lines, which are a projection on the plotted plane of the 3D current lines. The mid cross-section plane $z/d=0$ shows the penetration depth of the screening current [figure \ref{f.prismJ}(a)]. The lower plane $z/d=-0.09$ presents almost square current lines and higher penetration depth [figure \ref{f.prismJ} (b)]. The plane $z/d=-0.18$ shows smooth bending of the current lines at position 2 on figure \ref{f.prismJ}(c) and a flat path section marked as 1. The current path in position 1 is in the same $z$ plane but in position 2 the current path is going out of the shown $z$ plane and back, because of the $J_z$ component [figure \ref{f.prismJz}(c)]. The planes $z/d=-0.27$ and $z/d=-0.36$ start to make rounded current lines close to the diagonals [figure \ref{f.prismJ}(d)(e)] with higher penetration of the current density. The last plane $z/d=-0.45$ is close to the bottom surface, where the sample is almost fully penetrated. The current lines are circular at the middle of the plane and square close to the edges [figure \ref{f.prismJ}(f)]. The maps reveal that in the thinnest prism, $c=0.1$, there exist circular current lines closer to the top and bottom surfaces, as it was shown in the bulk sample section (\ref{s.cube_Jc}). 

\begin{figure}[tbp]
\centering
{\includegraphics[trim=0 0 0 10,clip,height=4.0 cm]{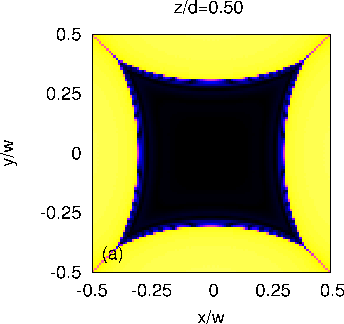}} 
{\includegraphics[trim=0 0 0 10,clip,height=4.0 cm]{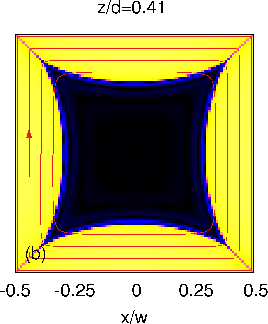}}
{\includegraphics[trim=0 0 0 10,clip,height=4.0 cm]{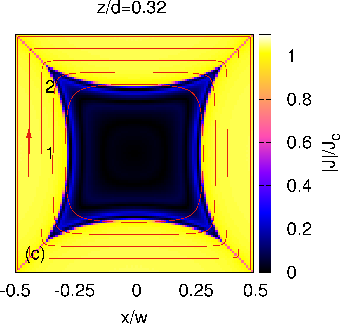}}\\ 
{\includegraphics[trim=0 0 0 10,clip,height=4.0 cm]{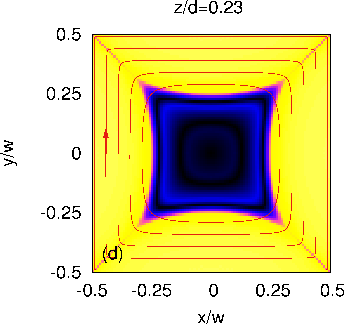}} 
{\includegraphics[trim=0 0 0 10,clip,height=4.0 cm]{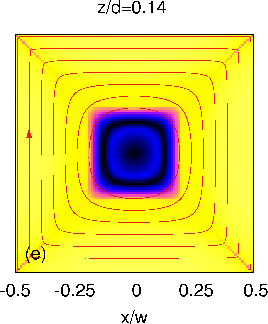}} 
{\includegraphics[trim=0 0 0 10,clip,height=4.0 cm]{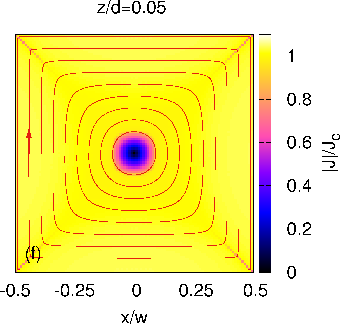}} 
\caption{The gradual penetration of the current modulus into the prism with aspect ratio $c$=0.1 and $n$=100 at the peak of the applied field 51.81 mT. The different positions of the $z$ plane are (a) $z/d$=0.0, (b) $z/d$=-0.09, (c) $z/d$=-0.18, (d) $z/d$=-0.27, (e) $z/d$=-0.36, (f) $z/d$=-0.45.}
\label{f.prismJ}
\end{figure}

The $J_z$ component of the current density is zero in plane $z/d=0$ [figure \ref{f.prismJz}(a)], because of the cube symmetry. Since the current can not flow outside the sample, $J_z$ is almost zero close to the bottom surface [figure \ref{f.prismJz}(f)]. The $J_z$ penetration front is a bit beyond the modulus penetration front with $J\approx J_c$[figure \ref{f.prismJz}(b)]. The penetration front contains the $J_z$ current density with higher value at the diagonals, such as position 2, and lower value at the straight parts, such as position 1 on figure \ref{f.prismJz}(c). The current path from the shown $z$ plane is bended down to a lower plane at the position 2 and then back. The complete 3D current lines are on figure \ref{f.prism3D}. The current lines are clearly bended at the diagonals. The highest bending is in the prism with aspect ratio $c=0.1$, being and smaller in the prism with $c=1$ (figure \ref{f.cube3D}). $J_z$ in the lower plane $z/d=-0.27$ is moving further into the prism with the highest value $0.3J_c$ [figure \ref{f.prismJz}(d)]. The next plane decreases $J_z$ [figure \ref{f.prismJz}(e)] since the $z$ planes are saturated with $J_x$ and $J_y$ components close to $J_c$. The prisms show that non-zero $J_z$ exists as well in the thinnest prism with aspect ratio $c=0.1$ and that its maximum value. 

\begin{figure}[tbp]
\centering
{\includegraphics[trim=0 0 0 10,clip,height=4.0 cm]{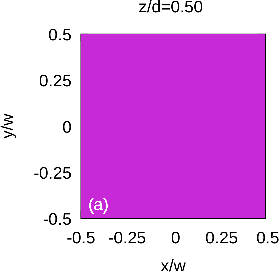}} 
{\includegraphics[trim=0 0 0 10,clip,height=4.0 cm]{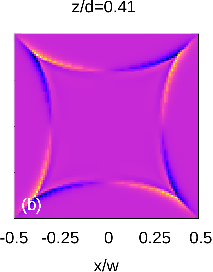}} 
{\includegraphics[trim=0 0 0 10,clip,height=4.0 cm]{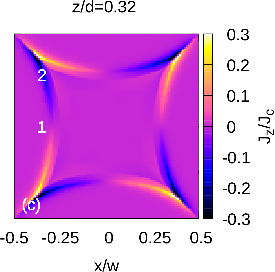}}\\
{\includegraphics[trim=0 0 0 10,clip,height=4.0 cm]{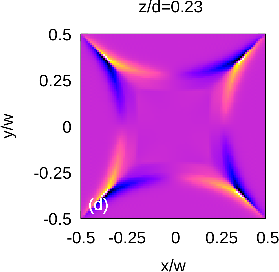}} 
{\includegraphics[trim=0 0 0 10,clip,height=4.0 cm]{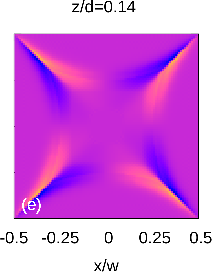}} 
{\includegraphics[trim=0 0 0 10,clip,height=4.0 cm]{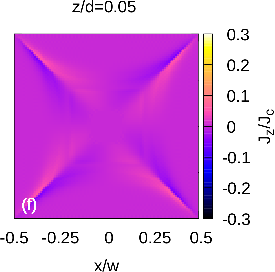}} 
\caption{$J_z$ in the prism with aspect ratio $c$=0.1 and $n$=100 at the peak of the applied field of 51.81 mT. The different positions of the $z$ plane are (a) $z/d$=0.0, (b) $z/d$=-0.09, (c) $z/d$=-0.18, (d) $z/d$=-0.27, (e) $z/d$=-0.36, (f) $z/d$=-0.45.}
\label{f.prismJz}
\end{figure}

\begin{figure}[tbp]
\centering
{\includegraphics[trim=10 20 25 20,clip,width=10.0 cm]{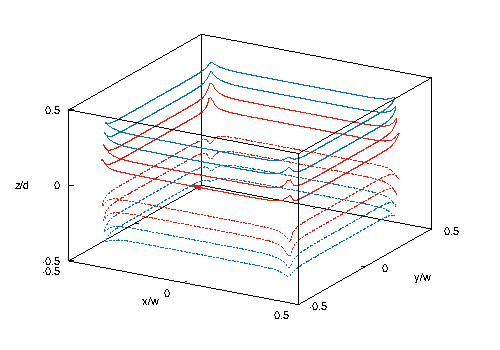}}
\caption{The 3D current lines in the prism with aspect ration $c$=0.1. The current path goes up and down in the places with the highest $J_z$ with the same behaviour like in bulk sample (figure \ref{f.cube3D}).}
\label{f.prism3D}
\end{figure}

The thickness-average current density of the prism with aspect ratio $c=0.1$ is on figure \ref{f.prism_ave}. The current lines are calculated from the average $\vT$, having only $z$ component. The penetration of critical current density and the shape of the current lines are similar to the thin film case [figure \ref{f.Jc}(a)]\cite{Pardo17SST}. It confirms that there must exist current lines with square and circular current paths in the cube or prism. Indeed, rounded current lines of the thickness average of $\vJ$ cannot be obtained by superposing rectangular current lines at every height $z$.  

\begin{figure}[tbp]
\centering
{\includegraphics[trim=15 4 15 10,clip,height=6 cm]{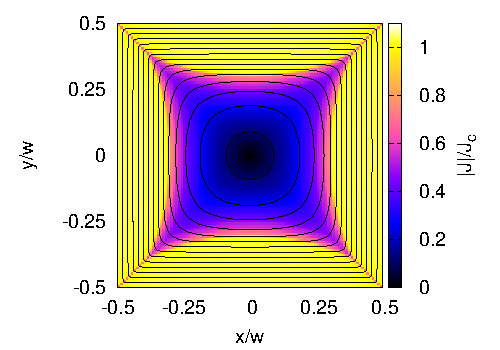}}
\caption{The average current density over the thickness of the prism with aspect ratio $c=0.1$. The current distribution is qualitatively similar to the thin film (figure \ref{f.Jc}).}
\label{f.prism_ave}
\end{figure}

%%%%%%%%%%%%%%%%%%%%%%%%%%%%%%%%%%%%%%%%%%%%%%%%%%%%%%%%%%%%%%%%%%%%%%%%%%%%%%%%%%%%%%%%%%%%%%%

\subsection{3D Magnetization currents with ${J_c(B)}$ dependence}
\label{s.cube_JcB}

The next calculated electromagnetic response is for a cube model that includes $J_c(B)$ dependence, using the same input parameters as for constant $J_c$ in section \ref{s.cube_Jc}. The critical current density is $J_{c0}=1\cdot 10^{8}$ A/m$^2$, being the same value as for the constant $J_c$ case, and $B_0=20$ mT. The $J_c(B)$ dependence is Kim model of (\ref{e.Kim}) with the following parameters $m$=0.5 and power law $n$=30. 

The distribution of $J_y$ in the mid plane $y/w=0$ at the applied field 178.2 mT is on figure \ref{f.cubeJcbJy}. The current distribution is similar to the $J_y$ of the constant $J_c$ (figure \ref{f.cubeJcJy}). The penetration front of $J_y$ reaches the $J_{c0}$ value and the current density slowly decreases when moving towards to the edges of the sample. The $J_y$ component decreases because the local magnetic field reduces $J_c$, and hence so does $J_y$. Since local $J_c(B)$ is lower than $J_{c0}$, the penetration depth is higher compared to the case of $J_c$ constant with value $J_{c0}$, in order to completely shield the applied magnetic field at the center.   

\begin{figure}[tbp]
\centering
{\includegraphics[trim=0 0 0 0,clip,height=6 cm]{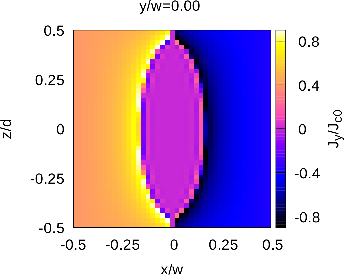}} 
\caption{The $J_y$ component at the mid cross-section plane $y=0$ at the applied magnetic field 178.2 mT with $J_c(B)$ dependence. The power law exponent is n=30. The penetration depth is higher compare to $J_c$ constant case (figure \ref{f.cubeJcJy}).}
\label{f.cubeJcbJy}
\end{figure}

The current density modulus in all planes shows the same penetration depth as $J_y$ [figure \ref{f.cubeJcBJ}(a,b,c)]. The current loops are square at the mid plane $z/d=0$ and $z/d=-0.39$ [figure \ref{f.cubeJcBJ}(a)(b)]. The current lines are a bit rounded close to the center in the plane $z/d=-0.49$ [figure \ref{f.cubeJcBJ}(c)]. The penetration front of the modulus shows a value around $J_{c0}$ with reduction when approaching to the edges like the $J_y$ component in figure \ref{f.cubeJcbJy}. The magnetic field dependence increases the penetration depth. The current loops close to the top and bottom are not fully rounded, because the sample is closer to the saturation state [figure \ref{f.cubeJcBJ}(c)]. The current lines are square in the square sample at the saturated state. The $J_z$ component decreases close to the edges, because of the local saturation state [figure \ref{f.cubeJcBJ}(e)]. $J_z$ is almost zero at the lateral surface $y/d=-0.49$ [figure \ref{f.cubeJcBJ}(d)]. The local magnetic field and the higher sample penetration decreases the maximum $J_z$ to the value of $0.2J_{c0}$.    

\begin{figure}[tbp]
\centering
{\includegraphics[trim=0 -10 0 0,clip,height=4.2 cm]{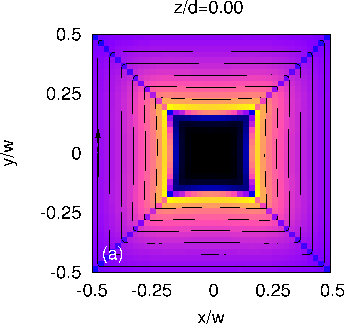}}
{\includegraphics[trim=0 -10 0 0,clip,height=4.2 cm]{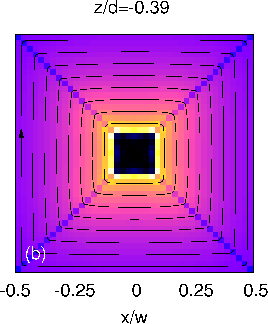}}  
{\includegraphics[trim=0 -10 0 0,clip,height=4.2 cm]{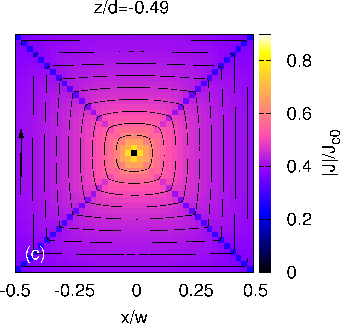}} \\
{\includegraphics[trim=0 0 0 0,clip,height=4.2 cm]{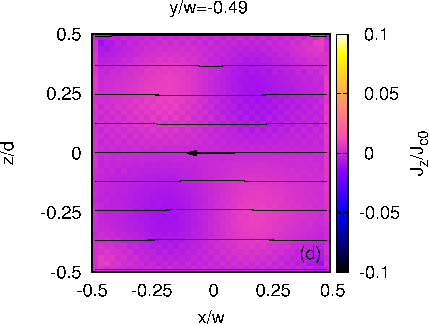}} 	
{\includegraphics[trim=0 0 0 0,clip,height=4.2 cm]{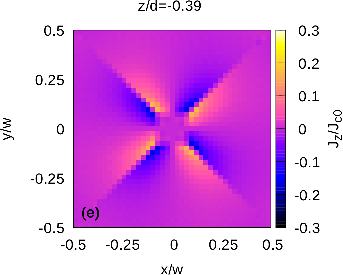}}
\caption{The penetration of the current density modulus and $J_z$ into the cube at the instantaneous applied field 178.2 mT with power law $n=$30 with $J_c(B)$ dependence. 
The cross-sectional planes are at (a) $z/d=0$, (b) $z/d=-0.39$, (c) $z/d=-0.49$, (d) $y/w=-0.49$ and (e) $z/d=-0.39$.}
\label{f.cubeJcBJ}
\end{figure}

%%%%%%%%%%%%%%%%%%%%%%%%%%%%%%%%%%%%%%%%%%%%%%%%%%%%%%%%%%%%%%%%%%%%%%%%%%%%%%%%%%%%%%%%%%%%%%%

\subsection{Magnetization loops and penetration field of rectangular prisms with constant $J_c$}
\label{s.prism_magnetization}

The magnetization loops for the prisms with all aspect ratios $c$ and the applied magnetic field 1 T is on figure \ref{f.prism_loop}. The saturation field increases with the thickness of the prism. Higher prisms can induce screening currents in a larger zone, and hence they shield higher applied fields at the sample center. 

We set the criterion that the saturation field (or penetration field) is 99\% of the magnetization of the sample, $M(B_p)=0.99M_s$. We choose that criterion, in order to unified the criterion for others shapes like cylinder $(M=0.9903M_s)$, slab $(M=0.75M_s)$ and strip $(M=0.9856M_s)$. Since the rectangular prisms contain similar behaviour from all of them. The penetration field dependence on the aspect ratio is on figure \ref{f.prism_Bp} and values are in the table \ref{t.prism_Bp}. The analytical fit of equation (\ref{e.fit}) reaches 97\% accuracy and it also includes the limits for infinite bar and thin square film. The analytical fit is 

\begin{equation}
{B_s(c)=\mu_0J_cwa_1\left[1+a_2e^{\frac{-\ln^2\left(a_3c\right)}{2a_4^2}}	\right]\tanh(a_5c)},
\label{e.fit}
\end{equation}
where the fit parameters are $a_1=0.3915,a_2=-0.26,a_3=2.56,a_4=0.75$ and $a_5=2.41$.

\begin{figure}[tbp]
\centering
{\includegraphics[trim=0 0 0 0,clip,width=10.5 cm]{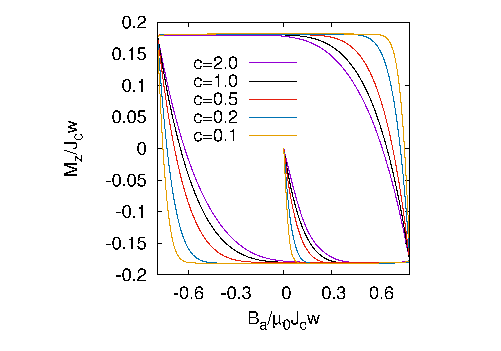}}
\caption{The hysteresis loops of the prisms with various aspect ratio $c=d/w$ and the applied field amplitude of 1 T. The saturation field increases with aspect ratio $c$.}
\label{f.prism_loop}
\end{figure}

\begin{table}[tpb]
\begin{center}
\begin{tabular}{llll}
\hline
\hline
{\bf Aspect} & {\bf Saturation } \\
{\bf ratio ${d/w}$} & {\bf field ${{B_s}/{J_{c}}w\mu_{0}}$} \\
\hline
2.0 & 0.38 \\
1.0 & 0.32 \\
0.5 & 0.25 \\
0.2 & 0.14 \\
0.1 & 0.08 \\
\hline
\hline
\end{tabular}
\caption{Calculated saturation field for the prisms with various aspect ratio $c$. The saturation field is assumed as the one that causes 99\% of the magnetization $M(B_s)=0.99M_s$.}
\label{t.prism_Bp}
\end{center}
\end{table}

\begin{figure}[tbp]
\centering
{\includegraphics[trim=0 0 0 0,clip,width=10.5 cm]{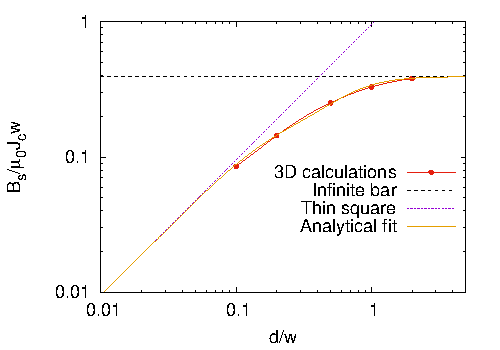}}
\caption{The saturation field increases with on the aspect ratio $c=d/w$. The analytical fit of equation (\ref{e.fit}) reaches 97\% accuracy and meets the thin square and infinite bar limits.}
\label{f.prism_Bp}
\end{figure}

%%%%%%%%%%%%%%%%%%%%%%%%%%%%%%%%%%%%%%%%%%%%%%%%%%%%%%%%%%%%%%%%%%%%%%%%%%%%%%%%%%%%%%%%%%%%%%%

\subsection{Benchmark of superconducting bulk with perpendicular applied magnetic field.}
\label{s.bulk_benchmark}

The state of the art of the modelling methods and tools is presented biannually at the International Workshop on Numerical Modelling of HTS \cite{HTS2016}. The HTS Work Group maintains the benchmark web site \cite{HTSworkgroup}, in order to evaluate many kinds of modelling tools. In conclusion, the workshop in 2016 showed that the 2D models reached the required level of development to model most power application.

The MEMEP 3D results for the cube magnetization have been chose as a simple benchmark test to develop 3D modelling tools. The simple modelling case serves as an example to test the speed and accuracy of new 3D modelling methods. 

The modelling situation is a superconducting cube with the same parameters as in section \ref{s.cube_Jc}, where all finite size effects are explained. 

The modelling mesh contains $41\times 41\times 41$ cells, which is 211806 degrees of freedom. The current tolerance is 10$^{-5}J_c$. The calculation time is less than 7 hours on desktop computer: Intel Core i7-4771 CPU@3.50GHz8, 8GB RAM, Linux Ubuntu 64 bit.

%%%%%%%%%%%%%%%%%%%%%%%%%%%%%%%%%%%%%%%%%%%%%%%%%%%%%%%%%%%%%%%%%%%%%%%%%%%%%%%%%%%%%%%%%%%%%%%

\section{Magnetization of stacks of tapes and bulks in tilted applied field}
\label{s.magnetization}

2D cross-sectional model cannot model finite size effects, which are important; and hence state-of-the art modelling tools starts to model full 3D cases \cite{Sirois15SST,Grilli16IES}. However, 3D modelling needs to be improved, since the accuracy of models is low \cite{grilli03SST,campbell09SST,farinon14SST,Pardo16IES} due to the required high number of elements.

The 3D modelling cases of the stack of tapes and homogeneous bulk with tilted applied magnetic field are interesting, since they can be seen in experiments by VSM and SQUID and for power applications such as motors and generators \cite{Patel13SST}. The 2D results have been already predicted by \cite{grilli07IESb,pardo12SSTb}. However, the full 3D models with all finite size effects are missing. The magnetic response of stacks of tapes and bulks is expected to be different.

%%%%%%%%%%%%%%%%%%%%%%%%%%%%%%%%%%%%%%%%%%%%%%%%%%%%%%%%%%%%%%%%%%%%%%%%%%%%%%%%%%%%%%%%%%%%%%%

\subsection{Screening currents}
\label{s.prism_benchmark}

The magnetization of stack and bulk by tilted applied field is modelled by MEMEP 3D, in order evaluate the method and see all finite size effects in full 3D models and compare to other methods. 

The modelling geometry is on figure \ref{f.bench_geometry}(a). The sample size is $10\times 10\times 1$ mm, where $w$ is width and $d$ is thickness of the sample. The sinusoidal applied magnetic field is of 200 mT amplitude and 50 Hz frequency with angle $\theta=60\degree$. The critical current density is $1\cdot 10^{8}$ A/m$^{2}$. We assume isotropic power law with n=25. We model both bulk and stack geometries. For the stack, we take the homogeneous approximation. The difference between them is that the bulk allows currents to flow in any 3D direction in contrast to the stack, where there is no current in the direction perpendicular to the tapes surface. The stack model assumes electrically isolated tapes, and hence $J_z$ always vanishes.    

\begin{figure}[tbp]
\centering
 \subfloat[][]
{\includegraphics[trim= 0 0 0 0,clip,width=6.0 cm]{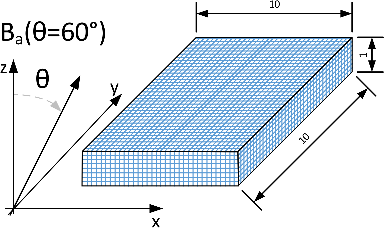}}
 \subfloat[][]
{\includegraphics[trim= 0 0 0 0,clip,width=6.0 cm]{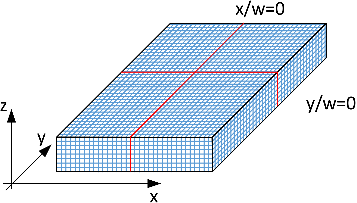}}
\caption{The geometry of the bulk and stack samples. (a) The size of the sample with the angle of the applied field. Dimensions are in mm. (b) The cross-sectional planes for the colour maps with $w$ as width of the sample.}
\label{f.bench_geometry}
\end{figure}

In the midplane with $y/w=0$ ($y$-midplane), the current density distribution in the bulk sample is fully saturated with $J_y$ [figure \ref{f.bench_J}(b)]. The border between positive and negative current density is tilted according to the applied field angle $\theta$. The border plane is not parallel to $\theta$ because of the effect of the self-field. For much larger applied fields, we except a border that is parallel to the applied field. In the midplane with $x/w=0$ ($x$-midplane), $J_x$ penetrates completely into the sample only in the mid-$z$ line [figure \ref{f.bench_J}(a)]. In contrast, the stack response to the applied magnetic field is different. The $J_x$ and $J_y$ components at the $x-$ and $y-$midplanes, respectively are symmetric, since $J_z$ is zero. The center zone is not fully saturated with any component of the current density [figure \ref{f.bench_J}(c) for $J_x$ and (d) for $J_y$], because the top and bottom ``tapes" partially shield the applied field. At the top and bottom planes, the prism is fully saturated. The screening current in the center, $z/d=0$, needs to shield smaller magnetic fields, and hence the current density penetrates less.

\begin{figure}[tbp]
\centering
 \subfloat[][]
{\includegraphics[trim= 0 0 -10 0,clip,width=6.0 cm]{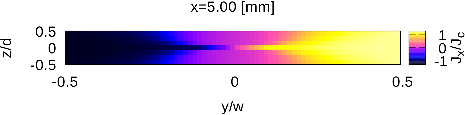}}
 \subfloat[][]
{\includegraphics[trim= -10 0 0 0,clip,width=6.0 cm]{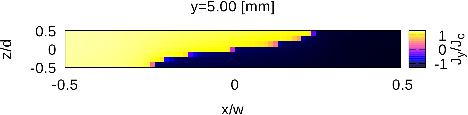}}\\
 \subfloat[][]
{\includegraphics[trim= 0 0 -10 0,clip,width=6.0 cm]{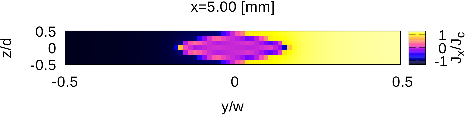}}
 \subfloat[][]
{\includegraphics[trim= -10 0 0 0,clip,width=6.0 cm]{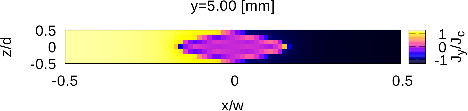}}
\caption{The current density penetration to the bulk (a, b) and stack (c, d) at the peak of the applied magnetic field 200 mT with angle $\theta=60\degree$ and $n$=25. The $J_x$ (a,c) and $J_y$ (b,d) components of the current density are calculated by MEMEP 3D.}
\label{f.bench_J}
\end{figure}

%%%%%%%%%%%%%%%%%%%%%%%%%%%%%%%%%%%%%%%%%%%%%%%%%%%%%%%%%%%%%%%%%%%%%%%%%%%%%%%%%%%%%%%%%%%%%%%

\subsection{AC loss, magnetization loops, and comparison with other methods}
\label{s.prism_comparison} 

Different 3D modelling methods can reproduce the electromagnetic response of our studied cases, which will further validate the MEMEP 3D method. In this sub-section, we focus on the comparison of three methods: the MEMEP 3D variational method, the Finite Element Method (FEM) based on $\vH$ formulation \cite{brambilla07SST,grilli13Cry,zermeno13SST}, and Volume Integral Method (VIEM) based on $\vA-\phi$ formulation \cite{Cristofolini02PCS,Albanese97AIP}. 

The instantaneous power dissipation for both cases is on figure \ref{f.bench_ACloss}. All three methods reach excellent agreement. The bulk sample generates higher AC loss than the stack, since bulk is more saturated with the screening current, and hence there is a bigger zone with $|\vJ|$ above $J_c$, which contributes to the AC losses. The total AC loss per cycle for bulk and stack \cite{Grilli14IESa} is in table \ref{t.bulk_ACloss}. The AC losses results agree very well between models. The power loss is calculated by instantaneous power dissipation as 
\begin{equation}
{Q_{JE}=2\int_{T/2}^T\int_{V}\vJ \cdot \vE dVdt}, 
\label{}
\end{equation}
where $V$ is the volume of the superconductor and the AC loss is calculated at a half cycle multiplied by 2, being $T$ the period. The magnetization loop is calculated by 
\begin{equation}
{Q_{MH}=\int m_a dH_a}, 
\label{}
\end{equation}
where $H_a$ is the applied field and $m_a$ is the magnetic moment in the applied field direction \cite{Grilli14IESa}.

\begin{figure}[tbp]
\centering
{\includegraphics[trim=0 0 0 0,clip,width=7.5 cm]{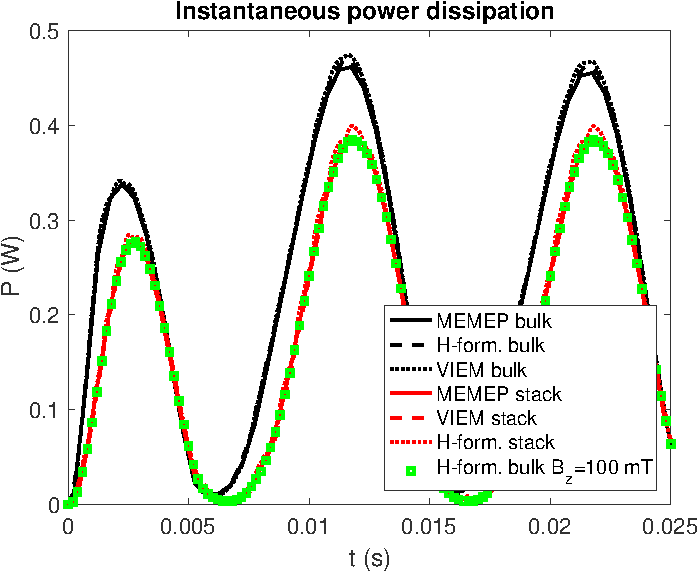}}
\caption{The instantaneous power dissipation calculated by MEMEP 3D, FEM and VIEM method for bulk and stack samples with tilted applied magnetic field. The results shown great accuracy between each method.}
\label{f.bench_ACloss}
\end{figure}

\begin{table}[tpb]
\begin{center}
\begin{tabular}{c c c c c}
\hline
\hline
{\bf Method}	&	{{\bf $Q_{JE}$} bulk}	& {{\bf $Q_{MH}$} bulk} & {{\bf $Q_{JE}$} stack} & {{\bf $Q_{MH}$} stack}\\
\hline
MEMEP 3D	& 4.58	&	4.62	&	3.48 & 3.50\\
H-formulation & 4.59	& 4.62	&	3.47 & 3.45\\
VIEM	& 4.67	& 4.70	&	3.56	& 3.56\\
\hline
\hline
\end{tabular}
\caption{Calculated AC loss per cycle for bulk and stack under tilted applied magnetic field by different methods MEMEP 3D, FEM, VIEM. The AC loss is calculated by integrating \cite{Grilli14IES} the instantaneous power dissipation $\vE\cdot\vJ$ (Q$_{JE}$) and magnetization loops (Q$_{MH}$).}
\label{t.bulk_ACloss}
\end{center}
\end{table}

The magnetization loops for bulks and stacks perfectly overlap by all three methods [figure \ref{f.bench_loop} (a) bulk (b) stack]. The stack of tapes does not have $M_x$ component because $J_z$ is equal to zero, and hence the magnetization loops close in the $xy$ plane. The magnetization component $M_z$ is slightly higher because the screening current is only in the $xy$ plane. The $M_x$ component in the bulk is small because of the thin sample.

\begin{figure}[tbp]
\centering
 \subfloat[][]
{\includegraphics[trim=0 0 0 0,clip,width=5.5 cm]{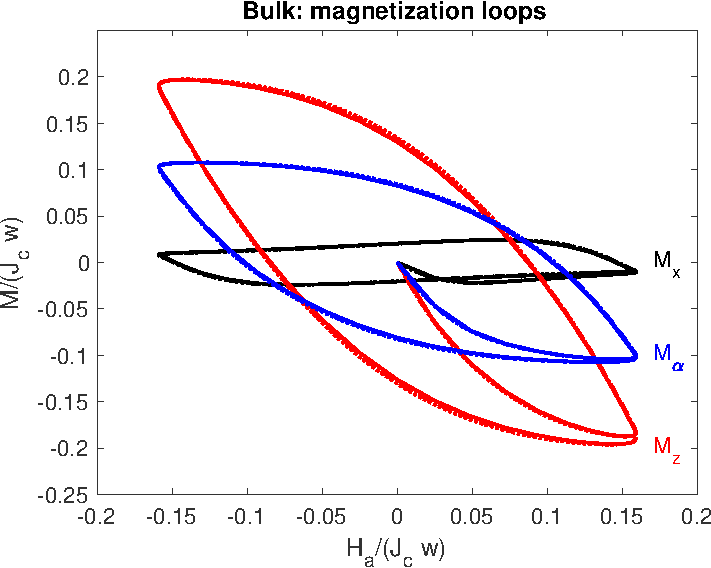}}
 \subfloat[][]
{\includegraphics[trim=0 0 0 0,clip,width=5.5 cm]{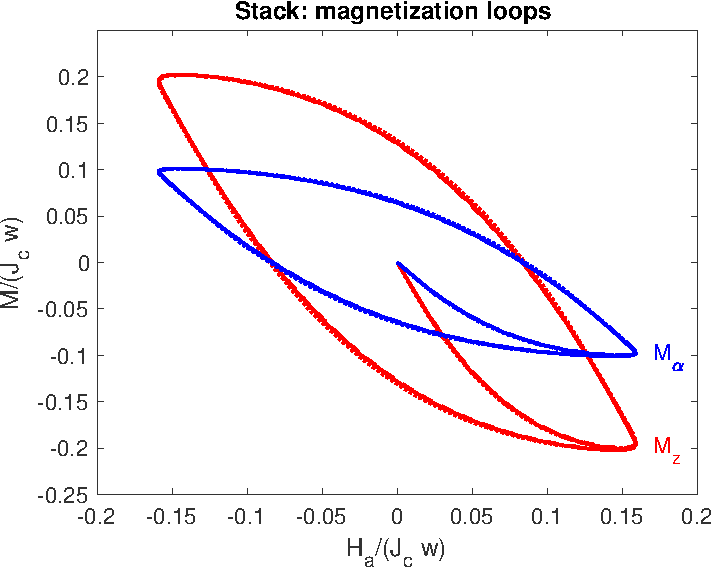}}
\caption{The magnetization loops of (a) bulk and (b) stack sample under tilted applied field.}
\label{f.bench_loop}
\end{figure}

The $J_y$ current density penetration into the bulk is on figure \ref{f.bench_Jy}. The current density distribution is in plane $z=0$ mm with three line positions $y=0, y=0.2w, y=0.4w$. The methods confirm again very nice agreement. 

\begin{figure}[tbp]
\centering
{\includegraphics[trim=0 0 0 0,clip,width=5.5 cm]{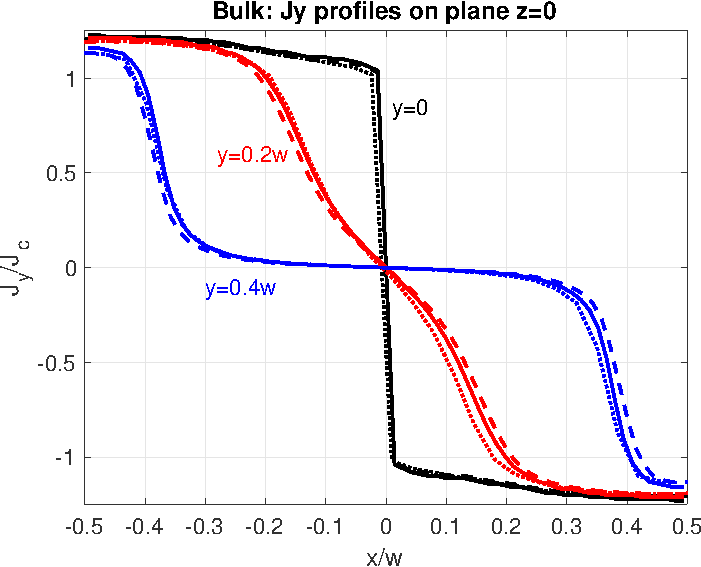}}
\caption{The $J_y$ profile at the plane $z=0$ at three different positions $y/w=0,0.2,0.4$ at the peak of the applied field of 200 mT.}
\label{f.bench_Jy}
\end{figure}

%%%%%%%%%%%%%%%%%%%%%%%%%%%%%%%%%%%%%%%%%%%%%%%%%%%%%%%%%%%%%%%%%%%%%%%%%%%%%%%%%%%%%%%%%%%%%%%

\section{Cross-field demagnetization of cubes.}
\label{s.cube_demagnetization}

Superconducting bulks are a potential alternative to permanent magnets. Bulk magnetization is higher compared to permanent magnets. The world record of trapped field, $B_t$, is 17.6 T \cite{Durrell14SST} with big effort to developed such material \cite{Namburi16JECS,Namburi16SST}. The transverse or cross-fields demagnetize the bulks, decrease the trapped field. There are demagnetization studies based on modelling by 2D $H$ formulation and comparison with experiments \cite{Vanderbemden03IES,vanderbemden07SST}. Experimental and 2D and 3D modelling study of hybrid structures such as ferromagnetic and superconducting material can be found in \cite{fagnard16SST}, where the superconductor is cylindrical. The comparison of 2D models based on $A$ and $H$ formulation with Brandth and Mikitik theory are in \cite{Campbell17SST}. However, demagnetization of a cube sample is still not well understood. Since the finite sample includes end effects, a full 3D accurate model is necessary.  

%%%%%%%%%%%%%%%%%%%%%%%%%%%%%%%%%%%%%%%%%%%%%%%%%%%%%%%%%%%%%%%%%%%%%%%%%%%%%%%%%%%%%%%%%%%%%%%

\subsection{Modelling of cross-field demagnetization}
\label{s.cube_dem_model}

The demagnetization model of a bulk by cross-fields follows the demagnetization measurements in section \ref{s.measurements_demag}. The input parameters are in table \ref{t.demag}. The critical current density is chosen as $J_c=2.6\cdot10^{8}$ A/m$^{2}$, in order to have the same trapped field as in the measurements. We magnetize the sample by an applied field in the $z$ axis with triangular wave-form and ramp rate of 13 mT/s. A transverse field of various amplitudes and frequencies is applied along the $x$ axes after the relaxation time of 900 s [figure \ref{f.demag_geometry}(a)].

\begin{figure}[tbp]
\centering
 \subfloat[][]
{\includegraphics[trim=0 0 -10 0,clip,height=4.5 cm]{fig52.png}} 
 \subfloat[][]
{\includegraphics[trim=-10 0 0 0,clip,height=4.5 cm]{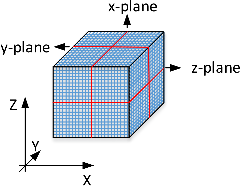}} 
\caption{The geometry of the cube for the demagnetization process. (a) The trapped field $B_t$ is along the $z$ axis and the applied cross-field ripples are along the $x$ axes. The Hall probe array is parallel to the ripples, sensing $\vB_z$ in the $z$ direction. (b) The cross mid planes for colour maps with current density components are at planes $x$=3 mm, $y$=3 mm and $z$=3 mm.}
\label{f.demag_geometry}
\end{figure}

\begin{table}[!t]
\renewcommand{\arraystretch}{1.3}
\centering
\begin{tabular}{|c||c|}
\hline
Size[mm] & 6x6x6\\
\hline
${J_c}$[A/m${^2}$] & 2.6${\cdot 10^8}$\\
\hline
${B_{az,max}}$[T] & 1.3\\
\hline
Ramp rate[mT/s] & 13\\
\hline
Relaxation[s] & 900\\
\hline
${E_c}$[V/m] & 1e-4\\
\hline
${f_{ax}}$[Hz] & 0.1,1\\
\hline
${B_{ax}}$[mT] & 35,73,130\\
\hline
n[-] & 30\\
\hline
\end{tabular}
\caption{The input parameters for the cross-field demagnetization modelling based on the real sample and measurement.}
\label{t.demag}
\end{table}

Figure \ref{f.demag_wave} shows the calculated magnetization inside the sample and trapped field 100 $\mu$m above the sample for the entire demagnetization process. The model assumes ripples with frequency 0.1 Hz and maximum amplitude, $B_{ax}$ of 130 mT. The ripples demagnetize the sample and decrease the trapped field, $B_t$. The $B_t$ value at the end of the relaxation time is 0.3 T and the peak of the trapped field is on figure \ref{f.demag_peak}. The peak decreases with the applied transverse field $B_{ax}$ and it is shifted aside from the center. The trapped field peak is shifted with the first positive peak of the transverse applied field on figure \ref{f.demag_peak} (blue line at time 1002.5 s). The second negative peak of $B_{ax}$ decreases the trapped field further and it shifts the trapped field peak to the opposite side along the $x$ axes [figure \ref{f.demag_peak} red line at time 1007.5 s]. 

The calculation result of MEMEP 3D is compared with Finite Element Method model made by Mark Ainslie in Comsol Multiphysics 5.2a based on $H$- formulation \cite{zhangM12SSTa,grilli13Cry,Ainslie14SST, Ainslie15SST}. Both methods agree very well and small discrepancy in FEM model comes from the linear (first-order) elements and coarse mesh (figure \ref{f.demag_peak}).  

\begin{figure}[tbp]
\centering
{\includegraphics[trim=2 45 20 55,clip,width=11.0 cm]{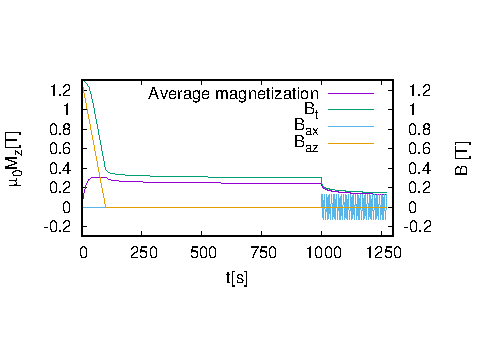}} 
\caption{The figure of the entire magnetization and demagnetization process over time with wave-forms of the applied field $B_{az}$ and ripple $B_{ax}$. The magnetization inside the sample and trapped field at 100 $\mu m$ above the sample are reduced by the applied ripples.}
\label{f.demag_wave}
\end{figure}

\begin{figure}[tbp]
\centering
{\includegraphics[trim=40 0 50 0,clip,width=7 cm]{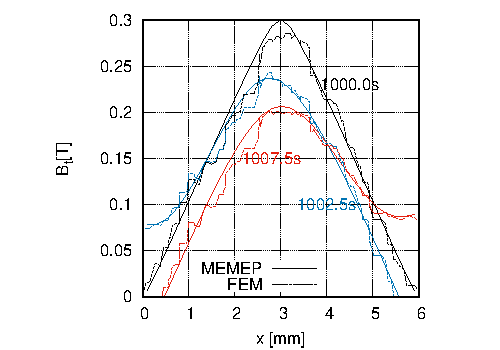}} 
\caption{The peak of the trapped field at the end of the relaxation time of 900 s, which follows the 100 s long magnetization by $B_{az}$. The red and blue lines are trapped fields at 1002.5 s 
(at first positive peak of ripple) and 1007.5 s (at first negative peak of ripple).}
\label{f.demag_peak}
\end{figure}

Figure \ref{f.demag_J} shows the current distribution crossing the mid-planes $x$=3mm, $y$=3mm, $z$=3mm [sketch \ref{f.demag_geometry}(b)] at the end of the relaxation time and at the first positive and tenth positive peak of the applied field ripples. 

The cube is fully saturated by screening current at the end of the relaxation time [figure \ref{f.demag_J}(a)(d)]. The screening current is perpendicular to the applied field $B_{az}$ according to the CSM. Since the sample is fully saturated, the $J_z$ component vanishes [figure \ref{f.demag_J}(g)], and hence there are only $J_x$ and $J_y$ components of the current density. The current density reduces its value to $0.8J_c$ during relaxation. 

The ripples are applied along the $x$ axis, and hence the new screening current penetrates into the sample with current loops in the $yz$ plane, inducing $J_z$ and $J_y$. The following situation is at the first positive peak of the ripple. $J_z$ smoothly penetrates the $z$ plane [figure \ref{f.demag_J}(h)] and it erases $J_x$ at the edges [figure \ref{f.demag_J}(b)]. $J_y$ changes the previous distribution in the $y$ plane and it creates an ``S" shaped front, which was already predicted by 2D models \cite{Campbell17SST}. This $J_y$ induced from the ripples increases the value at the corners up to $J_c$ [figure \ref{f.demag_J}(e)]. The ripple field does not only create the $S$-shaped current front in the $y$-plane but it also changes the direction of the screening current loops, from in the $xy$ plane for the initial magnetization currents to the $yz$ plane for the de-magnetization currents. Then, the ripple fields induce a $J_z$ component. 

The current distribution under the following ripples keep the same penetration behaviour. The $J_z$ penetration is sharper after ten cycles of the ripples $B_{ax}$ [figure \ref{f.demag_J}(c)], therefore $J_x$ in the plane $x=3$ mm is erased at the edges with sharp step to zero [figure \ref{f.demag_J}(c)]. After 10 cycles with bipolar peaks of $B_{ax}$, there appears a separate zone at the center with the first magnetization currents due to $B_{az}$ and a screening current zone due to $B_{ax}$ [figure \ref{f.demag_J}(f)]. The central zone is with value around $0.9J_c$, which corresponds to the value of $J_x$ at the plane $x=3$ mm [figure \ref{f.demag_J}(c)]. 

The entire demagnetization of the cube and the current distribution is the origin of the decreased trapped field and the asymmetry of the trapped peak \ref{f.demag_peak}. The FEM model confirmed the same current path and behaviour. The comparison of both methods with $J_y$ distribution in the plane $y=3$ mm at the first positive peak of the ripples is on figure \ref{f.demag_FEMJy}.            

\begin{figure}[tbp]
\centering
{\includegraphics[trim=0 0 48 0,clip,height=4.0 cm]{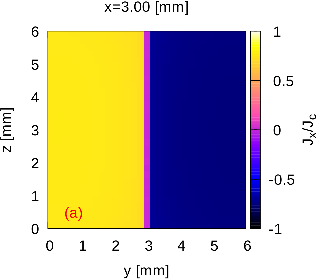}} 
{\includegraphics[trim=22 0 48 0,clip,height=4.0 cm]{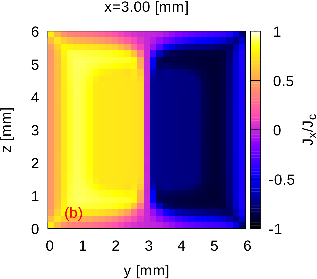}} 
{\includegraphics[trim=22 0 0 0,clip,height=4.0 cm]{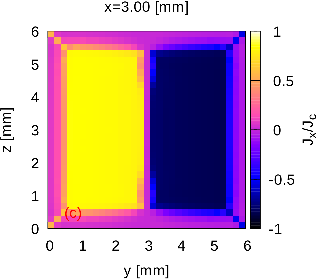}} \\
\vspace{1 mm}
{\includegraphics[trim=0 0 50 0,clip,height=4.0 cm]{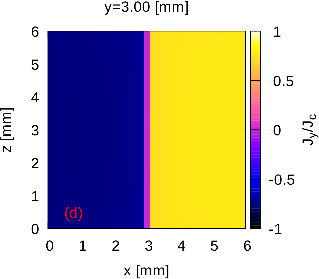}} 
{\includegraphics[trim=22 0 50 0,clip,height=4.0 cm]{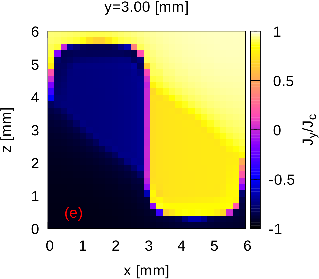}} 
{\includegraphics[trim=22 0 0 0,clip,height=4.0 cm]{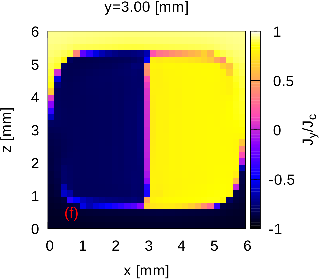}} \\
\vspace{1 mm}
{\includegraphics[trim=0 0 48 0,clip,height=4.0 cm]{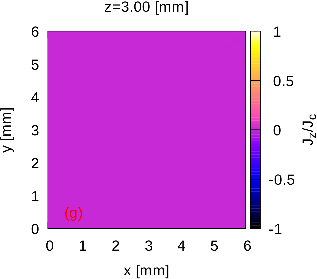}} 
{\includegraphics[trim=22 0 48 0,clip,height=4.0 cm]{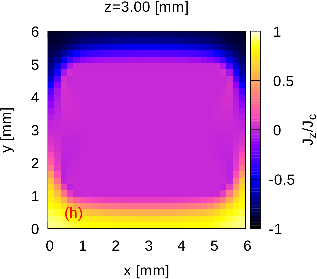}}
{\includegraphics[trim=22 0 0 0,clip,height=4.0 cm]{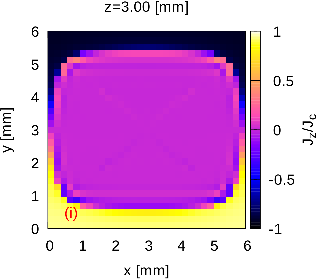}}
\caption{The current density distribution in the cube during the demagnetization process. The profiles at 1000 s (end of relaxation) (a),(d),(g), at 1002.5 s (end of first demagnetizing peak) (b),(e),(h) and at 1012.5 s (end of 10th demagnetizing peak) (c),(f),(i). The $J_x$ component is in the top row, $J_y$ in the mid row and ${J_z}$ at bottom row. The positions of the cross planes are on figure \ref{f.demag_geometry}(b).}
\label{f.demag_J}
\end{figure}

\begin{figure}[tbp]
\centering
 \subfloat[][]
{\includegraphics[trim=0 0 0 20,clip,height=5.0 cm]{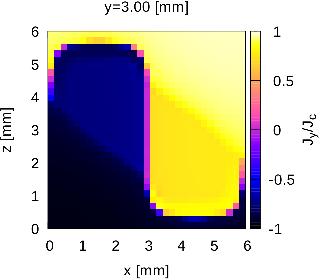}} 
 \subfloat[][]
{\includegraphics[trim=0 0 -10 0,clip,height=5.0 cm]{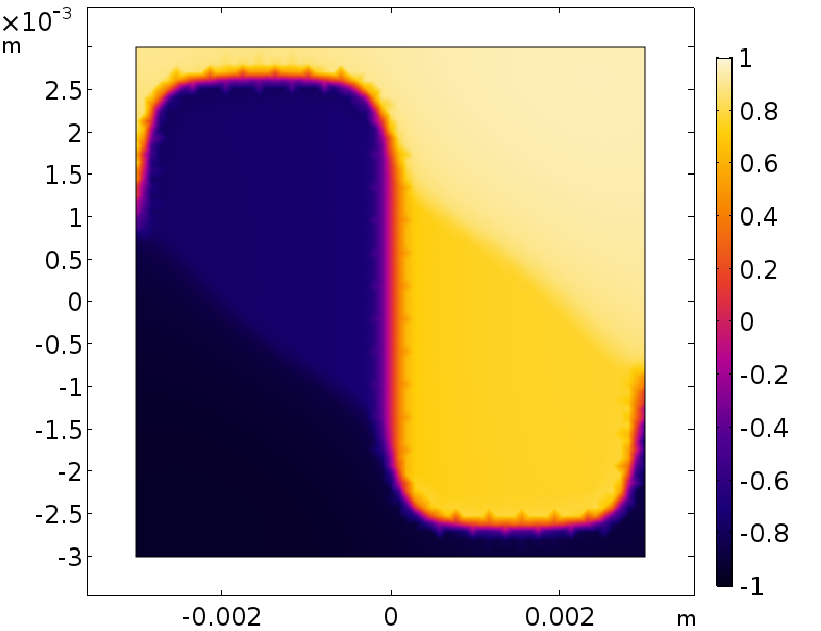}} 
\caption{The $J_y$ distribution in plane $y=3$ mm at the time 1002.5 s (after the first demagnetizing cycle). (a) The MEMEP 3D and (b) The FEM model, which confirmed the current distribution during demagnetization process.}
\label{f.demag_FEMJy}
\end{figure}

%%%%%%%%%%%%%%%%%%%%%%%%%%%%%%%%%%%%%%%%%%%%%%%%%%%%%%%%%%%%%%%%%%%%%%%%%%%%%%%%%%%%%%%%%%%%%%%

\subsection{Experiments and comparison to modelling}
\label{s.cube_dem_measurements}

The real sample is prepared and measured according to section \ref{s.measurements_demag}. The reduction of the trapped field by cross-field ripples is on figure \ref{f.demag_meas1}. The ripples are with frequency 0.1, 1 Hz and maximum applied fields $B_{ax}=B_t/2,B_t/4,B_t/8$ ($B_{ax}=130, 70, 35$ mT). The trapped field decreases rapidly during first a few cycles of ripples, because the current distribution changes mostly at that time. The higher frequency reduces faster the trapped field during the 10-minute measurements, due to the higher number of ripples compared to the case of 0.1 Hz. The trapped field decreases with increasing the applied ripple fields.

\begin{figure}[tbp]
\centering
 \subfloat[][]
{\includegraphics[trim=0 0 10 10,clip,width=6 cm]{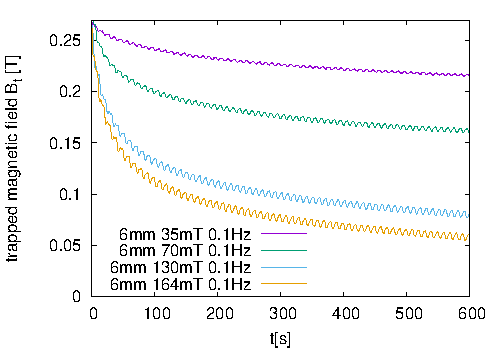}}
 \subfloat[][]
{\includegraphics[trim=0 0 10 10,clip,width=6 cm]{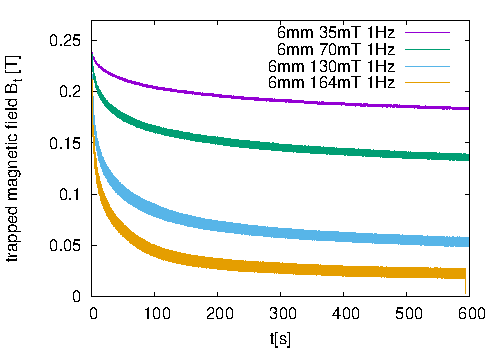}} 
\caption{The measurements of the trapped field with various applied fields ripples $B_{ax}=35, 70, 130, 164$ mT and frequency (a) 0.1 Hz, (b) 1.0 Hz. The measurements details in section \ref{s.measurements_demag}.}
\label{f.demag_meas1}
\end{figure}

The dependence of the reduced trapped field on the number of ripple cycles is on figure \ref{f.demag_meas2}. The curve for the ripples with amplitude 35 mT shows nice agreement for both frequencies. The applied field, $B_{ax}$, with higher amplitude presents 10\% frequency dependence caused by the finite power-law exponent. Higher frequency ripples induce higher electric fields, and hence the screening current is with higher amplitude and lower penetration depth \cite{Thakur11SST,Thakur11SSTa,Sander10JCS,Grilli14IESa}. 

\begin{figure}[tbp]
\centering
{\includegraphics[trim=0 10 10 18,clip,width=7 cm]{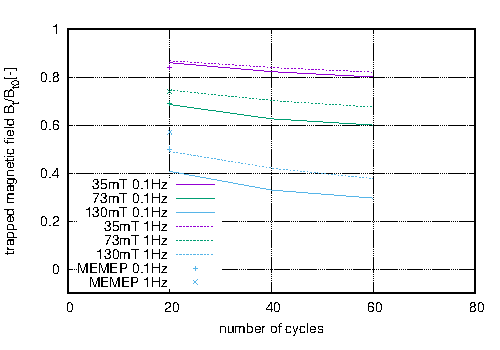}}
\caption{The trapped magnetic field dependence on the number of ripple cycles. The ripples of 35 mT are compared with MEMEP 3D model.}
\label{f.demag_meas2}
\end{figure}

The complete signal of the trapped field measured by the 7 Hall-probe sensors is on figure \ref{f.demag_meas3}. The position of the sensor array is on figure [\ref{f.demag_geometry}(a)]. The sensor 1 and 7, which are on both sides of the array confirms, the asymmetry of the trapped field, because the phase of the trapped field is shifted by 90$\degree$. 

\begin{figure}[tbp]
\centering
{\includegraphics[trim=40 0 40 0,clip,width=7 cm]{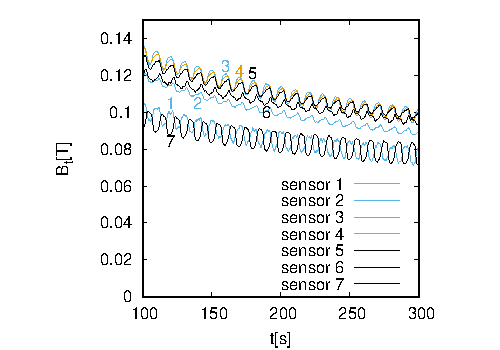}}
\caption{The trapped magnetic field profile during demagnetization measurement by the 7 Hall-probe sensor array. The measurements confirmed the asymmetry of the peak. The signal from Hall-probe sensors 
is marked by 1 to 7 from left to right side of the array.}
\label{f.demag_meas3}
\end{figure}

Finally, we compare the measurements with both models, MEMEP 3D and FEM (figure \ref{f.demag_meas4}). The models agree very well in low applied fields, $B_{ax}$, at frequency 0.1 Hz [figure \ref{f.demag_meas4}(a)] and 1 Hz [figure \ref{f.demag_meas4}(b)]. The applied field ripples with amplitude $B_{ax}=B_t/2$ show higher deviation, which can be explained by inaccurate power-law exponent and the  constant $J_c$ assumption. The FEM model shows even higher deviation compare to the MEMEP 3D. The deviation comes from coarse mesh and (first-order) elements.      

\begin{figure}[tbp]
\centering
 \subfloat[][]
{\includegraphics[trim=40 0 50 10,clip,height=5.85 cm]{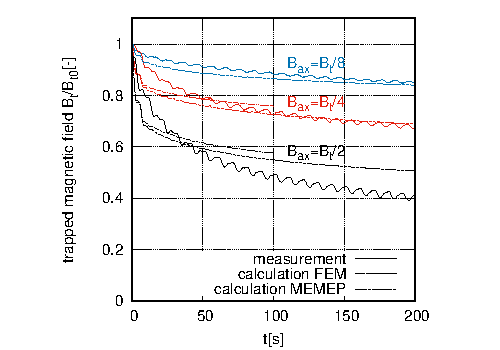}}
 \subfloat[][]
{\includegraphics[trim=90 0 50 10,clip,height=5.85 cm]{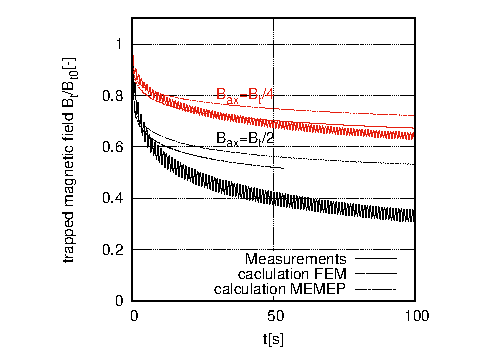}} 
\caption{The comparison of the cube demagnetization by cross-field $B_{ax}$ measurements with the MEMEP 3D and FEM models. The measurements with ripples of (a) 0.1 Hz and (b) 1 Hz. The models and measurements agree for low ripple amplitudes.}
\label{f.demag_meas4}
\end{figure}

%%%%%%%%%%%%%%%%%%%%%%%%%%%%%%%%%%%%%%%%%%%%%%%%%%%%%%%%%%%%%%%%%%%%%%%%%%%%%%%%%%%%%%%%%%%%%%%

\section{Anisotropic force-free effects}
\label{s.aniso}

Superconductors could be isotropic (LTS or MgB$_2$ wires, even thought MgB$_2$ is intrinsically anisotropic material) and anisotropic (YBCO or REBCO) materials. Isotropic superconductors are with homogeneous properties in any direction. However, anisotropic  superconductors change properties according the direction of the current flow. The anisotropy can be distinguished between ``intrinsic", ``de-pinning" and ``force-free" anisotropy. 

The intrinsic anisotropy is due to different internal material properties. For example, REBCO films and bulks present lower critical current density when current flows in the $c$-plane compared to the $ab$ plane. The second kind of anisotropy is de-pinning anisotropy, which comes from anisotropic pinning of vortices \cite{blatter94rmp} and depends on the applied filed direction. The de-pinning anisotropy is important to for increase $J_c$ by improving pinning centers at perpendicular applied field with $\vJ$. 

The other type of anisotropy is due to force-free effects, which depend on the angle between magnetic field $\vB$ and current density $\vJ$. Force-free effects are due to flux cutting and crossing \cite{Vlasko15FL,Clem82PRB,Clem11SST}, which appear when the local magnetic field is not perpendicular to the current density. They appear in all superconducting power applications with parallel magnetic fields. The macroscopic anisotropy study of superconductors developed new models such as the Double Critical State Model \cite{Clem82PRB} with a later improvement by the General Double Critical State Model \cite{clem84PRB}. Branth and Mikitik introduce the Extended DCSM \cite{Brandt07PRB} and Badia and Lopez introduce Elliptic CSM \cite{badia05APL,romerosalazar03APL}. The summarization of the previous models is in \cite{Clem11SST}. 

On the other hand, all models must be validated by experiments such as on REBCO tapes by de-pinning anisotropy \cite{Chepikov17SST,Lijima15IES,Lee14IES,Rosii16SST,Xu17IES,Lin17AIM,Miura17SST}, Bi223 \cite{Ayai08PCS,Goyal97JMR} or iron based superconductors \cite{Pallecchi15SST, Yi17APS, Ma17SST,Hecher18PRB}. oThere is a wide database of tapes characterizations with $J_c$ results \cite{HTSdatabase}. The correction of $I_c$ in measurements is in \cite{Pardo11SST,Zermeno16IES}. There are other more exotic causes of anisotropy, such as that due to flux channelling in vicinal films \cite{durrell03JAP,Lao17SST}, although this particular case could be considered as intrinsic anisotropy.

The force-free effects in infinite samples are still not fully understood, and hence light on this topic is important for material characterization and for optimization of power devices. The force-free effects are important for rotating machines containing bulks and magnets containing transposed cables like CORC or ROEBEL.

The 3D modelling tool with all finite size and force free effects is necessary. The following two sections are focused on modelling force-free effects in thin films and prisms with various applied magnetic fields angles. The anisotropic results are compared with the isotropic case.

%%%%%%%%%%%%%%%%%%%%%%%%%%%%%%%%%%%%%%%%%%%%%%%%%%%%%%%%%%%%%%%%%%%%%%%%%%%%%%%%%%%%%%%%%%%%%%%

\subsection{Finite superconducting thin film with anisotropic ${\vE(\vJ)}$ relation}
\label{s.film_anisotropy}

This section studies the magnetization of thin films with various angles of the applied magnetic field. The MEMEP 3D modelling tool can takes any $\vE(\vJ)$ relation into account, and hence the tool models force-free effects in thin films. The results here discuss all effects and current paths under several applied magnetic field angles $\theta$. 

The modelling situation of a square thin film magnetization is on figure \ref{f.ani_geometry}(a). The sample dimensions are 12mm$\times$ 1$\mu$m ($w\times d$), where $w$ is the width and $d$ is the thickness. The anisotropic power law parameters are the perpendicular critical current density $J_{c\perp}=3\cdot10^{10}$ A/m$^2$, parallel critical current density $J_{c\parallel}=9\cdot10^{10}$ A/m$^2$ and a realistic $n$ value of 30. The alternating applied magnetic field is with constant $z$ component of 50 mT for any applied field angle $\theta$. The applied field angles are $0\degree, 45\degree, 60\degree$ and $80\degree$. The minimization algorithm uses the magnetic field from the previous time step $B_{(t-\Delta t)}$, because the anisotropic power law is not well defined in very low or zero local magnetic field (section \ref{s.EJ}). Therefore, the results for the remanent state are shifted to the next time step with non-zero applied field. The error in calculation is decreased by higher number of total time steps per cycle, and hence lower $\Delta t$.    

\subsubsection{AC power device situation}

Next, we assume that the applied field is sinusoidal of 50 Hz frequency, in order to simulate the situation of a power device.

\begin{figure}[tbp]
\centering
 \subfloat[][]
{\includegraphics[trim=-30 0 0 0,clip,width=6.0 cm]{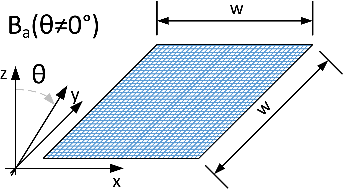}}
 \subfloat[][]
{\includegraphics[trim=-30 0 0 0,clip,width=6.0 cm]{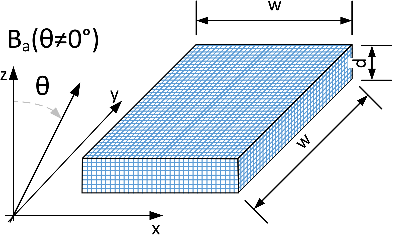}}
\caption{The geometry of the modelling sample with non-zero applied field angle $\theta$. (a) Thin film with width $w$. (b) Prism with thickness $d$ and width $w$.}
\label{f.ani_geometry}
\end{figure}

The first modelling case is with angle $\theta=0\degree$ and applied field amplitude of 50 mT. The screening current density gradually penetrates symmetrically into the sample from the edges already under the small applied field of 19.1 mT [figure \ref{f.ani_0J}(a)]. The modelling sample presents the same features as the isotropic case in figure \ref{f.Jc}. Actually, the current penetration process is identical. The almost saturated state is on figure [\ref{f.ani_0J}(b)] and the remanent state is on figure[\ref{f.ani_0J}(c)]. The screening current density is around $J_{c\perp}$, because the applied field angle and self-field are perpendicular to the sample surface.

\begin{figure}[tbp]
\centering
 \subfloat[][]
{\includegraphics[trim=30 0 50 0,clip,height=4.5 cm]{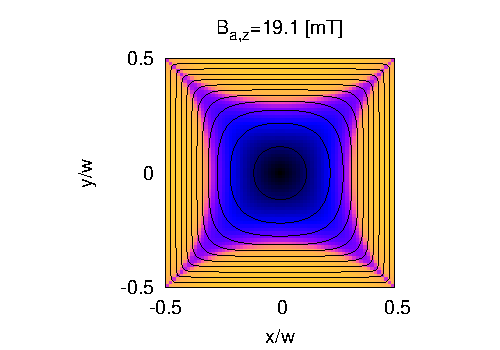}} 
 \subfloat[][]
{\includegraphics[trim=85 0 80 0,clip,height=4.5 cm]{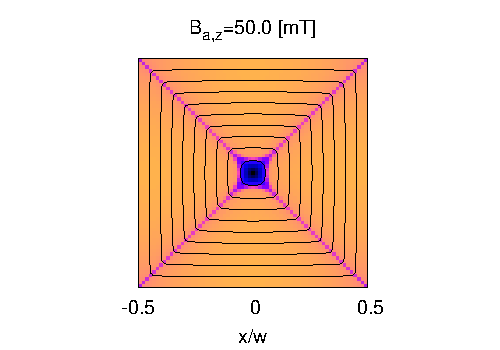}} 
 \subfloat[][]
{\includegraphics[trim=50 0 25 0,clip,height=4.5 cm]{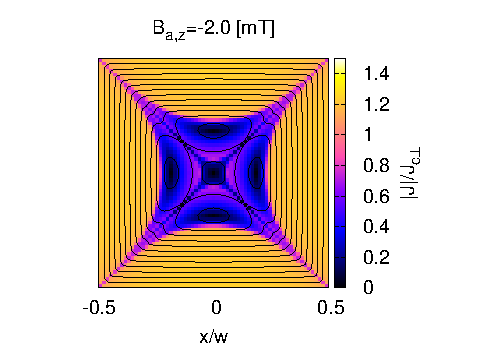}} 
\caption{The gradual penetration of the current density into the thin film with the sinusoidal applied field $B_{a}$=50 mT. The applied field angle is $\theta$=0$\degree$ and frequency 50. The anisotropic power law is with $n$ value 30 and $J_{c\parallel}=3J_{c\perp}$. The penetration $\vJ$ is the same as in isotropic case on figure \ref{f.Jc}.}
\label{f.ani_0J}
\end{figure}

The second modelling case is with angle $\theta=45\degree$ and the applied magnetic field amplitude is 70.7 mT, in order to keep $B_{a,z}$ component is equal 50 mT. The small $B_{az}$ of 19.1 mT creates significant penetration of the screening current into the sample [figure \ref{f.ani_45J}(a)]. In the colour maps there appear regions with $|\vJ|\approx J_c$ with $\vJ$ parallel to $x$ or $y$ axes. We refer  to the current density in these regions as $J_x$ and $J_y$ components, respectively. The penetration depth of $J_y$ is the same as the previous case with $\theta=0\degree$. However, the $J_x$ component of the current density is around 2$\times J_{c\perp}$. The reason is that the applied field is slightly aligned with the current density, enabling $|\vJ|$ between $J_{c\perp}$ and $J_{c\parallel}$. The higher $J_x$ density with lower penetration depth is enough to close the current loops with $J_y$ component with larger penetration depth. The sample is almost fully saturated with screening current density at the peak of applied field [figure \ref{f.ani_45J}(b)] with the same penetration depth of $J_y$ as in case $\theta=0\degree$. The $J_x$ component shows the highest value at the penetration front, where self-field, which is perpendicular to the sample is lower than behind the penetration front. The remanent state shows complex current paths with another penetration front due to the currents induced during the decrease of applied field. This front presents half the penetration depth [figure \ref{f.ani_45J}(c)]. The applied field is zero, which leaves as non-zero only the perpendicular self-field. Therefore, the previously enhanced current density by $J_{c\parallel}$ is reduced to a value around $J_{c\perp}$.

\begin{figure}[tbp]
\centering
 \subfloat[][]
{\includegraphics[trim=30 0 50 0,clip,height=4.5 cm]{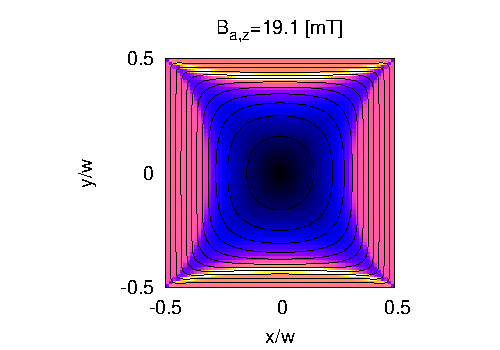}} 
 \subfloat[][]
{\includegraphics[trim=85 0 80 0,clip,height=4.5 cm]{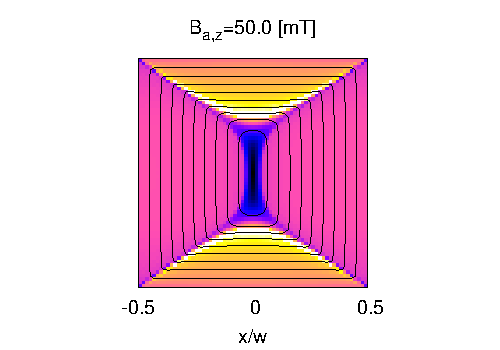}} 
 \subfloat[][]
{\includegraphics[trim=50 0 25 0,clip,height=4.5 cm]{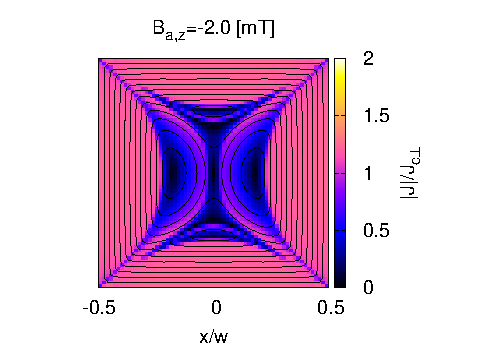}} 
\caption{The gradual penetration of the current density into the anisotropic thin film with the sinusoidal applied field amplitude $B_{a}$=70.7 mT. The applied field angle is $\theta$=45$\degree$.}
\label{f.ani_45J}
\end{figure}

The next cases with angles $\theta=60\degree$ and 80$\degree$ (applied fields 100 and 287.9 mT, respectively) show the same magnetization behaviour with the similar current paths. The $J_y$ screening current reaches the same penetration depth as for $\theta=0\degree$. For $B_{a,z}$=19.1 mT, [figures \ref{f.ani_60J}(a) and \ref{f.ani_80J}(a)]. The $J_x$ component increases the value with applied field angle $\theta$ and decreases the penetration depth. The saturation state [figure \ref{f.ani_60J}(b) and \ref{f.ani_80J}(b)] at the peak of applied field shows again the same penetration depth of $J_y$ up to the sample center, because of the $\theta$-independent $B_{a,z}$. $J_x$ increases with the angle $\theta$ and reaches a value around $J_{c\parallel}$. For both cases at the remanent state, there appears a reduction of $J_x$ component to $J_{c\perp}$, because the self-field of the thin film is perpendicular [figure \ref{f.ani_60J}(c) and \ref{f.ani_80J}(c)]. 

\begin{figure}[tbp]
\centering
 \subfloat[][]
{\includegraphics[trim=30 0 50 0,clip,height=4.5 cm]{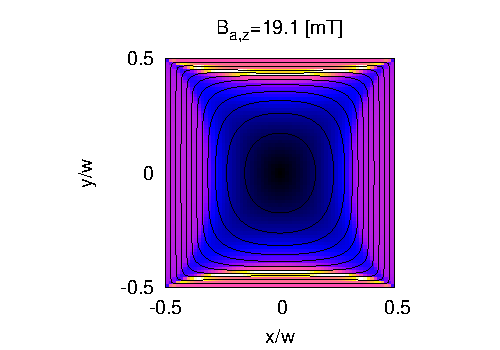}} 
 \subfloat[][]
{\includegraphics[trim=85 0 80 0,clip,height=4.5 cm]{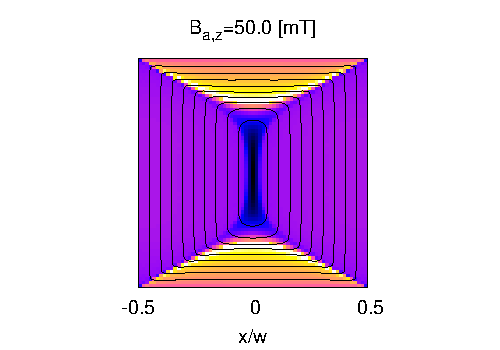}} 
 \subfloat[][]
{\includegraphics[trim=50 0 25 0,clip,height=4.5 cm]{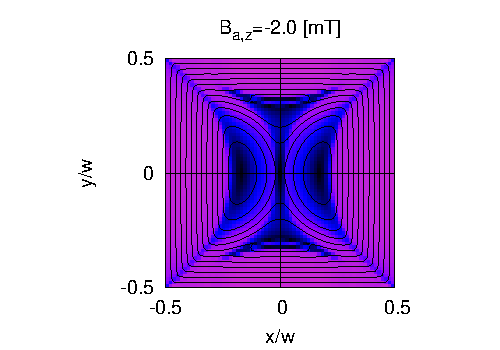}} 
\caption{The gradual penetration of the current density into the anisotropic thin film with the sinusoidal applied field of amplitude $B_{a}$=100 mT. The applied field angle is $\theta$=60$\degree$.}
\label{f.ani_60J}
\end{figure}

\begin{figure}[tbp]
\centering
 \subfloat[][]
{\includegraphics[trim=30 0 50 0,clip,height=4.5 cm]{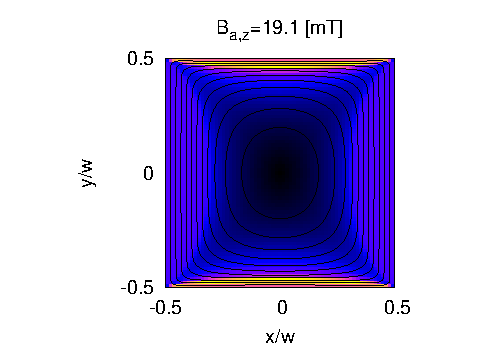}} 
 \subfloat[][]
{\includegraphics[trim=85 0 80 0,clip,height=4.5 cm]{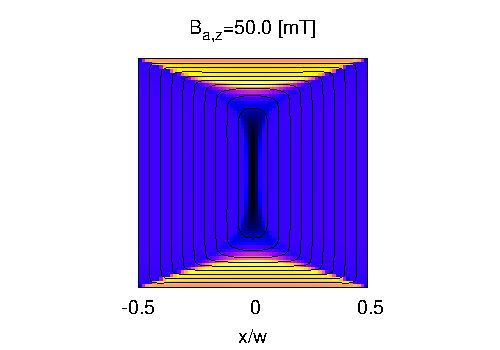}} 
 \subfloat[][]
{\includegraphics[trim=50 0 25 0,clip,height=4.5 cm]{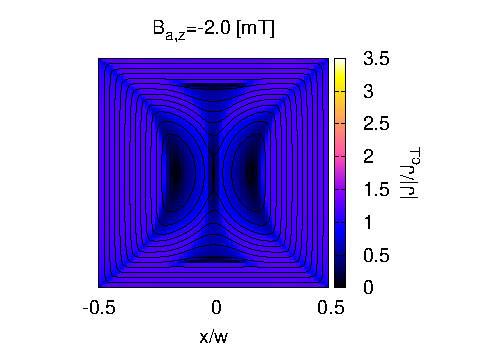}} 
\caption{The gradual penetration of the current density into the anisotropic thin film with the sinusoidal applied field of amplitude $B_{a}$=287.9 mT. The applied field angle is $\theta$=80$\degree$.}
\label{f.ani_80J}
\end{figure}

In the following, we analyse the hysteresis loops with anisotropic power law [figure \ref{f.ani_loop}(a)]. The magnetization increases with angle $\theta$, since this angle increases the current density in part of the sample. The remanent state is with with zero applied field and non-zero self-field, which reduces $|\vJ|$ to $J_{c\perp}$, since the self-field is perpendicular. Therefore, the hysteresis loops present a magnetization drop around the remanent state. The thin film model allows only the $J_x$ and $J_y$ components of the current density, and hence the only non-zero component of the magnetization is $M_z$.      

We also consider the case with isotropic power law, for comparison. Now, we assume frequency 50 Hz and applied field 50 mT. The magnetization loops are on figure \ref{f.ani_loop}(b). The magnetization loop is the same for any applied field angle $\theta$, since $J_c$ does not depend on the angle of local magnetic field and the $z$ component of the applied field is the same. The magnetization loop for any angle is the same as the anisotropic case with angle $\theta$=0$\degree$ [figure \ref{f.ani_loop}(a)]. 

\begin{figure}[tbp]
\centering
 \subfloat[][]
{\includegraphics[trim=70 0 70 0,clip,height=5.5 cm]{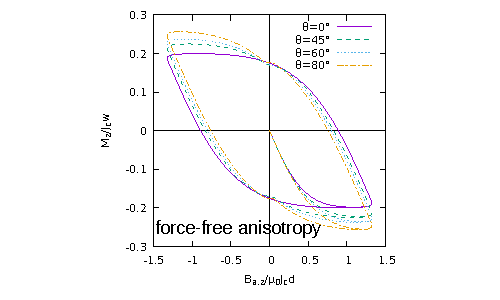}}
\centering
 \subfloat[][]
{\includegraphics[trim=70 0 70 0,clip,height=5.5 cm]{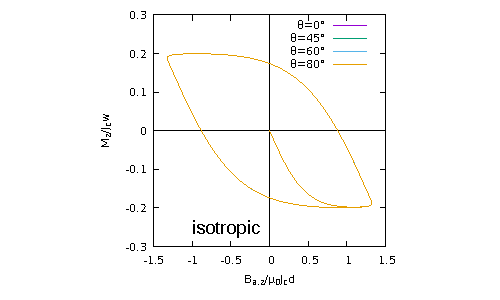}}
\caption{The magnetization loops with (a) anisotropic and (b) isotropic power law. The sinusoidal applied field amplitude is of $B_{a,z}$=50 mT with frequency of 50 Hz and $n$ value 30.}
\label{f.ani_loop}
\end{figure}

%%%%%%%%%%%%%%%%%%%%%%%%%%%%%%%%%%%%%%%%%%%%%%%%%%%%%%%%%%%%%%%%%%%%%%%%%%%%%%%%%%%%%%%%%%%%%%%

\subsubsection{Magnets situation}

Next, we discuss an applied field situation similar to magnets. The applied field frequency is 1 mHz with triangular waveform, which is the same as slow ramp up of the magnet. The applied field amplitude is with $B_{amz}$= 150 mT for any angle $\theta$, being the total amplitude $B_{am}=B_{amz}/\cos\theta$. The power law exponent $n$ is 100, and hence the $\vE(\vJ)$ curve is approximately the same as the Critical State model (section \ref{s.CSM}). The hysteresis loops are on figure \ref{f.ani_kimloop}. The magnetization increases with angle $\theta$ and the saturation state is with a flat curve. The flat part comes from the constant ramp, causing a constant induced electric field. In addition, the high $n$ value reduces current density to the value equal or below $J_{c\perp}$. The remanent state shows the same magnetization drop as previous cases, since the local magnetic field contains the self-field only.

\begin{figure}[tbp]
\centering
{\includegraphics[trim=70 0 70 0,clip,height=5.5 cm]{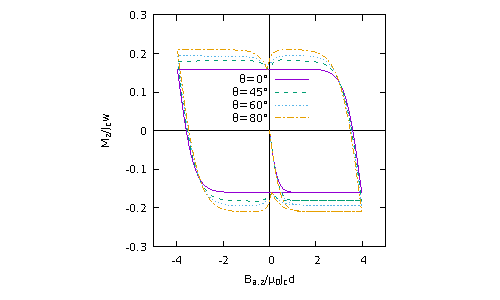}}
\caption{The magnetization loops with anisotropic power similar to the magnet situation. The input parameters of magnetization loops are the triangular applied magnetic field of amplitude $B_{a,z}$=150 mT with frequency 1 mHz and $n$ value 100.}
\label{f.ani_kimloop}
\end{figure}

The last thin film model includes a magnetic field dependence like Kim model. The Kim model is well defined even for low local magnetic fields, and hence the minimization process uses the applied field, $B_a(t)$, from the same time step to calculate $\vJ(t)$. Therefore, at the remanent state $B_a=0$ [figure \ref{f.ani_kloop}(b)]. We assume the Kim model parameters of $m$=0.5, $B_0$=20 mT and $J_{c0}=10^8$ A/m$^2$. The applied field amplitude is with constant $z$ component of 300 mT for all angles $\theta$. The magnetization increases with the angle $\theta$ [figure \ref{f.ani_kloop}(a)]. The $J_{c\perp}$ and $J_{c\parallel}$ depend on the local magnetic field, and hence the current density and magnetization decreases with the applied field, being the lowest at the remanent state. Increasing the applied field saturates the sample already at $B_{az}=$40 mT. The case of $\theta=80\degree$ shows the highest magnetization even with the highest applied field; since $\vB$ and $\vJ$ are more aligned, causing $J_c$ closer $J_{c\parallel}$. The perpendicular self-field at remanent state reduces $|\vJ|\approx J_c$, and hence in the magnetization loop there appears a drop down to the same value as the case of $\theta=0\degree$. The remanent state shows a smooth drop, because of the self-field and the high number of time steps per cycle of 480 [figure \ref{f.ani_kloop}(b)].

\begin{figure}[tbp]
\centering
 \subfloat[][]
{\includegraphics[trim=70 0 70 0,clip,height=5.5 cm]{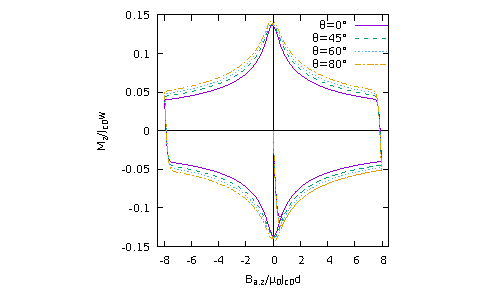}}
 \subfloat[][]
{\includegraphics[trim=60 0 60 0,clip,height=5.5 cm]{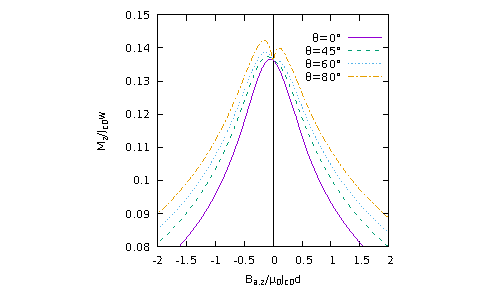}}
\caption{(a) The magnetization loops with anisotropic $\vE(\vJ)$ relation and magnetic depend like Kim model with parameters $m$=0.5, $B_0$=20 mT and $J_{c0}=10^8$ A/m$^2$. The triangular applied field is of 300 mT amplitude and frequency 1 mHz. (b) Zoom on the magnetization loop at the remanent state.}
\label{f.ani_kloop}
\end{figure}

%%%%%%%%%%%%%%%%%%%%%%%%%%%%%%%%%%%%%%%%%%%%%%%%%%%%%%%%%%%%%%%%%%%%%%%%%%%%%%%%%%%%%%%%%%%%%%%

\subsection{Prism with various thicknesses and anisotropic ${{\vE}(\vJ)}$ relation}
\label{s.prism_anisotropy}

The prism model contains all components of $\vT (T_x, T_y, T_z)$, compared to the thin film, where only $T_z$ was non-zero. Now the current path due to the force-free effects presents a more complex behaviour. 

%%%%%%%%%%%%%%%%%%%%%%%%%%%%%%%%%%%%%%%%%%%%%%%%%%%%%%%%%%%%%%%%%%%%%%%%%%%%%%%%%%%%%%%%%%%%%%%

\subsubsection{Detailed analysis for a given thickness}

The modelling sample size is $12\times 12 \times 1$mm with applied field angle $\theta$ in $x-z$ plane [figure \ref{f.ani_geometry}(b)]. In order to keep the magnetization comparable with the thin film, we take the same sheet current density $J_cd$, being $d$ the sample thickness. Then, the critical current densities of the 1 mm thick sample are $J_{c\perp}=3\cdot 10^{7}$ A/m$^{2}$ and $J_{c\parallel}=9\cdot 10^{7}$ A/m$^{2}$ with $n$ value 30. The sinusoidal applied magnetic field is the same as the thin film in the previous section, and hence the amplitude is $50,70.7,100,287.9$ mT for angles $0,45,60,80\degree$, respectively, with frequency 50 Hz. The minimization uses the anisotropic $\vE(\vJ)$ relation and the magnetic field from the previous time step $\vB_{(t-\Delta t)}$.

The first magnetization case is with the angle $\theta=0\degree$. The average current density over the thickness is on figure \ref{f.ani_prism0J} (b). The screening current density fully penetrates into the sample at the peak of applied field of 50 mT. The thin film model [figure \ref{f.ani_prism0J}(a)] agrees with the prism. The slight difference in the penetration depth comes from the relatively coarser mesh in he prism in the $z$-plane. The thin film mesh contains $65\times 65\times 1$ (4225) cells, which is 4096 degrees of freedom. However, the prism requires a is fully 3D model and the mesh contains $31\times 31\times 15$ (14415) cells, resulting in around 40000 degrees of freedom. The prism contains 10 times more degrees of freedom, but each $z$ plane contains 4.5 times less cells. A more precise comparison is by cross-sectional current density [figure \ref{f.ani_prism0J}(c),(d)]. The cross-section lines are in the middle of the sample and two time steps with instant applied field 19.1 and 50 mT. The current profiles agree with thin film with small deviation. The prism is fully saturated with the screening current density, and hence there is no $J_z$ component. The applied field is perpendicular to the sample surface, therefore magnitude of the current density is around $J_{c\perp}$. 

\begin{figure}[tbp]
\centering
 \subfloat[][]
{\includegraphics[trim=60 0 60 0,clip,height=5.5 cm]{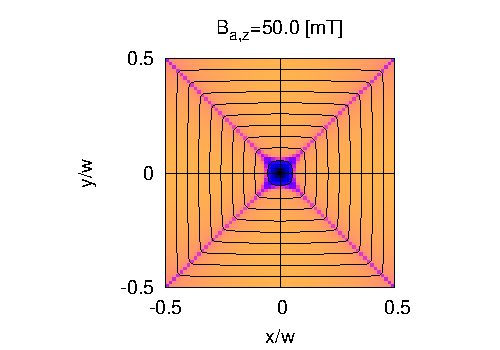}} 
 \subfloat[][]
{\includegraphics[trim=60 0 40 0,clip,height=5.5 cm]{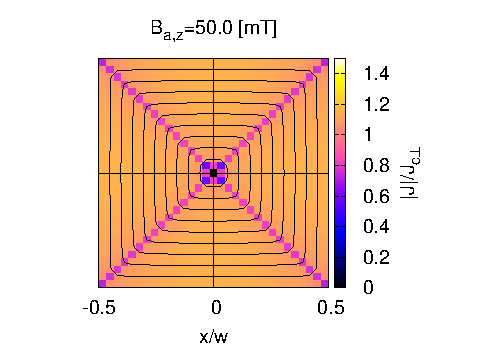}}\\
 \subfloat[][]
{\includegraphics[trim=40 0 50 0,clip,height=4.5 cm]{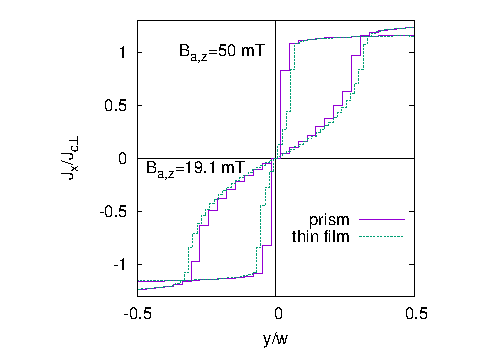}}  
 \subfloat[][]
{\includegraphics[trim=40 0 0 0,clip,height=4.5 cm]{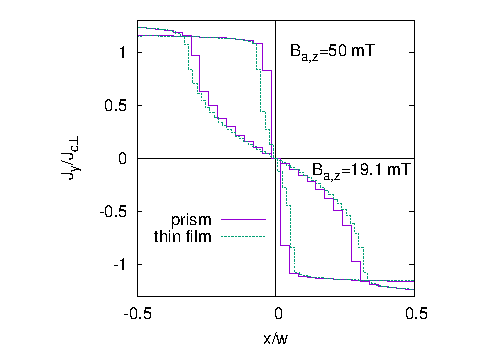}} 
\caption{The penetration of the current density at the peak of applied field of 50 mT in (a) thin film and (b) prism. The prism colour map is with average current density over thickness. The current density in the mid cross-section is shown with (c) $J_x$ and (d) $J_y$ components.}
\label{f.ani_prism0J}
\end{figure}

The second modelling case is with angle 45$\degree$ and applied field amplitude of 70.7 mT. The average current density at the peak of applied field [figure \ref{f.ani_prism45J}(b)] agrees very well with the thin film case [figure \ref{f.ani_prism45J}(a)]. The cross-sectional comparison confirms the same penetration depth. The $J_y$ component of the screening current is always perpendicular to the applied field, and hence the current density value is around $J_{c\perp}$[figure \ref{f.ani_prism45J}(d)]. However, the $J_x$ component is slightly parallel to the applied field, and hence force-free effects start to play a role. The alignment of $\vJ$ and $\vB$ in the $x$ direction increases the current density of $J_x$ to a value around 2$J_{c\perp}$ [figure \ref{f.ani_prism45J}(c)]. The cause is that for $\vJ||\vB$, $|\vJ|$ could reach up to $J_{c\parallel}$, being 3 times higher than $J_{c\perp}$ in our case. The thin film model shows $J_x$ at the penetration front with value around 2.8$J_{c\perp}$. The prism model does not shows the same sharp peak, because the prism mesh contains thicker cells, which smear out the results.

\begin{figure}[tbp]
\centering
 \subfloat[][]
{\includegraphics[trim=60 0 60 0,clip,height=5.5 cm]{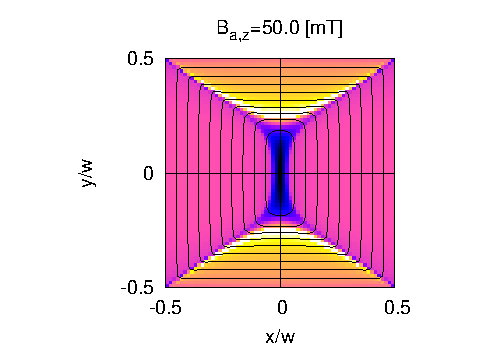}} 
 \subfloat[][]
{\includegraphics[trim=60 0 40 0,clip,height=5.5 cm]{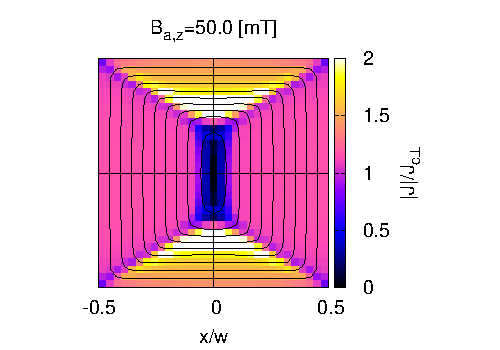}}\\
 \subfloat[][]
{\includegraphics[trim=40 0 50 0,clip,height=4.5 cm]{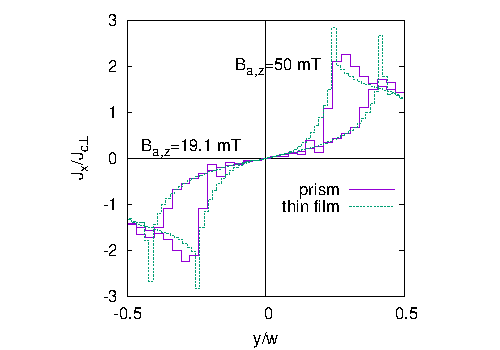}} 
 \subfloat[][]
{\includegraphics[trim=40 0 0 0,clip,height=4.5 cm]{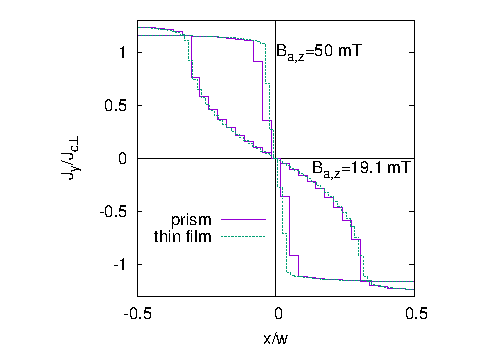}}  
\caption{The penetration of the thickness-average current density at the peak of applied field of 70.7 mT with angle $\theta=45\degree$ in (a) thin film and (b) prism. The current density in the mid cross-section is shown for the (c) $J_x$ and (d) $J_y$ components.}
\label{f.ani_prism45J}
\end{figure}

The third case is with $\theta=60\degree$ and applied field amplitude of 100 mT. The colour maps of current density show small deviation between the thin film [figure \ref{f.ani_prism60J}(a)] and the prism [figure \ref{f.ani_prism60J}(b)]. The $J_x$ component at the cross-sections [figure \ref{f.ani_prism60J}(c)] agrees very well with thin film. The $J_x$ value increases with angle $\theta$ and and reaches 2.5$J_{c\perp}$. The tilted applied field increases the $J_z$ component of the screening current density to a value around $J_{c\perp}$ and reduces $J_y$. The lower penetration depth of $J_y$ at the peak of applied field confirms the cross-sectional current density [figure \ref{f.ani_prism60J}(d)]. 

\begin{figure}[tbp]
\centering
 \subfloat[][]
{\includegraphics[trim=60 0 60 0,clip,height=5.5 cm]{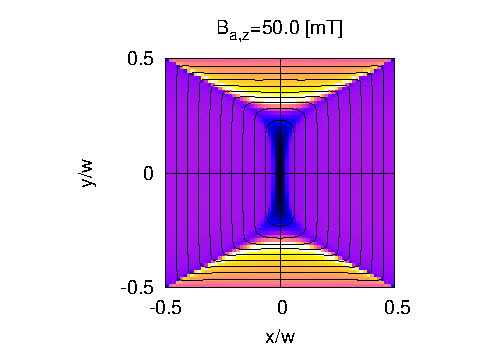}} 
 \subfloat[][]
{\includegraphics[trim=60 0 40 0,clip,height=5.5 cm]{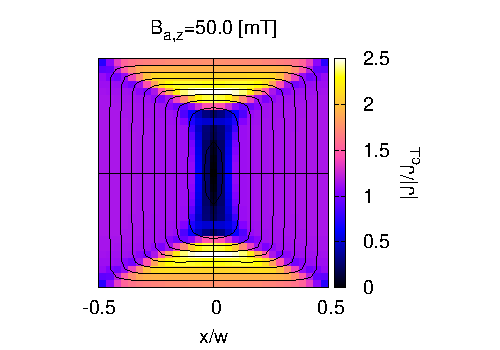}}\\ 
 \subfloat[][]
{\includegraphics[trim=40 0 50 0,clip,height=4.5 cm]{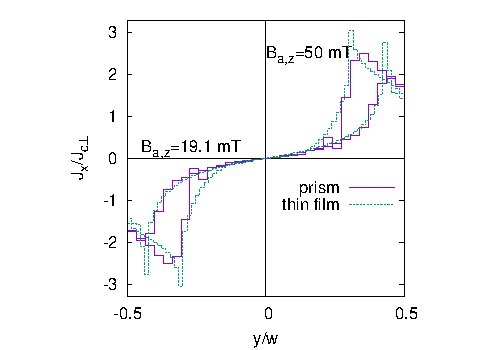}}
 \subfloat[][]
{\includegraphics[trim=40 0 0 0,clip,height=4.5 cm]{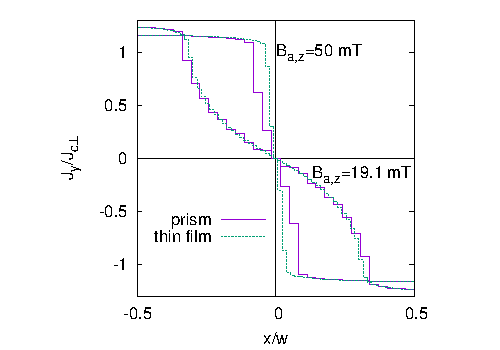}}
\caption{The penetration of the thickness-average current density at the peak of applied field of 100 mT with angle $\theta=60\degree$ in (a) thin film and (b) prism. The current density in the mid cross-section is shown for the (c) $J_x$ and (d) $J_y$ components.}
\label{f.ani_prism60J}
\end{figure}

The last calculation, for angle $\theta=80\degree$ and applied field of 287.9 mT, shows the same magnetization behaviour like the case of 60$\degree$. The model presents even lower penetration depth in $J_y$ [figure \ref{f.ani_prism80J}(b),(d)], because of tilted applied field angle. The $J_x$ component shows slightly lower penetration depth from the same reason [figure \ref{f.ani_prism80J}(c)]. 

\begin{figure}[tbp]
\centering
 \subfloat[][]
{\includegraphics[trim=60 0 60 0,clip,height=5.5 cm]{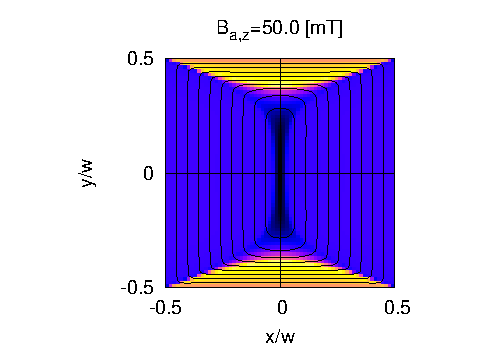}} 
 \subfloat[][]
{\includegraphics[trim=60 0 40 0,clip,height=5.5 cm]{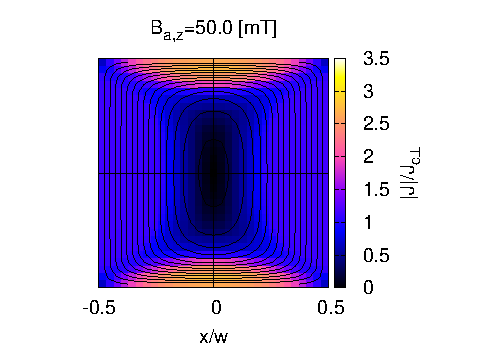}}\\
 \subfloat[][]
{\includegraphics[trim=40 0 50 0,clip,height=4.5 cm]{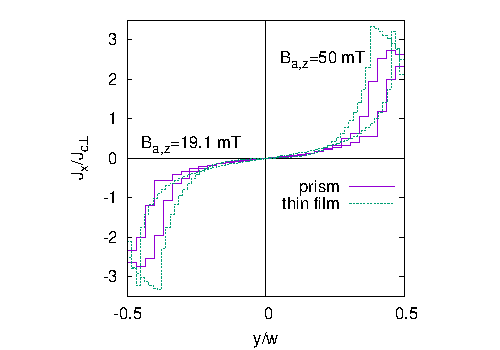}}
 \subfloat[][]
{\includegraphics[trim=40 0 0 0,clip,height=4.5 cm]{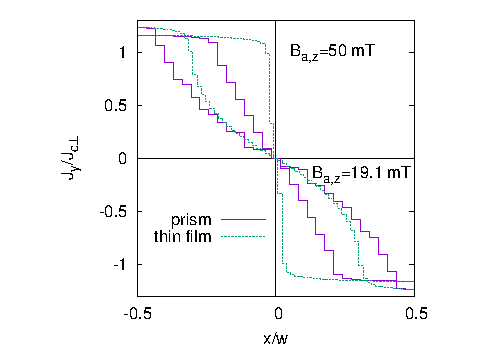}} 
\caption{The penetration of the thickness-average current density at the peak of applied field of 287.9 mT with angle $\theta=80\degree$ in (a) thin film and (b) prism. The current density in the mid cross-section is shown for the (c) $J_x$ and (d) $J_y$ components.}
\label{f.ani_prism80J}
\end{figure}

The 3D current distribution in the prism with perpendicular applied field is on figure \ref{f.ani_prism_3D0}. The sample is fully saturated with screening current [see figure \ref{f.ani_prism_3D0}(a),(b) central maps], and hence $J_z$ is zero [figure \ref{f.ani_prism_3D0}(c)]. The applied and local magnetic fields are perpendicular to the screening current flowing only in the $z$ plane, and hence there is no $|\vJ|\approx J_{c\parallel}$. 

The 3D current distribution with angle $\theta=80\degree$ is on figure \ref{f.ani_prism_3D80}. The penetration depth of $J_x$ is around 1 mm and the current density is around $J_{c\parallel}$ [figure \ref{f.ani_prism_3D80}(a)], because the current is almost parallel with the applied field. The highest penetration depth is for $J_y$. The $J_y$ value is around the $J_{c\perp}$, since the applied field is perpendicular with $\vJ$ and the border between positive and negative sign is aligned with the applied field direction [figure \ref{f.ani_prism_3D80}(b)]. The $J_z$ component is around the $J_{c\perp}$ with the same penetration depth as $J_x$[figure \ref{f.ani_prism_3D80}(c)]. 

The remanent state is on figure \ref{f.ani_prism_3D80_19}. The penetration depth of the $J_x$ component is lower than it was at the peak of applied field. $J_x$ is with opposite sign, because of the opposite sign of the ramp of the applied field. The value of $J_x$ is around $J_{c\parallel}$, since the applied field is almost zero and local magnetic field is parallel with $J_x$ [figure \ref{f.ani_prism_3D80_19}(a)]. The increase of $J_x$ close to remanent causes the peak in the hysteresis loop after the remanent state. Comparing to figure \ref{f.ani_prism_3D80}, $J_y$ swapped the sign with amplitude around $J_{c\perp}$ [figure \ref{f.ani_prism_3D80_19}(b)]. The lower penetration depth close to the corners is because of not complete reversal of the screening current, since the applied field is lower than the minus peak. $J_z$ also swaps the sign and keeps the value around $J_{c\perp}$.

\begin{figure}[tbp]
\centering
 \subfloat[][]
{\includegraphics[trim=0 0 0 0,clip,height=4.5 cm]{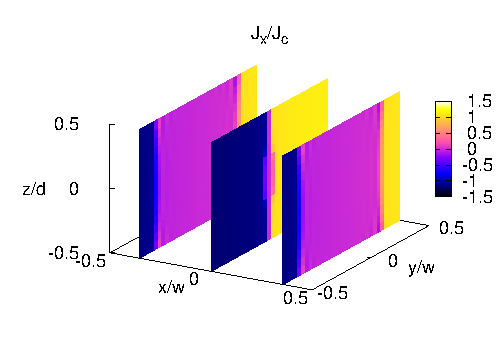}} 
 \subfloat[][]
{\includegraphics[trim=0 0 0 0,clip,height=4.5 cm]{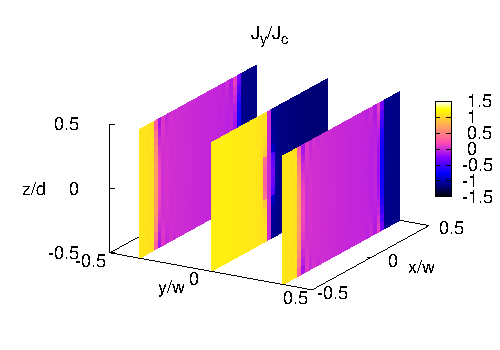}}\\ 
 \subfloat[][]
{\includegraphics[trim=0 0 0 0,clip,height=4.5 cm]{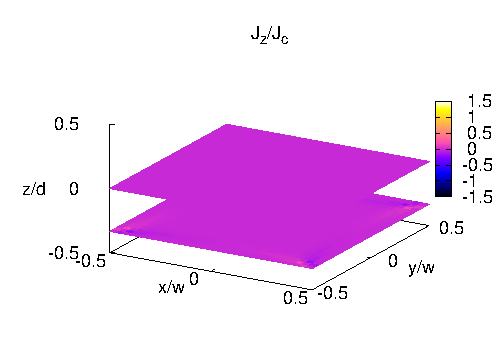}} 
\caption{The 3D current density in the prism with applied field 50 mT and angle $\theta=0\degree$. The maps are for screening current density components (a) $J_x$, (b) $J_y$ and (c) $J_z$.}
\label{f.ani_prism_3D0}
\end{figure}

\begin{figure}[tbp]
\centering
 \subfloat[][]
{\includegraphics[trim=0 0 0 0,clip,height=4.5 cm]{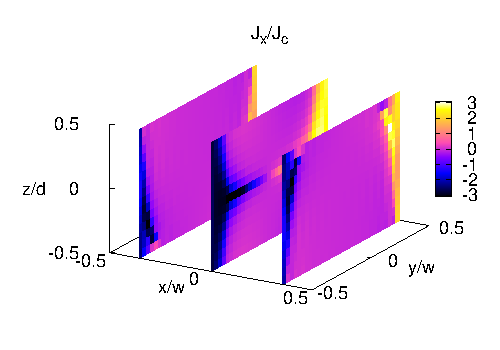}} 
 \subfloat[][]
{\includegraphics[trim=0 0 0 0,clip,height=4.5 cm]{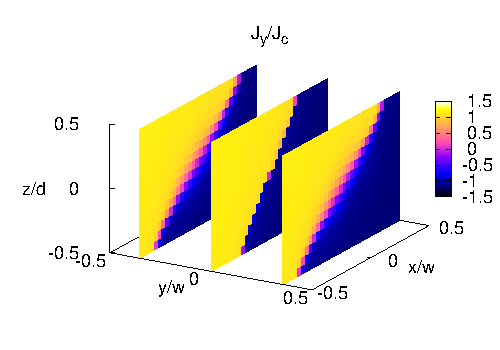}}\\ 
 \subfloat[][]
{\includegraphics[trim=0 0 0 0,clip,height=4.5 cm]{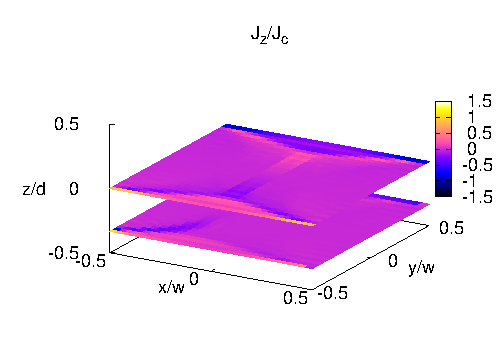}} 
\caption{The 3D current density in the prism with the applied field 287.9 mT and angle $\theta=80\degree$. The maps are for screening current density components (a) $J_x$, (b) $J_y$ and (c) $J_z$.}
\label{f.ani_prism_3D80}
\end{figure}

\begin{figure}[tbp]
\centering
 \subfloat[][]
{\includegraphics[trim=0 0 0 0,clip,height=4.5 cm]{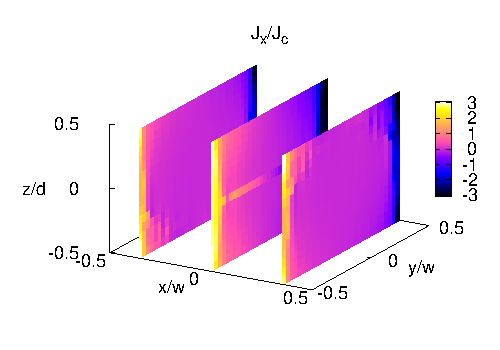}} 
 \subfloat[][]
{\includegraphics[trim=0 0 0 0,clip,height=4.5 cm]{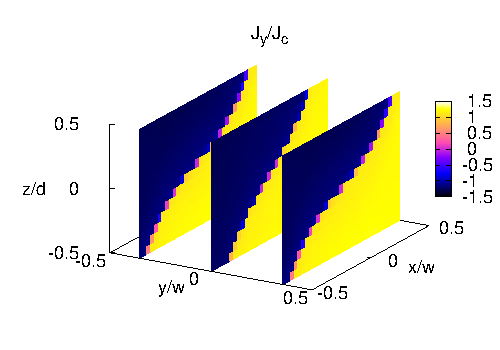}}\\ 
 \subfloat[][]
{\includegraphics[trim=0 0 0 0,clip,height=4.5 cm]{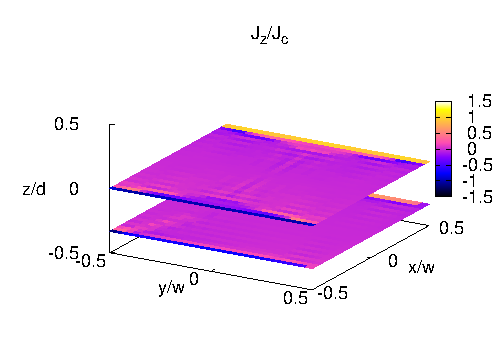}} 
\caption{The 3D current density in the prism with angle $\theta=80\degree$ at the remanent state. The maps are for the screening current density components (a) $J_x$, (b) $J_y$ and (c) $J_z$.}
\label{f.ani_prism_3D80_19}
\end{figure}

%%%%%%%%%%%%%%%%%%%%%%%%%%%%%%%%%%%%%%%%%%%%%%%%%%%%%%%%%%%%%%%%%%%%%%%%%%%%%%%%%%%%%%%%%%%%%%%

\subsubsection{Comparison between anisotropic and isotropic effects}

The comparison of the magnetization loops between isotropic and anisotropic $\vE(\vJ)$ relations reveals all finite size and force-free effects. 

The first loops are for isotropic power law [figure \ref{f.ani_prism_iloop}]. The magnetization $M_z$ decreases with the applied field angle [figure \ref{f.ani_prism_iloop}(b)] and $M_x$ increases with the angle [figure \ref{f.ani_prism_iloop}(a)]. The reason is that the magnetization loops are roughly normal to the applied field direction. Then, the angle $\theta$ tilts the magnetization loops, tilting also the magnetization. The saturation magnetization according the CSM for square slab is $M_s=J_cw/6$ \cite{Navau13IES}. For the prism, the saturation magnetization is $M_s\approx 0.17J_cw$. The model with power law exponent $n=$30 allows the current density value above $J_c$, and hence the saturation magnetization is slightly higher than the analytical prediction $M_s\approx 0.2J_cw$.

\begin{figure}[tbp]
\centering
 \subfloat[][]
{\includegraphics[trim=60 0 70 0,clip,width=6.5 cm]{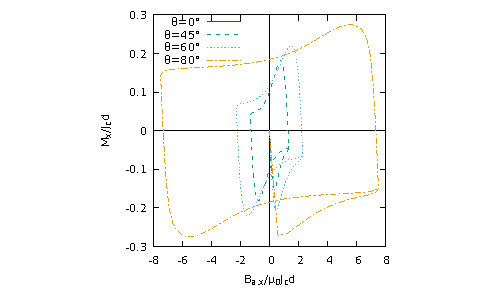}}
 \subfloat[][]
{\includegraphics[trim=60 0 70 0,clip,width=6.5 cm]{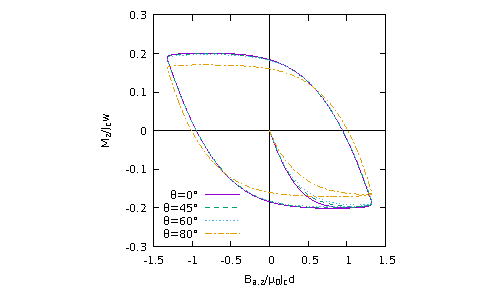}}
\caption{The magnetization loops with isotropic power law and various angles $\theta$ of the applied field. The magnetization components (a) $M_x$ and (b) $M_z$ are shown.}
\label{f.ani_prism_iloop}
\end{figure}

For the anisotropic case, $M_z$ increases with the applied field angle [figure \ref{f.ani_prism_aloop}(b)]. The magnetic field aligns with the current density in part of the sample. This increases $|\vJ|$ towards $J_{c\parallel}$, rising the magnetization. $M_x$ shows almost the same behaviour as for the isotropic case, increasing with applied field angle [figure \ref{f.ani_prism_aloop}(a)]. The anisotropic case presents a small peak for both magnetization components $M_x$ and $M_z$. The self-field is dominant at the remanent state and it is parallel to the top and bottom surface of the sample. The alignment of $\vJ$ and local $\vB$ increases $|\vJ|$ to $J_{c\parallel}$, and hence there appears a peak in the magnetization loop. The following increase of the applied field magnitude changes the local magnetic field, and hence reduces $|\vJ|$ to $J_{c\perp}$ and $M_x$, $M_z$ components [figure \ref{f.ani_prism_aloop}(a),(b)].  

\begin{figure}[tbp]
\centering
 \subfloat[][]
{\includegraphics[trim=60 0 70 0,clip,width=6.5 cm]{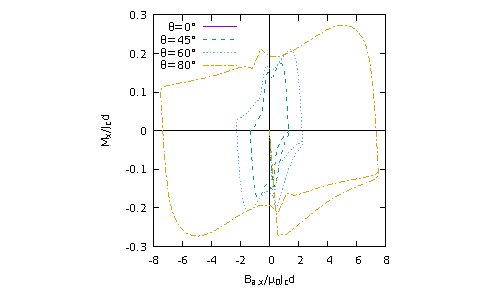}}
 \subfloat[][]
{\includegraphics[trim=60 0 70 0,clip,width=6.5 cm]{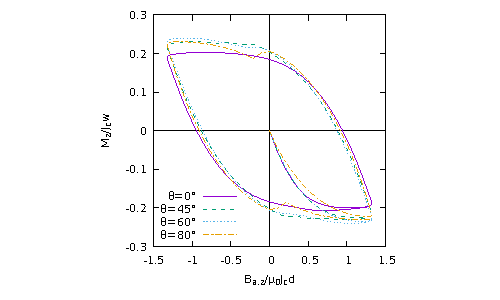}}
\caption{The magnetization loops with anisotropic power law and various angles $\theta$ of the applied field. The magnetization components (a) $M_x$ and (b) $M_z$ are shown.}
\label{f.ani_prism_aloop}
\end{figure}

%%%%%%%%%%%%%%%%%%%%%%%%%%%%%%%%%%%%%%%%%%%%%%%%%%%%%%%%%%%%%%%%%%%%%%%%%%%%%%%%%%%%%%%%%%%%%%%

\subsubsection{Detailed study of thickness effects}

The other comparison of magnetization loops is with the sample of various thickness and applied field angle $\theta=80\degree$. The prism with 1 $\mu$m thickness [the curve on figure \ref{f.ani_prism_tloop80}(b)] is modelled by the thin film approximation. $M_x$ increases with thickness [figure \ref{f.ani_prism_tloop80}(a)], because thicker sample contains higher $J_z$. The $M_z/J_cd$ component saturates to higher values for very low thickness of the sample. The normalized magnetization actually decreases with the thickness because of the tilt in the magnetization loops, raising $M_x$ but decreasing $M_z$ [figure \ref{f.ani_prism_tloop80}(b)]. The magnetization difference is only a finite size effect. This confirms the correctness of the MEMEP 3D model. All the samples with different thickness in the range 1$\mu$m-1mm show the same magnetization loop of $M_z$[figure \ref{f.ani_prism_tloop0}]. With small differences due to the finite thickness. The cases of 1 $\mu$m (thin film approach) and 0.1 mm (several cells over the thickness) present the same $M_z$ curve.

\begin{figure}[tbp]
\centering
 \subfloat[][]
{\includegraphics[trim=60 0 70 0,clip,width=6.5 cm]{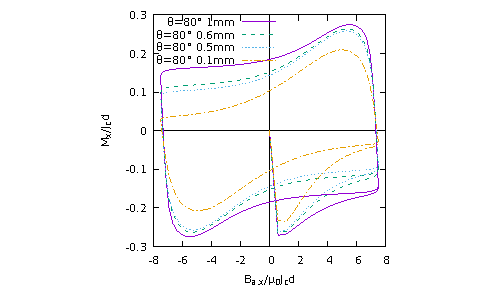}}
 \subfloat[][]
{\includegraphics[trim=60 0 70 0,clip,width=6.5 cm]{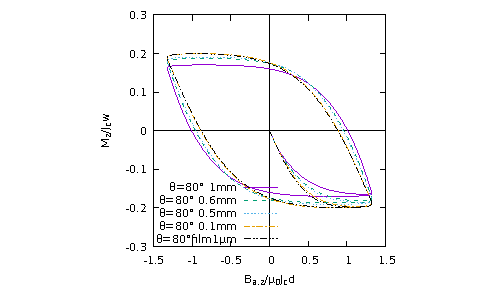}}
\caption{The magnetization loops with isotropic power law and various thicknesses for an applied field angle $\theta=80\degree$. The magnetization components (a) $M_x$ and (b) $M_z$ are shown.}
\label{f.ani_prism_tloop80}
\end{figure}

\begin{figure}[tbp]
\centering
{\includegraphics[trim=60 0 70 0,clip,width=6.5 cm]{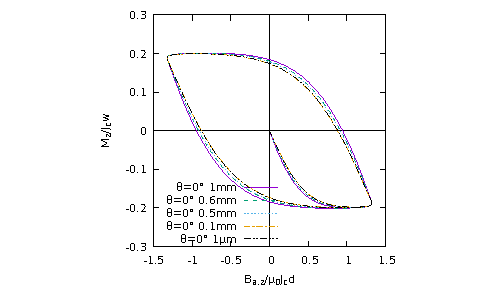}}
\caption{The magnetization loops with isotropic power law and various thicknesses for perpendicular applied field ($\theta=0\degree$).}
\label{f.ani_prism_tloop0}
\end{figure}

%%%%%%%%%%%%%%%%%%%%%%%%%%%%%%%%%%%%%%%%%%%%%%%%%%%%%%%%%%%%%%%%%%%%%%%%%%%%%%%%%%%%%%%%%%%%%%%
%%%%%%%%%%%%%%%%%%%%%%%%%%%%%%%%%%%%%%%%%%%%%%%%%%%%%%%%%%%%%%%%%%%%%%%%%%%%%%%%%%%%%%%%%%%%%%%
%%%%%%%%%%%%%%%%%%%%%%%%%%%%%%%%%%%%%%%%%%%%%%%%%%%%%%%%%%%%%%%%%%%%%%%%%%%%%%%%%%%%%%%%%%%%%%%
%%%%%%%%%%%%%%%%%%%%%%%%%%%%%%%%%%%%%%%%%%%%%%%%%%%%%%%%%%%%%%%%%%%%%%%%%%%%%%%%%%%%%%%%%%%%%%%

\chapter{Conclusion}

The full 3D calculation methods for non-linear superconducting materials is necessary, in order to explain all finite size effects. The finite size effects help to understand and optimize superconductors in power applications, since power devices are of finite size. The finite size effects are also important to interpret characterization measurements. Modelling methods needs to include any dependence like $J_c(B)$ and any $\vE(\vJ)$ relation in order to provide realistic predictions, which is very demanding. Moreover, it must handle full 3D mesh with a lot of elements to fulfil accurate results in relatively short times.

This thesis developed and implemented a novel 3D modelling method for the electromagnetic response of superconductors. We developed a new variational method in $\vT$ formulation, where $\vT$ is the effective magnetization. The variational method is called Minimum Electro Magnetic Entropy Production MEMEP 3D. The entire calculation method is implemented in a modelling tool written in C++ language with parallel computing hierarchy. The parallel computing efficiency is tested on a computer cluster with efficiency of 80\%. Parallel computing enabled modelling cases with high resolution of elements in the mesh, higher number of cycles and total time steps per cycle. The numerical method included sectors, in order to drastically reduce the computing time, even without parallel programming, and hence the electromagnetic response of the superconducting sample is modelled in a relatively short time. The results of the models confirmed the correctness of the 3D functional, which is the core of the entire MEMEP 3D method, the solution being unique without saddle points.

The electromagnetic response of the superconducting sample was verified by analytical solutions by 2D cross-sectional models for infinitely long thin films and thin disks. We also modelled a square film with constant $J_c$ and $J_c(B)$. The predictions were for the current density profiles, hysteresis loops and qualitative gradual penetration of the current density under perpendicular applied magnetic field.

The modelling geometry is focused on thin films and cubic samples, which are of high importance. Nowadays superconducting commercial tapes are around 1 $\mu$m thick and stacks of tapes are an alternative to permanent magnets. The comparison between the AC loss measurement on two soldered superconducting tapes and model showed high accuracy of 96\%. The coupling loss is the dominant part at low applied fields below 10 mT. The current state of the modelling tool can accurately model AC loss in tapes of up to 10 filaments. Elongated cells allowed to prolong the sample length far beyond the length of the usual sample for characterization measurements of the tapes. 

The general magnetization model of the cubic bulk found the existence of the $J_z$ component, which does not appear in superconducting cylinders or fully saturated square samples. The fundamental study improved the general knowledge of magnetization on rectangular prism samples, which is still not fully understood. The HTS modelling work group chose our magnetization model of cubic bulk as a benchmark model for other modelling method on the international HTS modelling workshop 2016. Further study about prism magnetization of different thickness proofed a remaining $J_z$ component of $0.3J_c$ even for aspect ratios as low as 0.1. Study presented analytical fit of magnetization on aspect ratio dependence between infinitely thin film and slab of 97\% accuracy. 

International collaboration resulted in a further comparison of three modelling methods. The modelling cases were bulk and stack magnetization with tilted applied fields. The comparison further validated the MEMEP 3D method. The calculation of AC loss, hysteresis loops and 3D current path have been performed by MEMEP 3D, FEM and VIEM with great accuracy of results. 

Demagnetization by cross-fields of cubic bulks are important for power applications as potential alternative to permanent magnets. The model showed the asymmetry of the trapped magnetic field during demagnetization process, which confirmed the measurements on my stay in Cambridge University. The comparison agreed very well for low fields but showed a small deviation at high cross-fields of 150 mT. The origin of the asymmetry in the trapped field can be explained by the 3D current penetration, which was partially predicted by 2D cross-sectional models. The full 3D current path was confirmed with a FEM model. The MEMEP 3D method modelled the entire magnetization and demagnetization process with more than 500 time steps.        

The modelling method can take any $\vE(\vJ)$ relation, and hence the last study in this Thesis was on force-free effects in thin films and bulks with anisotropic power law. The force-free effects are important for power applications where superconducting tapes are expose to rotating magnetic fields or in magnets with 3D transpose cables, and also for sample characterisation. The force-free effects appear when current density and local magnetic fields are not perpendicular as it is explained by the Elliptic Double Critical State Model. The force-free effects study on the thin film with tilted applied field revealed zones of value $|\vJ|\approx J_{c\parallel}$ and reduction at remanent state back to $|\vJ|\approx J_{c\perp}$. The reduction came from the self-field, which was always perpendicular to the film samples without applied field, and hence no ``enhanced" $J_c$ is present. The magnetization of the $M_z$ component increased with the applied field angle, because of the $J_{c\parallel}$ influenced by the alignment of $\vJ$ and $\vB$. The anisotropic bulk presented even more complex force-free effects. The enhanced current density created a peak after the remanent state in the hysteresis loop. The thickness study on the isotropic prism showed an increase of $M_z$ with decreasing the thickness, since the screening current is squeezed towards the $xy$ plane, and hence there is a reduction of the $M_x$ component.

In conclusion, the variational method was verified by analytical predictions and measurements. The MEMEP 3D method showed very good results of new findings with high speed calculation time. However, there is still room for improvement of the calculation time. The method can includes any $\vE(\vJ)$ relation, and hence the method is promising for another fundamental studies of the 3D superconductors in not fully understood electromagnetic configurations.  

%%%%%%%%%%%%%%%%%%%%%%%%%%%%%%%%%%%%%%%%%%%%%%%%%%%%%%%%%%%%%%%%%%%%%%%%%%%%%%%%%%%%%%%%%%%%%%%
%%%%%%%%%%%%%%%%%%%%%%%%%%%%%%%%%%%%%%%%%%%%%%%%%%%%%%%%%%%%%%%%%%%%%%%%%%%%%%%%%%%%%%%%%%%%%%%
%%%%%%%%%%%%%%%%%%%%%%%%%%%%%%%%%%%%%%%%%%%%%%%%%%%%%%%%%%%%%%%%%%%%%%%%%%%%%%%%%%%%%%%%%%%%%%%
%%%%%%%%%%%%%%%%%%%%%%%%%%%%%%%%%%%%%%%%%%%%%%%%%%%%%%%%%%%%%%%%%%%%%%%%%%%%%%%%%%%%%%%%%%%%%%%

\chapter{Appendix}

\section{Average vector potential of the self-point interaction.}
\label{s.appendix_A}

The minimization process requires the evaluation of the average vector potential \ref{e.asij}. The self-interaction average vector potential contains a double volume integral, whose solution has to be found. This appendix is focused on the analytical solution of the double volume integral and its correctness.   

The elements of the vector-potential interaction matrix between surfaces $i$ and $j$ of the $s$ type ($s\in{x,y,z}$) is
\begin{eqnarray}
a_{sij}=\frac{\mu_{0}}{4\pi V_{si}V_{sj}}\int_{V}d^3r\int_{V}d^3r'\frac{h_{si}(r)h_{sj}(r')}{|\vr-\vr'|},    
\label{}
\end{eqnarray} 
where $\vr, \vr'$ are vector positions of the surfaces. The approximation of the analytical formula for the case of not overlapping surfaces $i\ne j$ are equations (\ref{e.ap1}) and (\ref{e.ap2}). The undefined double volume integral related to equations (\ref{e.as1}) and (\ref{e.as2}) is 
\begin{equation}
f=\int_{V}\dif^3r\int_{V}\dif^3r'\frac{1}{|\vr-\vr'|}.
\label{}
\end{equation}
The full length analytical formula is
\begin{eqnarray}
f= - \frac{\left(x^2+y^2\right)\sqrt{x^2+y^2+z^2}^3}{216} + \frac{x^2yz^2}{4}\ln\left(y+\sqrt{x^2+y^2+z^2}\right)	\nonumber\\
	 + \frac{xy^2z^2}{4}\ln\left(x+\sqrt{x^2+y^2+z^2}\right) + \frac{x^2y^2z}{4}\ln\left(z+\sqrt{x^2+y^2+z^2}\right)  \nonumber\\
	 + \frac{\left(5x^4+5y^4+4x^2y^2-x^2z^2-y^2z^2\right)\sqrt{x^2+y^2+z^2}}{72} - \frac{x^2y^2z}{48} - \frac{x^2yz^2}{48} \nonumber\\
	 + \frac{x^4y}{12}\ln\left(x^2+z^2\right) + \frac{yz^4}{12}\ln\left(x^2+z^2\right) + \frac{x^4z}{12}\ln\left(x^2+y^2\right) + \frac{y^{4}z}{12}\ln\left(x^2+y^2\right) \nonumber\\
	 + \frac{xy^4}{16}\ln\left(y^2 + z^2\right) + \frac{xz^4}{48}\ln\left(y^2 + z^2\right) + \frac{x^4y}{24}\ln\frac{y + \sqrt{x^2+y^2+z^2}}{-y + \sqrt{x^2+y^2+z^2}} \nonumber\\
	 + \frac{xy^4}{24}\ln\frac{x + \sqrt{x^2+y^2+z^2}}{-x + \sqrt{x^2+y^2+z^2}} + \frac{x^4z}{24}\ln\frac{z + \sqrt{x^2+y^2+z^2}}{-z + \sqrt{x^2+y^2+z^2}} \nonumber\\
	 + \frac{xyz^3}{6}\arctan\frac{y^2 + z^2 + y\sqrt{x^2+y^2+z^2}}{xz} + \frac{xy^3z}{6}\arctan\frac{y^2 + z^2 + z\sqrt{x^2+y^2+z^2}}{xy} \nonumber\\
	 - \frac{xyz^3}{3}\arctan\frac{xy}{z\sqrt{x^2+y^2+z^2}} - \frac{xy^3z}{3}\arctan\frac{xz}{y\sqrt{x^2+y^2+z^2}} \nonumber\\
	 - \frac{x^3yz}{6}\arctan\frac{yz}{x\sqrt{x^2+y^2+z^2}} \nonumber \\
	 - \frac{xy^4}{16}\ln\left[\left(x^2 + y^2 + x\sqrt{x^2+y^2+z^2}\right)^2+y^2z^2\right] \nonumber\\
	 - \frac{x^4y}{16}\ln\left[\left(x^2 + y^2 + y\sqrt{x^2+y^2+z^2}\right)^2+x^2z^2\right] \nonumber\\
	 - \frac{x^4z}{16}\ln\left[\left(x^2 + z^2 + z\sqrt{x^2+y^2+z^2}\right)^2+x^2y^2\right] \nonumber\\
   - \frac{xz^4}{48}\ln\left[\left(x^2 + z^2 + x\sqrt{x^2+y^2+z^2}\right)^2+y^2z^2\right] \nonumber\\
	 - \frac{yz^4}{48}\ln\left[\left(y^2 + z^2 + y\sqrt{x^2+y^2+z^2}\right)^2+x^2z^2\right] \nonumber\\
	 - \frac{y^4z}{48}\ln\left[\left(y^2 + z^2 + z\sqrt{x^2+y^2+z^2}\right)^2+x^2y^2\right] \nonumber\\
	 - \frac{\sqrt{x^2+y^2+z^2}^5}{45} + \frac{7\sqrt{x^2+y^2+z^2}^3\left(-2x^2-2y^2+3z^2\right)}{540},
\label{long_form}
\end{eqnarray} 
where $(x,y,z)$ are defined by the vector positions $r$ and $r'$ as $x=r-r'_x$, $y=r-r'_y$ and $z=r-r'_z$. We checked that the analytical solution follows mirror symmetry [$f(x,y,z)=f(-x,y,z)=f(x,-y,z)=f(x,y,-z)$] and permutation [$f(x,y,z)=f(y,x,z)=f(x,z,y)=f(z,y,x)$] even if this is not evident in the expression. The solution was checked several times by comparison with other formulas. However, there exist simplified analytical solutions for a cube element \cite{ciftja11PLA} and a square surface \cite{ciftja10PLA}. The cube formula is 
\begin{eqnarray}
f=\frac{1}{l_{si}}\frac{1+\sqrt{2}-2\sqrt{3}}{5}-\frac{\pi}{3}+\ln\left[ \left( 1+\sqrt{2} \right)\left( 2+\sqrt{3} \right) \right],
\label{}
\end{eqnarray}
where $l_{si}$ is the side of the cube. The thin square surface formula is
\begin{eqnarray} 
f=\frac{1}{l_{si}}\frac{1-\sqrt{2}}{3}+\ln\left( 1+\sqrt{2} \right),
\label{}
\end{eqnarray}
where $l_{si}$ is the side of the surface bigger than its thickness.

%%%%%%%%%%%%%%%%%%%%%%%%%%%%%%%%%%%%%%%%%%%%%%%%%%%%%%%%%%%%%%%%%%%%%%%%%%%%%%%%%%%%%%%%%%%%%%%

\section{Euler equations of the functional}
\label{s.appendix_B}

The variational method is based on a certain functional, whose solution is found by minimization. The minimum of the functional should be the same as the solution of Maxwell differential equations.  
One needs to proof that the Euler equations of the functional of the extreme correspond to the Maxwell equations and that the extreme is a unique minimum. 

A general formalism for this does not exist for functionals of double volume integrals. Since the functional contains double volume integral the Euler an equation needs to be found. The following appendix presents this formalism of the general functional with the double volume integral and it explains the Euler-equations.  

The general form of a functional with double volume integral \cite{Pardo17JCP} is
\begin{equation}
L[\{u_i\}]=\int_\Omega\dvoln\int_\Omega\dvoln'\ f(\{r_\alpha\},\{r_\alpha'\},\{u_i\},\{u_i'\},\{u_i^{(\alpha)}\},\{{u_i'}^{(\alpha)}\}) ,
\end{equation}
where the integer $\alpha\in[1,n]$, $u_i'$ and $u_i$ are functions with variables $\{r_\alpha'\}$ and $\{r_\alpha\}$, $u_i'=u_i(\{r_\alpha'\})$,${u_i}^{(\alpha)}\equiv\partial_\alpha u_i\equiv\partial u_i/\partial r_\alpha$, ${u_i'}^{(\alpha)}\equiv\partial_\alpha'u_i'\equiv\partial u_i'/\partial r_\alpha'$. 

After Taylor expansion, (\cite{Pardo17JCP} A.3) the first variation of double volume integral becomes, using that $\partial_\alpha(u_i+\epsilon g_i)=u_i^{(\alpha)}+\epsilon g_i^{(\alpha)}$
\begin{eqnarray}
\delta L[\{u_i\}] & = & \epsilon\int_\Omega\dvoln\int_\Omega\dvoln' \Bigg( \frac{\dif }{\dif\epsilon} f(\{r_\alpha\},\{r_\alpha'\},\{u_i+\epsilon g_i\},\{u_i'+\epsilon g_i'\}, \nonumber \\
&&  \{u_i^{(\alpha)}+\epsilon g_i^{(\alpha)}\},\{{u_i'}^{(\alpha)}+\epsilon {g_i'}^{(\alpha)}\} ) \Bigg)_{\epsilon=0} ,
\label{e.dvi}
\end{eqnarray}
and with
\begin{equation}
\left ( \frac{\dif f}{\dif \epsilon} \right )_{\epsilon=0}=\fui g_i + \fuia g_i^{(\alpha)} + \fuip g_i' + \fuiap {g_i'}^{(\alpha)}.
\end{equation}
Integration by parts rewrite it to
\begin{eqnarray}
\delta L[\{u_i\}] & = & \epsilon\int_\Omega\dvoln\int_\Omega\dvoln' \left [ g_i \left ( \fui - \partial_\alpha\fuia \right ) + \right . \nonumber\\
&& \left . g_i'\left ( \fuip-\partial_\alpha'\fuiap \right ) \right].
\end{eqnarray}
If the functional density is symmetric
\begin{eqnarray}
&& f(\{r_\alpha\},\{r_\alpha'\},\{u_i\},\{u_i'\},\{u_i^{(\alpha)}\},\{{u_i'}^{(\alpha)}\}) 
= \nonumber \\ && 
f(\{r_\alpha'\},\{r_\alpha\},\{u_i'\},\{u_i\},\{{u_i'}^{(\alpha)}\},\{{u_i}^{(\alpha)}\}) , 
\label{Lsym'}
\end{eqnarray}
the first variation becomes
\begin{equation}
\delta L[\{u_i\}] = 2\epsilon\int_\Omega\dvoln g_i \int_\Omega\dvoln' \left [ \fui - \partial_\alpha\fuia \right ].
\end{equation}
The extreme of the functional appears when for any arbitrary $g_i$ the variation is $\delta L=0$ and
\begin{equation}
2\int_\Omega\dvoln' \left [ \fui - \partial\alpha\fuia \right ]=0.
\end{equation}
This is the Euler equation of the functional. A functional with both single and double volume integrals, is 
\begin{eqnarray}
&& L[\{u_i\}] = \int_\Omega\dvoln h(\{r_\alpha\},\{u_i\},\{u_i^{(\alpha)}\}) \nonumber\\
&& + \int_\Omega\dvoln\int_\Omega\dvoln'\ f(\{r_\alpha\},\{r_\alpha'\},\{u_i\},\{u_i'\},\{u_i^{(\alpha)}\},\{{u_i'}^{(\alpha)}\}),
\label{Lhf}
\end{eqnarray}
Following the same step as for the previous functional, the variation of the functional is 
\begin{eqnarray}
\delta L[\{u_i\}] & = & \epsilon\int_\Omega\dvoln g_i \Bigg[ \hui - \partial_\alpha\huia \nonumber\\
& + & \left . 2\int_\Omega\dvoln' \left ( \fui - \partial_\alpha\fuia \right ) \right ].
\label{dLboth}
\end{eqnarray}
Then, the Euler equations are 
\begin{eqnarray}
 \hui - \partial_\alpha\huia + \left [ 2\int_\Omega\dvoln' \left ( \fui - \partial_\alpha\fuia \right ) \right ] = 0.
\label{EulerdV2}
\end{eqnarray}
The second variation proves, if the functional extreme, where $\delta L=0$, is a minimum or not. The second variation of the double volume integral is
\begin{eqnarray}
\delta^2 L[\{u_i\}] & = & \half\epsilon^2\int_\Omega\dvoln\int_\Omega\dvoln' \Bigg( \frac{\dif^2 }{\dif\epsilon^2} f(\{r_\alpha\},\{r_\alpha'\},\{u_i+\epsilon g_i\},\{u_i'+\epsilon g_i'\}, \nonumber \\
&&  \{u_i^{(\alpha)}+\epsilon g_i^{(\alpha)}\},\{{u_i'}^{(\alpha)}+\epsilon {g_i'}^{(\alpha)}\} ) \Bigg)_{\epsilon=0} ,
\end{eqnarray}
where the second derivative is
\begin{eqnarray}
\left ( \frac{\dif^2 f}{\dif \epsilon^2} \right )_{\epsilon=0} & = & \fuij g_ig_j + f^{(u_i'u_j')}g_i'g_j' + 2\fuijb g_ig_i^{(\beta)} +
2f^{(u_i'{u_j'}^{(\beta)})} g_i'{g_j'}^{(\beta)} \nonumber\\
& + & \fuiajb g_i^{(\alpha)}g_j^{(\beta)} + f^{( {u_i'}^{(\alpha)}{u_j'}^{(\beta)} )} {g_i'}^{(\alpha)}{g_j'}^{(\beta)} \nonumber\\
& + & 2f^{( {u_i}{u_j'}^{(\beta)} )}g_i{g_j'}^{(\beta)} + 2f^{( {u_i'}{u_j}^{(\beta)} )}{g_i'}{g_j}^{(\beta)} \nonumber \\
& + & 2f^{( {u_i}{u_j'} )}g_i{g_j'} + 2f^{( {u_i}^{(\alpha)}{u_j'}^{(\beta)} )} {g_i}^{(\alpha)}{g_j'}^{(\beta)},
\end{eqnarray}
with notation $f^{(u_i{u_j'}^{(\beta)})}\equiv \partial^2f/(\partial u_i\partial {u_j'}^{(\beta)})$. If the functional is symmetric, the second variation is rewritten as
\begin{eqnarray}
\delta^2 L[\{u_i\}] & = & \half\epsilon^2\int_\Omega\dvoln\int_\Omega\dvoln' \left [ 2\fuij g_ig_j + 2\fuiajb g_i^{(\alpha)}g_j^{(\beta)} \right . \nonumber \\
& + & 2f^{( {u_i}{u_j'} )}g_i{g_j'} + 2f^{( {u_i}^{(\alpha)}{u_j'}^{(\beta)} )} {g_i}^{(\alpha)}{g_j'}^{(\beta)} \nonumber\\
& + & \left . 4\fuijb g_ig_i^{(\beta)} + 4f^{( {u_i}{u_j'}^{(\beta)} )}g_i{g_j'}^{(\beta)} \right ],
\label{d2LdV2}
\end{eqnarray}
If the functional contains both single and double volume integrals, $\delta^2 L$ becomes
\begin{eqnarray}
\delta^2 L[\{u_i\}] & = & \half\epsilon^2\int_\Omega\dvoln \nonumber \left [ h^{(u_iu_j)}g_ig_j + h^{(u_i^{(\alpha)}u_j^{(\beta)})}g_i^{(\alpha)}g_j^{(\beta)} + 2h^{(u_iu_j^{(\beta)})}g_ig_j^{(\beta)} \right ] \\
& + & \half\epsilon^2\int_\Omega\dvoln\int_\Omega\dvoln' \left [ 2\fuij g_ig_j + 2\fuiajb g_i^{(\alpha)}g_j^{(\beta)} \right . \nonumber \\
& + & 2f^{( {u_i}{u_j'} )}g_i{g_j'} + 2f^{( {u_i}^{(\alpha)}{u_j'}^{(\beta)} )} {g_i}^{(\alpha)}{g_j'}^{(\beta)} \nonumber\\
& + & \left . 4\fuijb g_ig_i^{(\beta)} + 4f^{( {u_i}{u_j'}^{(\beta)} )}g_i{g_j'}^{(\beta)} \right ].
\label{d2Lboth}
\end{eqnarray}

%%%%%%%%%%%%%%%%%%%%%%%%%%%%%%%%%%%%%%%%%%%%%%%%%%%%%%%%%%%%%%%%%%%%%%%%%%%%%%%%%%%%%%%%%%%%%%%

\chapter{Parameters of the input file}
\label{c.input_file}

The modelling tool of any method has a lot of input options, and hence user friendly interface is an essential feature. The MEMEP 3D modelling tool contains an input.txt file, which is loaded to the tool at the beginning of the calculation with all possible combinations of the input options. The example of input parameters with short explanation are:\\
	x[m]:	- width of the sample\\
xl[m]: -	width of the metallic part in the sample between two filaments\\
y[m]: - length of the sample\\
z[m]: - thickness of the sample\\
nsucx[-]: - number of the cells along the $x$ axes in the superconducting material\\
nncx[-]: - number of the cells along the $x$ axes in the metal material (tape with filaments)\\
ncy[-]: - number of the cells along the $y$ axes\\
ncz[-]: - number of the cells along the $z$ axes (thin film approximation ncz=1)\\
elc[-]: - 0 disable/1 enable, elongated cells in the long sample with aspect ratio $>$2\\
tol elc[-]:	- tolerance criterion for average vector potential of elongated cells (0.001-default)\\
Bamax[T]: - maximum amplitude of the applied magnetic field\\  
Bamax1[T]: - maximum amplitude of the applied magnetic cross-field\\
Bshape[-]: - waveform of the applied field : 0-sinusoidal, 1-ramp down follows by cross-field of Bamax1 and fi1, 2-constant ramp (triangular)\\
Btrape[-]: - 0 disable/1 enable, calculation of the magnetic field outside of the sample in a certain plane (B-plane)\\
rcx plane[m]: - x component of the center position of the B-plane (see Btrape)\\
rcy plane[m]: - y component of the center position of the B-plane (see Btrape)\\
rcz plane[m]: - z component of the center position of the B-plane (see Btrape)\\
x plane[m]: -	width of the B plane (see Btrape)\\
y plane[m]: -	length of the B plane (see Btrape)\\
z plane[m]: - thickness of the B plane (see Btrape)\\
ncx plane[-]:	-	number of the cells in the B plane along the x axes (see Btrape)\\
ncy plane[-]:	-	number of the cells in the B plane along the y axes (see Btrape)\\
ncz plane[-]:	-	number of the cells in the B plane along the z axes (see Btrape)\\
theta[degree]: - angle of the applied magnetic field from the $x$ axes to the $y$ axes\\
fi[degree]: - angle of the applied magnetic field from $z$ axes to the $x$ axes\\
fi1[degree]: - angle of the cross-applied magnetic field from $z$ axes to the $x$ axes for Bamax1 and f1\\
uni[-]: -	type of the mesh: 0-non-uniform, 1-uniform (default), 2-semi-uniform\\
rel[-]: - $\vE(\vJ)$ relation: 1-isotropic, 2-$J_c(B)$ Kim model, 3-anisotropic, 4-multi-valued CSM\\
sym[-]: -	type of minimization: 0 no speed up, 1-symmetry (odd input), 2-sectors, 3-sectors with symmetry\\
Ec[V/m]: -	critical electric field\\
Jo[A/m2]:	-	critical current density\\
Jol[A/m2]: - current density for metallic material\\
rhoR[ohm*m]: - effective resistivity of the metallic material between filaments\\
dl[m]: -	width of the metallic joint\\
Jcpa[A/m2]:	-	parallel critical current density\\
Jcpe[A/m2]:	-	perpendicular critical current density\\
Bo[T]: -	characteristic magnetic field for the Kim model\\
N[-]: -	power law exponent\\
Nl[-]: -	power law exponent for metallic material (1-default)\\
m[-]: -	Kim model exponent\\
f[Hz]: -	frequency of the applied field\\
f1[Hz]: -	frequency of the applied cross-field Bamax1\\
ns[-]: -	number of the time steps per cycle\\
step[-]: -	total number of the time steps\\
tolJ[-]: -	tolerance of the current density(1e-5 default)\\ 
shape[-]: -	geometry shape of the sample: 0-square/rectangular, 1-disk/ball, 2-cylinder, 3-tape with filaments, 4-stack of tapes\\
num threads[-]: -	number of the threads in the computer\\
nscx[-]: -	number of cells in one sector along the $x$ axes\\
nscy[-]: -	number of cells in one sector along the $y$ axes\\
nscz[-]: -	number of cells in one sector along the $z$ axes (thin film approximation nscz=1)\\
shift1[-]: -	shift by the cells between the first and second set of sectors along the $x$ and $y$ axes\\ 
shift2[-]: -	shift by the cells between the second and third set of sectors along the $x$ and $y$ axes\\
shiftz1[-]: -	shift by the cells between the first and second set of sectors along the $z$ axes\\
shiftz2[-]: -	shift by the cells between the second and third set of sectors along the $z$ axes\\

%%%%%%%%%%%%%%%%%%%%%%%%%%%%%%%%%%%%%%%%%%%%%%%%%%%%%%%%%%%%%%%%%%%%%%%%%%%%%%%%%%%%%%%%%%%%%%%

\chapter*{Publications}

\begin{itemize}
\item ${\bf{M. Kapolka}}$, E. Pardo``Three-dimensional modeling of macroscopic force-free effects in superconducting thin films and rectangular prisms", (preprint) arXiv:1803.06342
\end{itemize}

\begin{itemize}
\item ${\bf{M. Kapolka}}$, V. M. R. Zermeno, S. Zou, A. Morandi, P. L. Ribani, E. Pardo, F. Grilli ``Three-Dimensional Modeling of the Magnetization of Superconducting Rectangular-Based Bulks and Tape Stacks", IEEE Transactions on Applied Superconductivity (Volume:28, Issue:4), Article number: 8201206, 2018 \\${\bf{DOI}}$: 10.1109/TASC.2018.2801322
\end{itemize}

\begin{itemize}
\item ${\bf{M. Kapolka}}$, J. Srpcic, D. Zhou, M. Ainslie, E. Pardo, A. Dennis ``Demagnetization of cubic Gd-Ba-Cu-O bulk superconductor by cross-fields: measurements and 3D modelling", IEEE Transactions on Applied Superconductivity (Volume:28, Issue:4), Article number: 6801405, 2018 \\${\bf{DOI}}$: 10.1109/TASC.2018.2808401
\end{itemize}

\begin{itemize}
\item E. Pardo, ${\bf{M. Kapolka}}$, ``3D computation of non-linear eddy currents: Variational method and superconducting cubic bulk", "J. Comput. Phys." (Volume:344, Issue:), Article number:, 2017 \\${\bf{DOI}}$: 10.1016/j.jcp.2017.05.001
\end{itemize}

\begin{itemize}
\item E. Pardo, ${\bf{M. Kapolka}}$, ``{3D} magnetization currents, magnetization loop, and saturation field in superconducting rectangular prisms", "Supercond. Sci. Technol." (Volume:30, Issue:), Article number: 064007, 2017 \\${\bf{DOI}}$: 10.1088/1361-6668/aaa5db
\end{itemize}

\begin{itemize}
\item E. Pardo, ${\bf{M. Kapolka}}$, J. Kovac, J. Souc, F. Grilli, A. Pique ``3D modeling and measurement of coupling AC loss in soldered tapes and striated coated conductors", ${\bf{Invited paper}}$ IEEE Transactions on Applied Superconductivity (Volume:26, Issue:3), Article number: 4700607, 2015 \\${\bf{DOI}}$: 10.1109/TASC.2016.2523758
\end{itemize}
	
%%%%%%%%%%%%%%%%%%%%%%%%%%%%%%%%%%%%%%%%%%%%%%%%%%%%%%%%%%%%%%%%%%%%%%%%%%%%%%%%%%%%%%%%%%%%%%%
%%%%%%%%%%%%%%%%%%%%%%%%%%%%%%%%%%%%%%%%%%%%%%%%%%%%%%%%%%%%%%%%%%%%%%%%%%%%%%%%%%%%%%%%%%%%%%%
%%%%%%%%%%%%%%%%%%%%%%%%%%%%%%%%%%%%%%%%%%%%%%%%%%%%%%%%%%%%%%%%%%%%%%%%%%%%%%%%%%%%%%%%%%%%%%%
%%%%%%%%%%%%%%%%%%%%%%%%%%%%%%%%%%%%%%%%%%%%%%%%%%%%%%%%%%%%%%%%%%%%%%%%%%%%%%%%%%%%%%%%%%%%%%%

%\bibliographystyle{IEEEtran}	
%\bibliography{all.bib}


%%%%%%%%%%%%%%%%%%%%%%%%%%%%%%%%%%%%%%%%%%%%%%%%%%%%%%%%%%%%%%%%%%%%%%%%%%%%%%%%%%%%%%%%%%%%%%%

\end{document}